\documentclass[10pt]{article}

\usepackage{array}
\usepackage{color} 
\usepackage{mathrsfs}
\usepackage{graphicx}
\usepackage{amsmath}
\usepackage{bbm}
\usepackage{amsfonts}
\usepackage{amssymb}
\usepackage{amsbsy}
\usepackage[T1]{fontenc}
\usepackage{theorem}
\usepackage{stmaryrd}
\usepackage{tipa}
\usepackage{textcomp}
\usepackage{subfigure}
\usepackage{epsfig}
\usepackage[sort,colon]{natbib}
\usepackage{lmodern}
\usepackage{framed}
\usepackage[subfigure]{tocloft}
\usepackage{subeqnarray}
\usepackage{alphalph}
\usepackage[refpage,cfg,prefix]{nomencl}
\usepackage{multirow}
\usepackage{xifthen}
\usepackage{enumerate}
\usepackage{color}
\usepackage{wasysym}
\usepackage{bibentry}
\usepackage{todonotes}
\nobibliography*
\usepackage{verbatim}
\usepackage{booktabs}
\usepackage[hyperindex,breaklinks]{hyperref}
\hypersetup{
colorlinks=true,
urlcolor=blue,
linkcolor=black,
citecolor=black,
}
\usepackage{breakurl}
\usepackage{url}
\usepackage{setspace}
\usepackage{xr}
\usepackage{wrapfig}
\usepackage{mathtools}
\usepackage{dsfont}
\usepackage[toc,page]{appendix}

\makenomenclature

\usepackage[labelfont=bf,labelsep=period,justification=raggedright]{caption}

\makeatletter
\renewcommand{\@biblabel}[1]{\quad#1.}
\makeatother

\date{}

\newcommand{\ud}{\mbox{d}}

\newcommand{\bs}[1]{\boldsymbol{#1}}

\newcommand{\Var}{\mathbb{V}\text{ar}}
\newcommand{\partder}[2]{\frac{\partial #1}{\partial #2}}
\newcommand{\ordder}[2]{\frac{\text{d} #1}{\text{d} #2}}
\newcommand{\secpartder}[2]{\frac{\partial^2 #1}{\partial #2 ^2}}
\newcommand{\set}[2]{\left\{#1:#2\right\}}
\newcommand{\smallsub}[2]{#1_{\text{{\tiny #2}}}}
\newcommand{\Deltat}{\Delta t}

\newcommand{\listofalgorithms}{\textbf{\Huge{List of Algorithms}}}
\newlistof{Algorithm}{exp}{\listofalgorithms}
\newcounter{instructioncounter}
 
\newenvironment{Algorithm}[2][]
{\refstepcounter{Algorithm}
\begin{framed}\addcontentsline{exp}{Algorithm}{\protect\numberline{\theAlgorithm} #1}\par\begin{center}\textbf{Algorithm \theAlgorithm : #2}\end{center}\begin{list}
{\bf{(\arabic{Algorithm}\alph{instructioncounter}})}{\usecounter{instructioncounter}}}{\end{list}\end{framed}}
\graphicspath{{./Images/}}

\begin{document}

\begin{flushleft}
{\Large
\textbf{The auxiliary region method: A hybrid method for coupling PDE- and Brownian-based dynamics for reaction-diffusion systems}
}
\\
Cameron A. Smith$^{1,\ast}$, 
Christian A. Yates$^{1,\ast\ast}$
\\
$^1$\textbf{Centre for Mathematical Biology, Department of Mathematical Sciences, University of Bath, Claverton Down, Bath, BA2 7AY, United Kingdom \\
$\ast$ E-mail: c.smith3@bath.ac.uk \\
$\ast\ast$ E-mail: c.yates@bath.ac.uk}
\end{flushleft}


\section*{Abstract}
Reaction-diffusion systems are used to represent many biological and physical phenomena. They model the random motion of particles (diffusion) and interactions between them (reactions). Such systems can be modelled at multiple scales with varying degrees of accuracy and computational efficiency. When representing genuinely multiscale phenomena, fine-scale models can be prohibitively expensive, whereas coarser models, although cheaper, often lack sufficient detail to accurately represent the phenomenon at hand.  Spatial hybrid methods couple two or more of these representations in order to improve efficiency without compromising accuracy. 

In this paper, we present a novel spatial hybrid method, which we call the auxiliary region method (ARM), which couples PDE and Brownian-based representations of reaction-diffusion systems. Numerical PDE solutions on one side of an interface are coupled to Brownian-based dynamics on the other side using compartment-based ``auxiliary regions''. We demonstrate that the hybrid method is able to simulate reaction-diffusion dynamics for a number of different test problems with high accuracy. Further, we undertake error analysis on the ARM which demonstrates that it is robust to changes in the free parameters in the model, where previous coupling algorithms are not. In particular, we envisage that the method will be applicable for a wide range of spatial multi-scales problems including, filopodial dynamics, intracellular signalling, embryogenesis and travelling wave phenomena.  

\section{Introduction} \label{sect:Introduction}
%Some biological motivation. 
Reaction-diffusion models are important mathematical tools that are used to represent and understand complex biological and physical behaviours. They model the random movement of the particles (diffusion) and the interactions between particles (reactions), giving them a wide array of applications across multiple spatial scales. These applications range from the large-scale representation of striped vegetation patterns in semi-arid landscapes \citep{sherratt2005avs} and the spread of epidemics \citep{volpert2009rdw} to smaller-scale studies of pattern formation during embryogenesis \citep{turing1952cbm,mort2016rdm} and, at even smaller scales, to the study of actin dynamics inside a cell's filopodia \citep{erban2014msr} and intracellular dynamics \citep{khan2011scd,andasari2012iid,zhuge2000dsb}. \label{xr:Refs}

%Introduction to all three methods although playing up the PDE and the Brownian methods.
Reaction-diffusion models can be specified at different levels of detail depending on the temporal, spatial and concentration scales involved in the application (see Table \ref{tab:Comparison}). At the finest scale that we will consider are microscopic dynamics. These models and methods \label{xr:GFRD} (which include Brownian motion for purely diffusive systems and Smoluchowski dynamics \citep{andrews2004ssc, smoluchowski1917vem} \label{xr:Andrews} or Green's function reaction dynamics (GFRD) \citep{van2005gfr} for reaction-diffusion systems) are amongst the more detailed representations of such systems, but consequently are relatively computationally expensive\footnote{Throughout this paper, regardless of whether we have interactions between particles or not, we shall refer to models at this microscopic scale as ``Brownian dynamics''.}. \label{xr:GFRD_Smoldyn} They require not only the knowledge of the location of all particles at all times, but in the case of second- and higher-order reactions, the pairwise distances between particles, which requires large memory, and are expensive to calculate for many time-steps. In the case of diffusion-limited reactions, time-steps must be taken to be extremely small to ensure that reactive particles do not jump past each other and that the attendant reaction events are not missed.  All update steps also require the production of a normally distributed random number for each co-ordinate of each particle which can be computationally expensive depending on the reaction system that is being modelled.  However, some of these expensive steps can be accelerated by considering event-driven algorithms or employing approximate algorithms with longer time-steps. GFRD is an event-driven algorithm differs from the standard method for simulating Brownian motion. It uses a maximum time-step so that only single particles, or pairs of particles, need to be considered. It then utilises the exact solution to the Smoluchowski equation in order to combine movement of, and interactions between, particles. If particles are far apart, the event-based time-steps are large. Smoldyn uses relatively long time-steps, and accounts for the error that this causes (due to possible reactant pairs passing by one another without the possibility of reacting) by making the effective particle sizes larger. \label{xr:Ind_Diff} Micro-scale modelling is particularly useful when fine scale detail is required, for example, when considering the binding of particles to receptors \citep{dobramysl2015pbd,english2004bad,moy2000tct}. An even finer scale representation, in which atomistic dynamics can be represented, is available, if required. Typically, modelling at this scale is known as molecular dynamics, and we direct the reader to \citet{holley1971mhp} and \citet{durr1981mmb} for more information about modelling at this scale. \label{xr:Even_Finer_Scale}

%This type of hybrid method is important for many biological application areas, such as that of ion channels \citep{moy2000tct}. Channels open and close when others bind to and unbind from receptors. This action requires detailed knowledge of the locations of all particles in a region around the receptors. However, such detailed knowledge is not required in the bulk of the computational domain, and would cause inefficiencies. This is where a hybrid method would prove to be a useful tool. \citet{dobramysl2015pbd} have used the two-regime method \citep{flegg2012trm}, a mesoscopic-to-microscopic method, in order to simulate a calcium ion channel. However the method presented here may in fact provide even larger computational savings by using a PDE rather that the compartment-based method in the bulk.

At a coarser scale we have compartment-based or mesoscopic models. Like the fine-scale, microscopic models, these also account for stochastic variation, however particles are now considered to belong to compartments rather than having their exact locations tracked. Particles can either react with one another within a compartment, or can jump between adjacent compartments with given rates, simulating diffusion. Compartment-based models can be simulated using either exact but computationally expensive \citep{gillespie1977ess,gibson2000ees,elf2004ssb} or inexact but computationally cheaper \citep{gillespie2001aas} stochastic simulation algorithms (SSAs). The exact methods (so-called because they produce sample paths consistent with the associated chemical master equation) effectively assign exponential waiting times to every possible event (diffusive jump or reaction) and then choose the event with the shortest waiting time to enact. In general, they are faster than the microscopic methods, since pairwise reaction distances do not need to be calculated for bi-molecular reactions and individual particle identities are not tracked, but are less accurate, since they only record a particle's location up to the accuracy of the compartment size, and generally particles are only allowed to react with others in the same compartment \citep{isaacson2013crd}.

Finally, at the coarsest scale lie continuum or macroscopic models. The most commonly employed macroscopic models for reaction-diffusion systems comprise partial differential equations (PDEs)\footnote{However, with the increasing awareness of the importance of randomness, stochastic partial differential equations (SPDEs) are also becoming popular macroscopic representations.}. These methods are generally only valid for high particle numbers. The stochastic variations, which are considered small enough to be neglected at high copy numbers, play a pivotal role in the dynamics at low copy numbers, leading the PDE solutions to diverge from the true underlying dynamics. There is a wealth of well established numerical methods that can quickly simulate an approximate solution to a PDE. These include finite-difference methods, finite-volume methods and finite-element methods (see for example, \citep{smith1985nsp,morton2005nsp,eymard2000fvm,brenner2004fem}).

\begin{table}[h!]
\centering
\begin{tabular}{|p{1.75cm}|p{6.5cm}|p{6.5cm}|}\hline  
\textbf{Scale} & \textbf{Advantages} & \textbf{Disadvantages} \\\hline
Micro & Most accurate representation.\newline Can be used for low copy numbers. & Slow to compute reactions.\newline Impractical for large numbers of particles.\\ \hline
Meso & Fast for low particle numbers. \newline
Represents individual-level behaviour. &
 Can be slow for large copy numbers.\newline
Does not retain precise location or particle identity.\\\hline
Macro & Fast to compute solutions.\newline Suitable for high copy numbers.\newline Often amenable to analytical solutions. & Inaccurate for low copy numbers.\newline Mean-field models diverge from individual dynamics for higher-order reactions.\\  \hline
\end{tabular}
\caption{A comparison of the advantages and disadvantages of the three most prominent scales at which reaction-diffusion processes are modelled.}
\label{tab:Comparison}
\end{table} 

Often though, important biological and physical phenomena are genuinely multiscale \citep{markevich2004ssb,black2012sfe,gillespie2013psa,robinson2014atr}. In spatial reaction-diffusion systems, concentration may vary over orders of magnitude. In regions of low concentration it is often important to employ detailed individual-based models in order to correctly represent the dynamics. If these models were to be employed indiscriminately throughout the domain, however, the regions of high concentration, in which there are many individual particles to be evolved, might render the system computationally intractable. In these regions, it might be acceptable to employ a coarser and less computationally expensive model. A canonical example of this phenomenon is the stochastic Fisher wave \citep{breuer1995mls,breuer1994few}. The wave speed is determined by the stochastic activity at the pulled front, so it is important to employ an accurate individual-based representation of the dynamics in this region. Conversely, behind the wave front, the detailed dynamics are of little importance. It is possible, therefore, to employ a coarser, cheaper representation of the dynamics in this region.

% Other spatially extended hybrid algorithms.
% Include more detail on the GC and PC methods - not report level detail however (1 para between both).
Spatially coupled hybrid methods have been developed for precisely this purpose: to simulate spatially inhomogeneous domains both accurately and efficiently. In general, such methods are designed to accelerate expensive computations whilst maintaining reasonable levels of accuracy. The majority of spatially coupled hybrid methods divide the computational domain into distinct regions using interfaces. The dynamics of adjacent regions are represented using different methods.  Regions in which detailed representations of the dynamics are required for accuracy are simulated using a fine-scale method, whereas regions in which less detail is required are modelled with a coarser, less computationally expensive method. There can be two reasons for this. The first is in order to resolve a particular region of the spatial domain in more detail, such as when looking at the behaviour of ions around gated channels \citep{dobramysl2015pbd}, or when building a model for the energy in a liquid crystal \citep{robinson2017mcm}. Both of these examples have a prohibitively slow but accurate model that is required in certain regions of space, but which is too computationally expensive be used everywhere. The second reason is to simply segregate a region of the domain in which there are very few particle numbers. In these regions a coarse method (for example a continuum model) may be too inaccurate. \label{xr:Usefulness}

There exist hybrid methods that couple each of the different scales described above to one another (and indeed many more, see \citet{smith2018seh} for a comprehensive review of such methods). Macroscopic-to-mesoscopic methods have been proposed which employ averaged fluxes in order to calculate appropriate boundary conditions for each regime at the interface(s) \citep{wagner2004hcf,moro2004hms,harrison2016hac}, as well as using an extra compartment within the macroscopic region \citep{yates2015pcm}. Mesoscopic-to-microscopic methods, which also employ extra compartments, this time in the microscopic regime, have been developed \citep{flegg2015cmc}, and a class of methods using adapted rates of diffusion from the mesoscopic to the microscopic domains have been proposed and successfully applied to represent biological processes \citep{flegg2012trm,robinson2014atr,flegg2014atr,dobramysl2015pbd,erban2014msr}. There are fewer macroscopic-to-microscopic hybrid methods in the literature. Macro-to-micro methods that allow mass to flow over the interface in both directions in order to initialise particles \citep{franz2012mrd} or that average solutions on either side of the interface to find a flux \citep{alexander2002ars} can be found in the literature. For a more detailed review of spatially extended hybrid methods, see \citep{smith2018seh}.

\begin{figure}[h!!!!!]
	\begin{center} 
	\subfigure[][]{
		\includegraphics[width=0.48\textwidth,trim={150pt 120pt 100pt 150pt},clip]{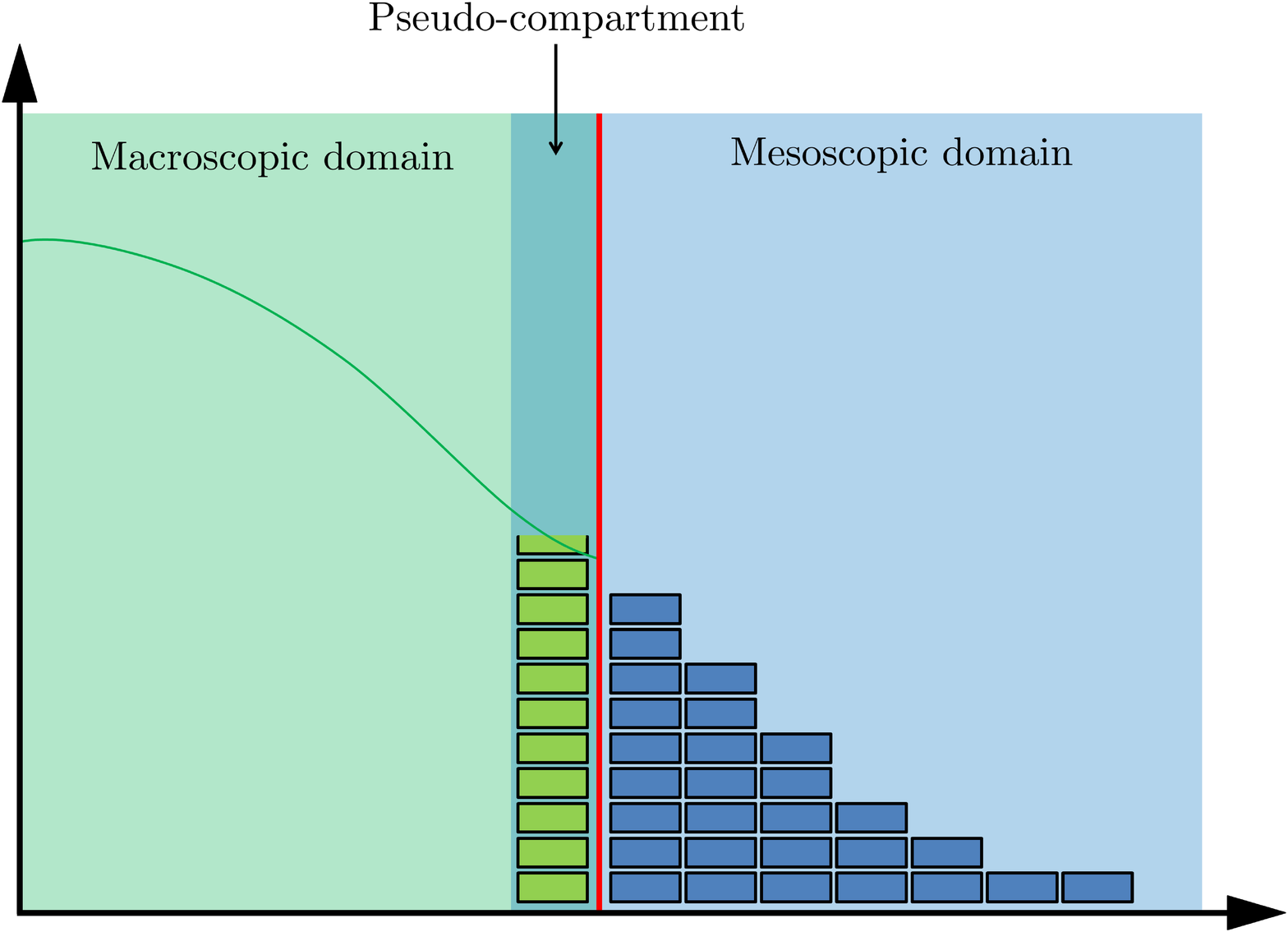}
		\label{fig:yates2015pcm}
	}
	\subfigure[][]{
		\includegraphics[width=0.48\textwidth,trim={150pt 120pt 100pt 150pt},clip]{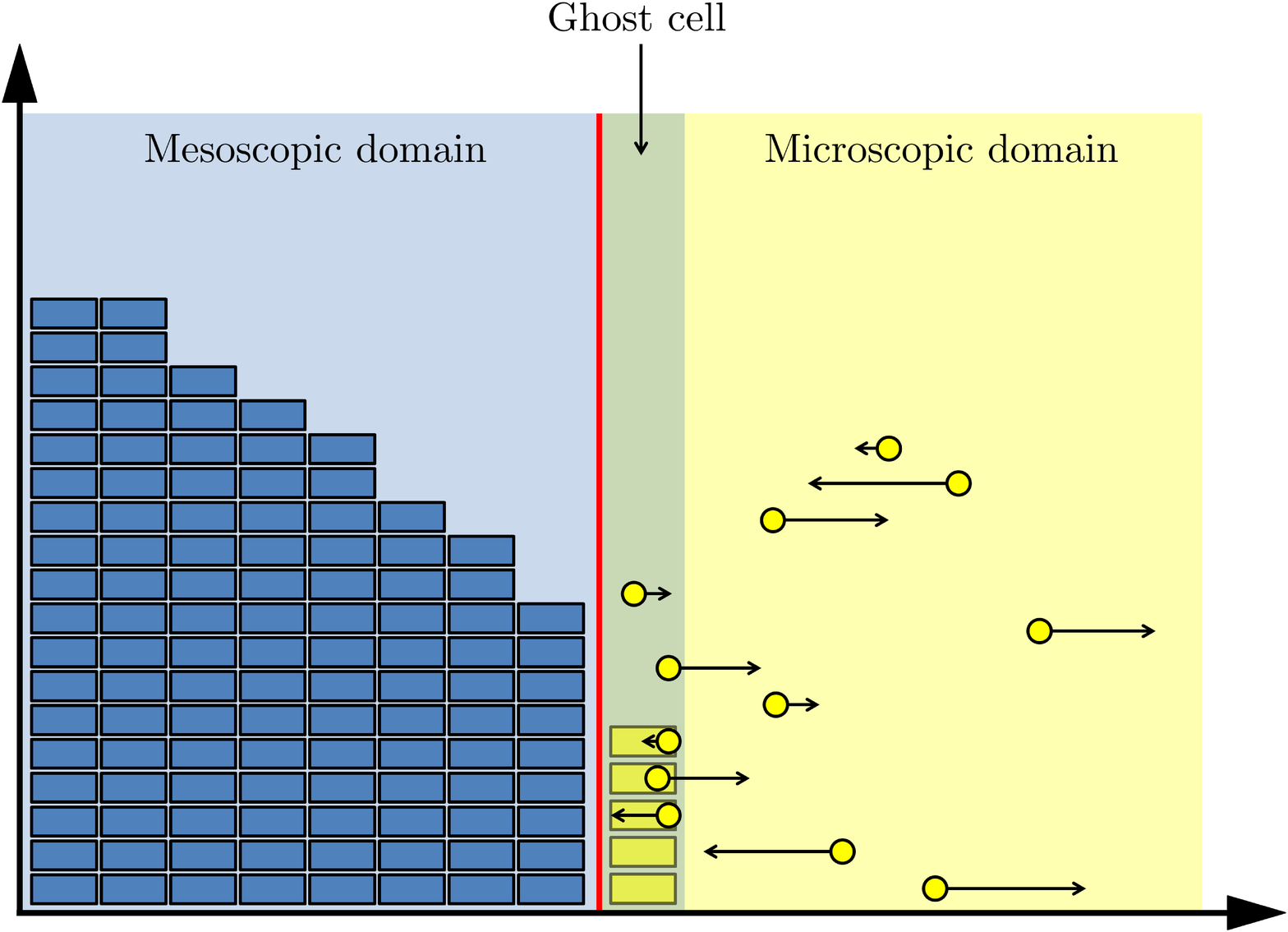}
		\label{fig:flegg2015cmc}
	}
	\caption{Schematics for \subref{fig:yates2015pcm} the pseudo-compartment method and the \subref{fig:flegg2015cmc} the ghost cell method. The green line represents the PDE solution, the blue boxes the number of particles in the mesoscopic region of the respective hybrid methods, and the yellow dots denote the Brownian particles. These particles are shown with a volume, but in the simulations do not have a mass or volume. In the scenario illustrated, the particles reside on the one-dimensional line, but have been illustrated in the plane in order to show the directions and magnitudes of their next movement clearly (black arrows). The green boxes in \subref{fig:yates2015pcm} denote the number of particles in the pseudo-compartment, and similarly, the yellow boxes in \subref{fig:flegg2015cmc} are the number of particles in the ghost cell, with each box representing a single particle. In each case, the red line denotes the point interface between the two regimes.}
	\label{fig:Plots_yates_flegg}
\end{center}
\end{figure}

Two of the above-mentioned hybrid methods are of particular relevance for the purposes of this paper. The pseudo-compartment method, presented by \citet{yates2015pcm}, is a macroscopic-to-mesoscopic (specifically PDE-to-compartment) method in which the coupling is achieved using an extra compartment, known as the ``pseudo-compartment'', adjacent to the interface within the macroscopic domain. In this compartment, mass is represented using both the PDE solution and the compartment-based method (with particle numbers found by direct integration of the PDE over this region). Particles are then allowed to cross the interface in both directions using the compartment-based method. We give a schematic representation of this method in Figure \ref{fig:Plots_yates_flegg} \subref{fig:yates2015pcm}.

The ghost cell method proposed by \citet{flegg2015cmc} is a mesoscopic-to-microscopic method which uses an extra compartment in the microscopic domain. The number of particles in this ``ghost cell'' is simply the number of Brownian particles which reside in this region. Again, particles are allowed to jump across the interface using the compartment-based mesoscopic method. A schematic representation of the method is given in Figure \ref{fig:Plots_yates_flegg} \subref{fig:flegg2015cmc}.

In this paper, we employ these two methods in order to couple a macroscopic PDE description for reaction-diffusion systems to a corresponding microscopic Brownian dynamics representation through the use of ``auxiliary regions''. These regions are compartments, which lie either side of the interface, and allow mass to pass between the two regimes via a mesoscopic jump process (see Figure \ref{fig:Schematic} on page \pageref{fig:Schematic} for a schematic representation). Within the auxiliary regions, mass is simultaneously represented using both the description for the region in which they reside (i.e. PDE or Brownian) and the mesoscopic description. Changes (i.e. reactions or diffusion events) implemented under one modelling paradigm (e.g. the compartment-based representation of the auxiliary region) are simultaneously implemented in the other (e.g. the PDE or Brownian representations in these regions). The interface, which divides the two modelling paradigms, can either be static, in which case it remains in its initial position, or adaptive, in which case it moves with the density profile in order to ensure that regions of space with few particles are simulated using the finest scale. Through a series of test cases, we demonstrate our algorithm to be more accurate and more robust to model parameters than previous PDE-to-Brownian coupling algorithms. 

%Methods that couple Brownian dynamics to a PDE are particularly useful in several scenarios. For example, if there is a need to resolve a region of space at a very fine level (such as the region of space in close proximity to an ion channel \citep{dobramysl2015pbd}) or the coarser scale dynamics fail to resolve some of the details of the physical process, such as nucleation of liquid crystals \citep{robinson2017mcm}, we can split space and use the appropriate modelling paradigm. Furthermore, if there is a steep concentration gradient, for example, with a travelling wave \citep{moro2004hms}, we may want to use a finer scale method in front of the wave in order to resolve the wave speed correctly. \label{xr:Sharp_Interface} \textbf{*** Most of this has already been said four paragraphs ago in the paragraph beginning ``Spatially coupled hybrid methods...'' ***}

% Summary... what happens in paper.
The paper is organised as follows. In Section \ref{sect:Previous}, a previous attempt at hybridising a Brownian dynamics model to its corresponding mean-field PDE description is evaluated in more detail \citep{franz2012mrd}. A description of our novel auxiliary region method (ARM) is presented in Section \ref{sect:ARM} alongside the relevant justifications and pseudocode. Numerical results, verifying the accuracy of our hybrid method, are presented in Section \ref{sect:Results}. Numerical error analysis is conducted in Section \ref{sect:Error}, where we also discuss restrictions on the model parameters for the effective functioning of the coupling algorithm. We conclude with a discussion of the effectiveness of our new hybrid method and suggest avenues for further exploration in Section \ref{sect:Discussion}. 

\section{An existing PDE-to-Brownian coupling} \label{sect:Previous}
%Replicating Franz results.
In this section we summarise the pioneering work of \citet{franz2012mrd}, who were among the first to couple PDE and Brownian dynamics representations of reaction-diffusion. By replicating their results, we demonstrate that their ``PDE-assisted Brownian dynamics'' algorithm is not robust to simulation parameter choice, even for simple diffusive processes. This motivates the need for a more robust coupling method, which we provide in the form of the ARM in Section \ref{sect:ARM}.

\subsection{PDE-assisted Brownian dynamics}\label{section:_PDE_assisted_brownian_dyanamics}

Hybrid methods that couple the PDE description of a reaction-diffusion system to its corresponding Brownian dynamics representation have been relatively poorly investigated in comparison to PDE-to-compartment-based and compartment-based-to-Brownian couplings. In part, this is a result of the fact that such hybrid algorithms neglect meso-scale representations of particle dynamics, meaning that they must bridge a greater scale separation than either of the other two hybrid paradigms. Mainly though, the absence of many examples of PDE-to-Brownian hybrid methods is due to the inherent difficulty when converting PDE mass to individual particles (and vice-versa) when coupling Brownian dynamics models to continuum PDE representations. Below, we describe two algorithms proposed by \citet{franz2012mrd}, but focus on the first, a method with an interfacial coupling. We choose to focus on this coupling because our ARM coupling method, described in Section \ref{sect:ARM}, also utilises an interface. \label{xr:Franz_Interface}

\begin{figure}[h!!!!!!!!!!!]
\centering
\includegraphics[width=0.8\textwidth,trim={150pt 120pt 100pt 150pt},clip]{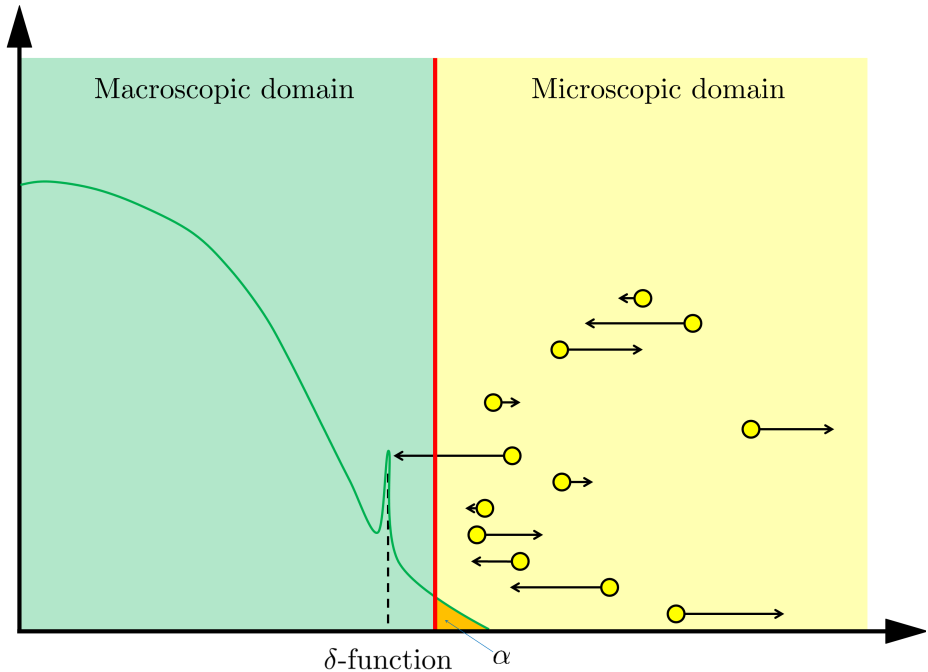}
\caption{A schematic for the method proposed by \citet{franz2012mrd}. Descriptions are as in Figure \ref{fig:Plots_yates_flegg}. The PDE mass labelled $\alpha$ (orange) is the density of PDE mass that has flowed over the interface in the Brownian update-step. The peak in the PDE curve near the interface represents the addition of a $\delta$-function corresponding to a Brownian particle that crosses the interface.}
\label{fig:franz2012mrd}
\end{figure}

\citet{franz2012mrd} present two related algorithms. In the first, the non-overlapping PDE and Brownian domains are separated by an interface (see Figure \ref{fig:franz2012mrd}). Both PDE and Brownian representations are updated using a time-driven algorithm, with the PDE time-step much smaller than the Brownian time-step. The discretised PDE is evolved (until the time reaches the next Brownian time-step) using a centred finite-difference scheme with implicit Euler time-stepping, and PDE mass is allowed to cross the interface between the two regimes. Provided that the Brownian time-step is sufficiently small, the amount of mass that crosses the interface between Brownian time-steps gives the probability that a new particle is placed within the Brownian domain. A uniformly distributed random number is used to determine whether a particle is initialised in the Brownian regime or not. If it is, this particle's position is randomly initialised according to the normalised density profile of the PDE mass that crossed the interface in the previous Brownian time-step. If a Brownian particle crosses into the PDE domain, a particle's worth of mass is added to the PDE solution at its new location as a $\delta$-function and the individual particle is removed. We have illustrated this method schematically in Figure \ref{fig:franz2012mrd}.

\citet{franz2012mrd} found the variance in particle numbers in the Brownian region of the hybrid domain to be altered in comparison to the variance that would be expected in a fully Brownian simulation. In order to counteract this problem, they introduced a second algorithm, in which an overlap region replaces the interface. Within the overlap region, mass can be simulated as either Brownian particles or as part of the PDE. The coupling works in the same way as in the interfacing algorithm, however the Brownian particles are subsumed into the PDE only once they have crossed the boundary of the overlap region closest to the fully-PDE domain. Similarly, PDE mass can only be converted to Brownian particles once it has flowed over the overlap boundary adjacent to the fully-Brownian domain. \label{xr:Franz_Overlap}

The Brownian time-step in the algorithm is required to be small, in order that the total probability of initialising a particle in the Brownian regime is less than one. However, the algorithm runs into difficulties if the time-step is chosen to be too small. Specifically, the amount of mass that flows over the interface between updates of the Brownian dynamics is too small in comparison to that which would be predicted theoretically using the exact diffusion kernel. This gives rise to inaccuracies in the algorithm, particularly if long simulation times are required. This sensitivity to the choice of Brownian time-step restricts the physical scenarios to which the algorithm can be applied. 

\begin{figure}[h!!!!!]
	\begin{center} 
	\subfigure[][]{
		\includegraphics[width=0.31\textwidth,trim={20pt 0pt 5pt 0pt},clip]{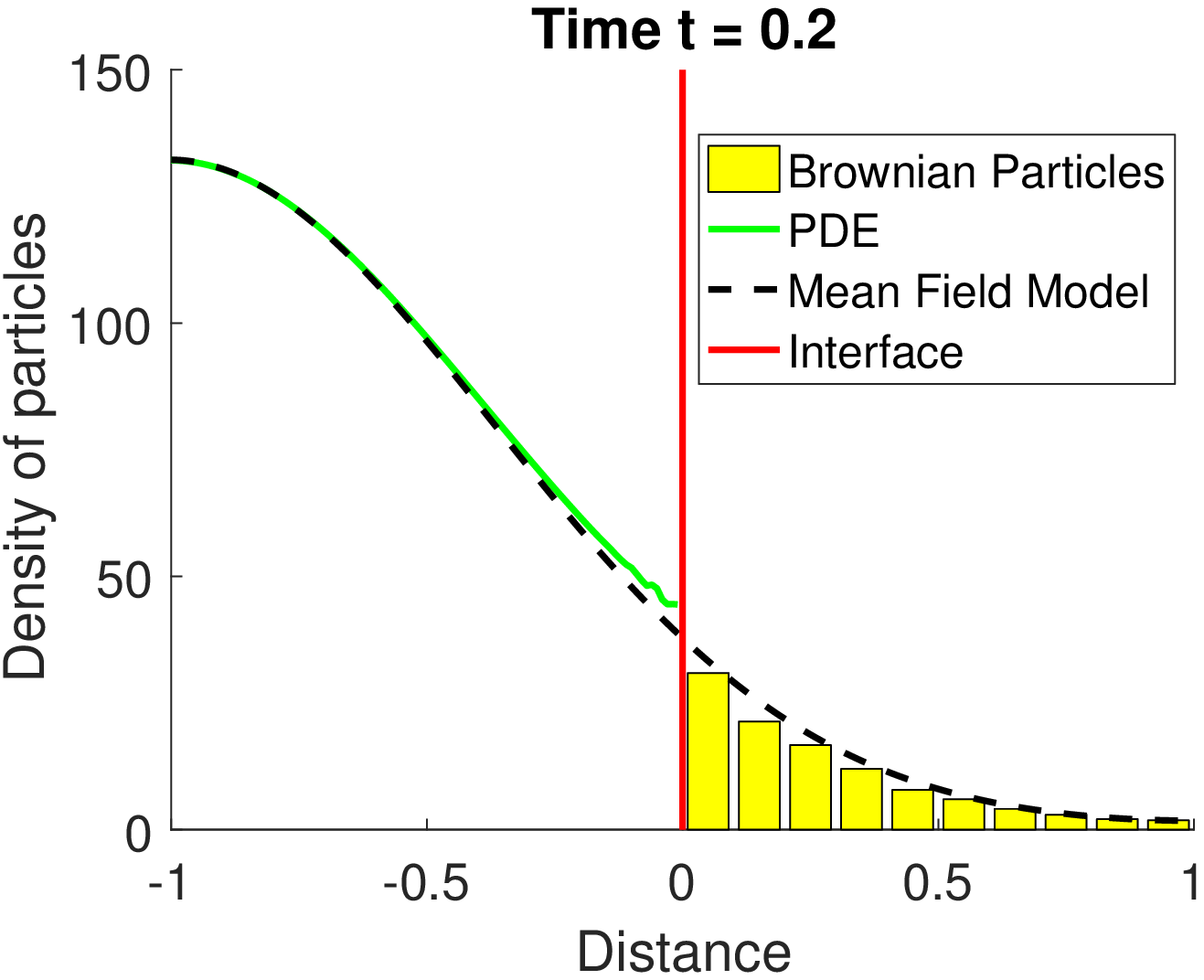}
		\label{fig:franz_0.20}
	}
	\subfigure[][]{
		\includegraphics[width=0.31\textwidth,trim={20pt 0pt 5pt 0pt},clip]{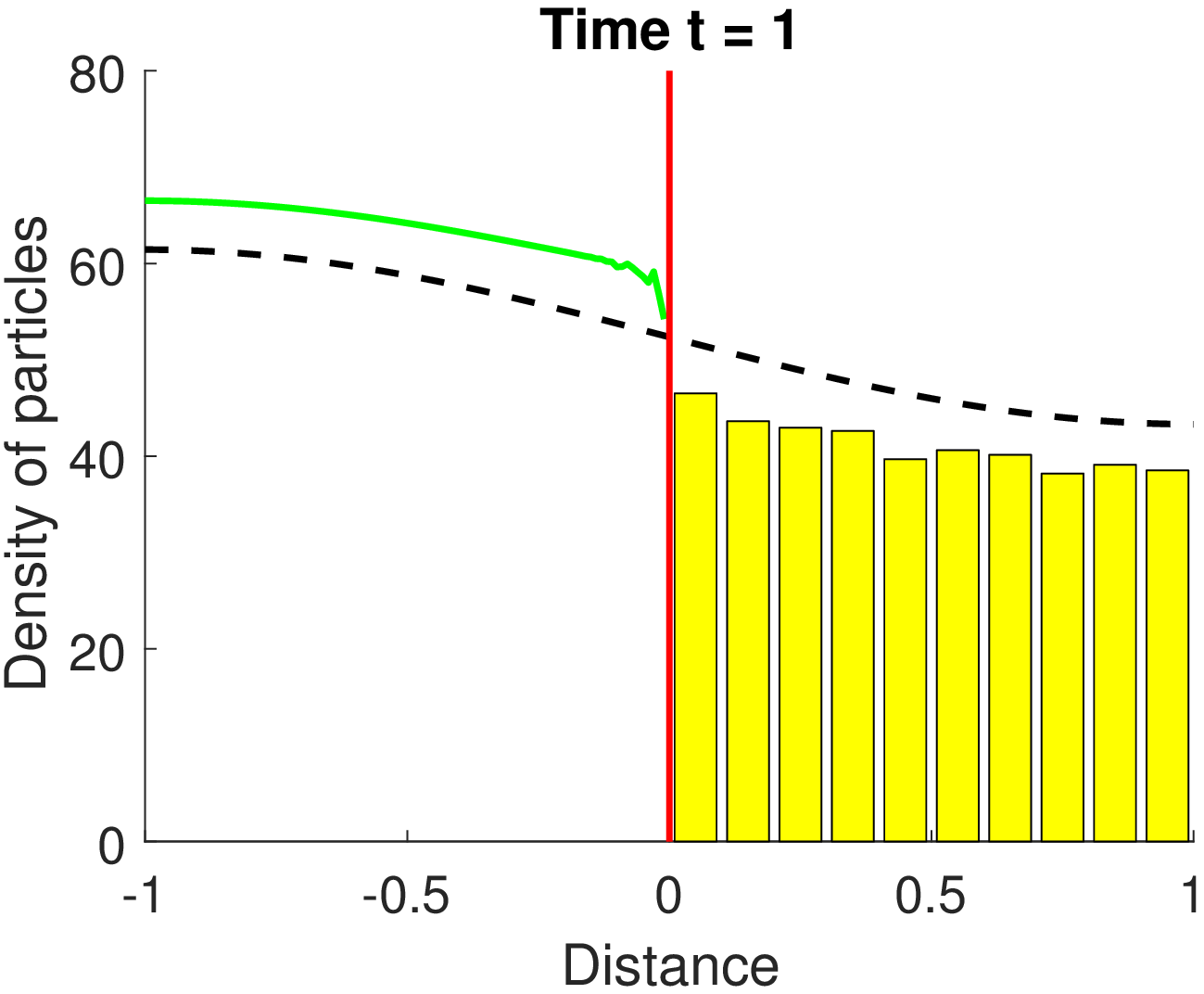}
		\label{fig:franz_1.00}
	}
	\subfigure[][]{
		\includegraphics[width=0.31\textwidth,trim={20pt 0pt 5pt 0pt},clip]{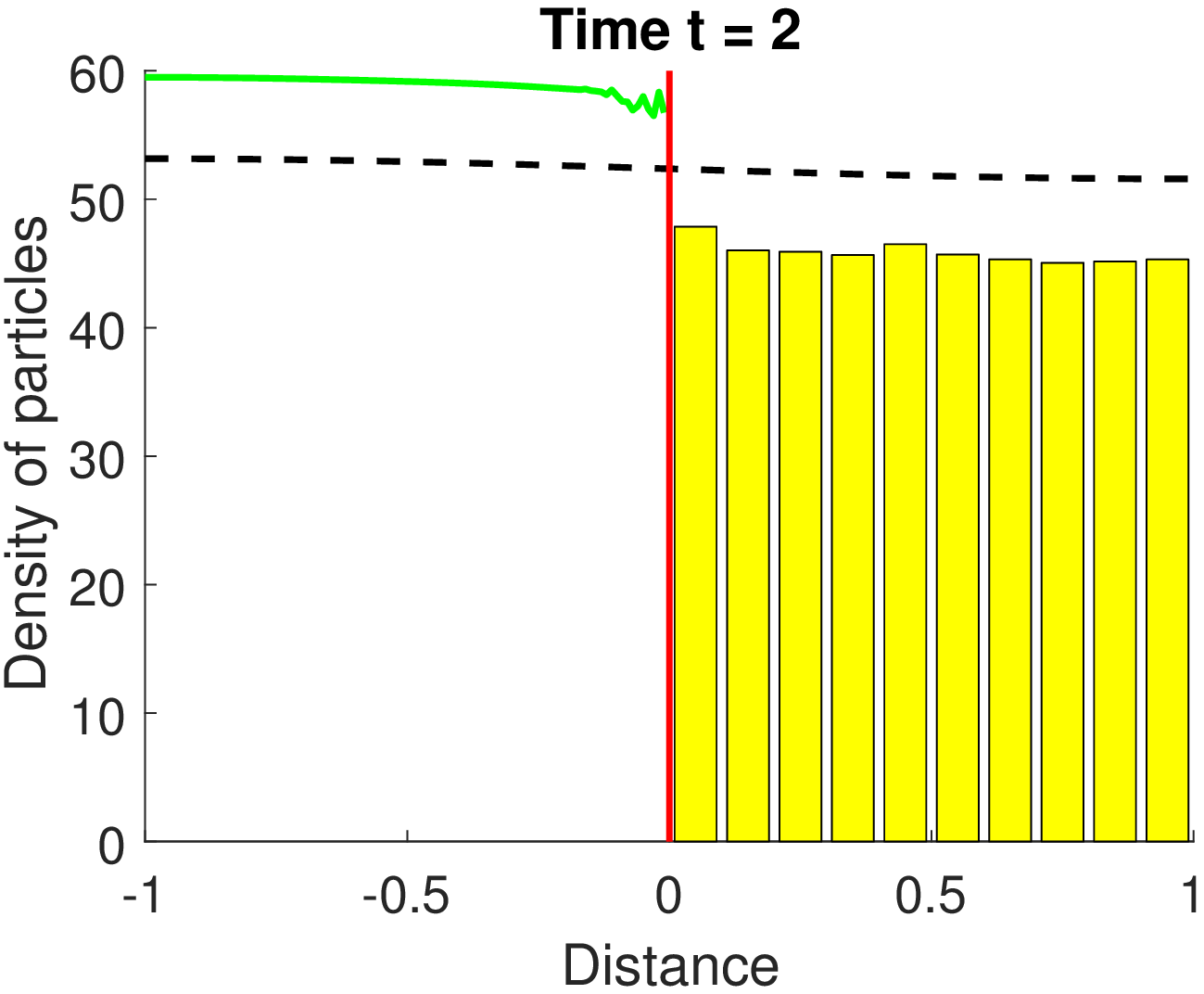}
		\label{fig:franz_2.00}
	}
	\caption{The evolution of 100 particles worth of mass initialised at a single PDE point at position $x=-0.95$, at times \subref{fig:Plots_Unif_1} 0.2, \subref{fig:Plots_Unif_2} 1 and \subref{fig:Plots_Unif_3} 2 under the first hybrid method of \citet{franz2012mrd}. The green line is the PDE part of the hybrid method, the yellow bars represent the Brownian dynamics for the hybrid method (appropriately binned for visualisation purposes), the red line is the interface and the black dashed line is the solution of the mean-field diffusion equation (equation \eqref{eqn:pure_diffusion_mean_field_model}). Results shown are averaged over 100 repeats.}
	\label{fig:Plots_franz}
\end{center}
\end{figure}

In figure \ref{fig:Plots_franz} we present three snapshots of the evolution of the first version of the algorithm (interface rather than overlap region) which illustrate this problem. By time $t=2$, in Figure \ref{fig:Plots_franz} \subref{fig:franz_2.00}, there is a clear disparity between the hybrid method and the mean field solution (black dotted line).  Disparities of this nature are not acceptable when modelling real reaction-diffusion systems, irrespective of the computational savings the algorithm is able to produce.

%Showing that there is a problem with larger values of the final time.
%Where does this leave us? - need a methodology.
%Geyer and Gorba papers and their limitations (i.e. thermal bath).
%Alexander

\section{The auxiliary region method} \label{sect:ARM}
In this section we present our novel ``auxiliary region method'' (ARM) for coupling PDE and Brownian representations of reaction-diffusion. For simplicity we will present a version of the method with a single interface separating two regimes. However, the method can be easily generalised to multiple interfaces which separate alternating PDE and Brownian regions. Sequentially, we describe the composition of the domain and the models we employ in each region; the nature of the auxiliary regions; the implementation of movement of mass across the boundary; the implementation of reactions; and finally the specific details required for the simulation of the algorithm, including pseudocode for its implementation. All code, which has been written in MATLAB, can be found in the electronic supplementary material online. 

\subsection{The domain composition} \label{sect:ARM_description}
%Verbal description of how the algorithm works.
Recall that, for our coupling method, space is partitioned into two regions within which we use different modelling paradigms (PDE and Brownian dynamics) to simulate the underlying reaction-diffusion system. Separating the two regions is a point interface, over which particles can jump according to a compartment-based method. 

Consider a one-dimensional domain\footnote{Note that the method can be extended to higher dimensions with (hyper-)planar interfaces in a straight-forward manner.} $\Omega=(L_1,L_2)\subseteq\mathbb{R}$ for some $L_1<0<L_2$. We split $\Omega$ into two regions, $\smallsub{\Omega}{P}=(L_1,0)$ and $\smallsub{\Omega}{B}=(0,L_2)$ (separated by an interface $I$ at position $0$), within which the evolution of the system will be represented using a PDE description and Brownian dynamics, respectively. 
%$\Omega_{\text{P}}$ and $\Omega_{\text{B}}$ are such that $\Omega_{\text{P}}\cap\Omega_{\text{B}}=\emptyset$ and $\bar{\Omega}_P\cup\bar{\Omega}_B=\Omega$. We then define the interface as the single point $I$ contained in $\bar{\Omega}_P\cap\bar{\Omega}_B$. 

\begin{figure}[h!!!!!!!!!]
\centering
\includegraphics[width=0.8\textwidth,trim={150pt 120pt 100pt 150pt},clip]{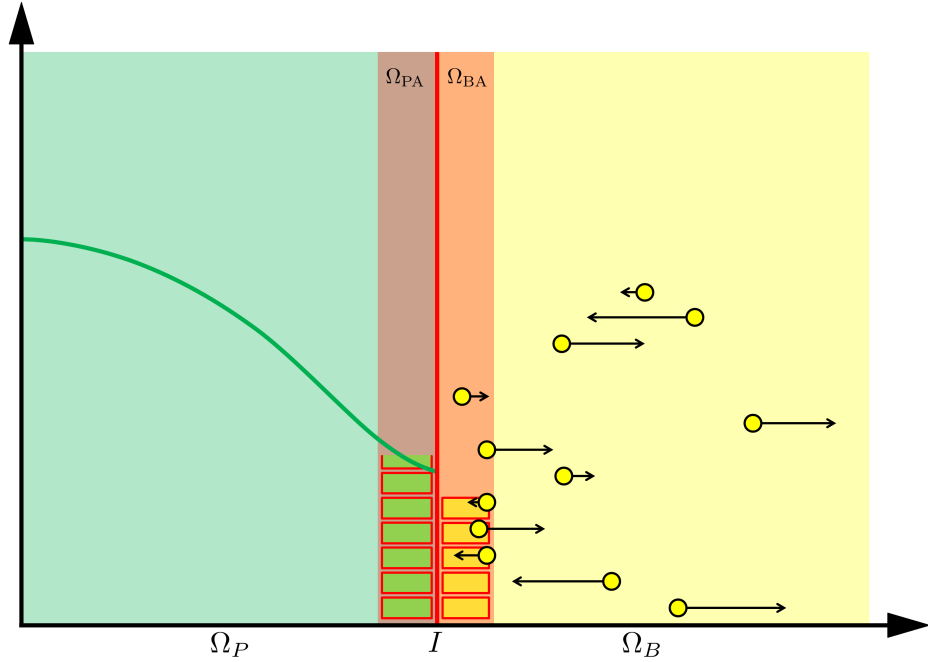}
\caption{A schematic for the auxiliary region method (ARM). Descriptions as in Figure \ref{fig:Plots_yates_flegg}. The interface is the red line in the centre and the two auxiliary regions are shown with blocks to indicate the number of particles residing within them. In the PDE and Brownian auxiliary regions, each block represents a particle in the compartment-based representation and the number of blocks is determined by integrating the PDE over the auxiliary region $\smallsub{\Omega}{PA}$, and counting the number of Brownian particles in $\smallsub{\Omega}{BA}$, respectively.}
\label{fig:Schematic}
\end{figure}

\subsection{The auxiliary regions}
Particles can move between the two domains ($\smallsub{\Omega}{P}$ and $\smallsub{\Omega}{B}$) via the auxiliary regions $\smallsub{\Omega}{PA}$ and $\smallsub{\Omega}{BA}$; subsets of $\smallsub{\Omega}{P}$ and $\smallsub{\Omega}{B}$ respectively, each of width $h_a>0$. Within these regions, mass/particles are simultaneously represented according to the default methodology for their domain (either PDE in $\smallsub{\Omega}{P}$ or Brownian dynamics in $\smallsub{\Omega}{B}$), but also as well-mixed particles in their respective auxiliary regions $\smallsub{\Omega}{PA}$ and $\smallsub{\Omega}{BA}$. These auxiliary regions act as a bridge between the fine- and coarse-scale descriptions. A schematic representation of domain's composition is given in Figure \ref{fig:Schematic}.

We justify the use of the Brownian auxiliary region by following the methodology set out in \citet{flegg2015cmc}. The entire Brownian domain can be simulated using a mesoscopic compartment-based regime, and equivalently using a microscopic simulation. In the absence of reactions, if the particles in the microscopic simulation are ``binned'' into the same compartments as the mesoscopic simulation, the expected numbers in each compartment for each simulation would be the same. At this scale, the two methods are equivalent ways of simulating the same diffusive process \citep{flegg2015cmc}. 

To justify the use of the PDE auxiliary region, we appeal to the arguments of \citet{yates2015pcm}. We note that the PDE density can be thought of as the probability of finding a particle at a particular position and time, scaled by the number of particles within the PDE domain. Provided that the auxiliary region is sufficiently narrow, the PDE density within the auxiliary region can be thought of as being approximately uniformly distributed across the region with the appropriate number of particles. This is precisely the interpretation of the contents of a compartment within the mesoscopic, compartment-based framework.

\subsection{The PDE regime, $\smallsub{\Omega}{P}$} \label{sect:ARM_PDE}

Within $\Omega_{\text{P}}$, we represent the mass of particles using: 
\begin{align}
\text{PDE}&:\quad \partder{\bs{c}}{t}(x,t)=D\secpartder{\bs{c}}{x}(x,t) + \bs{f}(\bs{c}(x,t));\quad x\in\Omega_{\text{P}};\quad t\in(0,T),\label{eqn:PDE}\\
\text{BCs}&:\quad \partder{\bs{c}}{x}(x,t) = 0; \quad x\in\partial\smallsub{\Omega}{P}; \quad t\in(0,T), \notag \\
\text{IC}&:\quad \bs{c}(x,0) = \bs{c}_0(x); \quad x\in \smallsub{\Omega}{P}. \notag
\end{align} 
Here, $\bs{c}(x,t)=(c_1(x,t),...,\smallsub{c}{K}(x,t))^T$, denotes the density of species $k={1,\dots,K}$ at position $x$ and time $t$, $D$ is a diagonal matrix containing the Fickian diffusion constants for each species, and $\bs{f}$ is a function that encapsulates the effect of any reactions on each species. We also use the notation $\partial\smallsub{\Omega}{P}$ to represent the boundary of $\smallsub{\Omega}{P}$, and $\bs{c}_0(x)$ is the initial condition. 
%As an example, consider a single species evolving according to the Fisher-Kolmogorov-Petrovsky-Piscounov (FKPP) equation \citep{fisher1937waa} which describes the spread of an advantageous gene, \begin{equation*}
%\bf{f}(c(x,t))=\lambda c(x,t)(1-c(x,t)),
%\end{equation*} where $\lambda$ is a measure of spread.
For all the simulations presented in this paper we employ the finite-difference $\theta$-method (a general family of finite-difference methods)\footnote{Note that this PDE can be simulated using any appropriate numerical solver, including the finite-element method or finite-volume method.}. Although the Crank-Nicolson method  ($\theta=0.5$) is second-order accurate and unconditionally stable, we use $\theta=0.51$ since the Crank-Nicolson method can give rise to spurious oscillations when implemented on step-function initial conditions of the sort we will consider \citep{smith1985nsp}. \label{xr:Instabilities}

\subsection{The Brownian regime, $\smallsub{\Omega}{B}$}

Within $\smallsub{\Omega}{B}$, all particles are tracked and their positions updated according to the following stochastic differential equation (SDE) which simulates Brownian motion: 
\begin{equation}
y_i^k(t+\Deltat) = y_i^k(t) + \sqrt{2D_k\Deltat}\, \xi_i^k; \quad \xi_i^k\sim N(0,1); \quad \text{ for } \quad i\in\{1,...,\smallsub{N}{HB}^k(t)\}  \quad\text{ and } \quad \ k\in\{1,...,K\}, 
\label{eqn:BD}
\end{equation}
where $y_i^k(t)$ denotes the location of particle $i$ of species $k$ within $\smallsub{\Omega}{B}$, $\Deltat$ is the time-step for both the PDE and Brownian dynamics simulators\footnote{Note that there is no requirement for the PDE and Brownian time steps to be the same. In many situation it may be useful to have a significantly finer Brownian time-step than PDE time step in order to accurately resolve the individual-based dynamics. We employ the same time-step in our simulations for simplicity.} and $\smallsub{N}{HB}^k(t)$ is the number of particles of species $k$ in $\smallsub{\Omega}{B}$ at time $t$. Once again, we set reflective boundary conditions at both ends of $\smallsub{\Omega}{B}$ to ensure that no particles can leave this domain via a Brownian diffusion event. The zero-flux boundary conditions at the interface for both PDE and Brownian regimes ensure that mass can only cross the interface according to the compartment-based method.

\subsection{Movement across the interface}

Since both domains, $\smallsub{\Omega}{P}$ and $\smallsub{\Omega}{B}$, have zero-flux boundaries at the interface, particles can only cross over the interface via the auxiliary regions. In effect, these regions comprise a two-compartment reaction-diffusion master equation (RDME) model. Each particle in each auxiliary region jumps to its neighbouring region on the other side of the interface with a rate $d_k$ (for species $k$), which is related to the macroscopic diffusion coefficient (for species $k$), $D_k$, via\begin{equation}
d_k = \frac{D_k}{h_a^2}.
\label{eqn:link_d_D}
\end{equation} 
Here, $h_a$ is the width of each auxiliary region, which is assumed to be the same for both the Brownian and PDE auxiliary regions. In order to implement jumps (or reactions, where necessary) according to the RDME, we require particle numbers.

Borrowing terminology from \citet{yates2015pcm}, the number of ``pseudo-particles'' of species $k$ within the PDE auxiliary region, $\smallsub{\Omega}{PA}$, at time $t$, denoted $\smallsub{N}{PA}^k(t)$, is calculated as 
\begin{equation}
\smallsub{N}{PA}^k(t) = \displaystyle\int_{\smallsub{\Omega}{PA}}{c_k(x,t)\ dx}.
\label{eqn:particle_numbers_PDE_AR}
\end{equation} 
The number of particles of species $k$ in the Brownian auxiliary region, $\smallsub{\Omega}{BA}$, is given by 
\begin{equation}
\smallsub{N}{BA}^k(t) = \left|\set{j}{y_j^k(t)\in \smallsub{\Omega}{BA}}\right|.
\label{eqn:particle_numbers_BD_AR}
\end{equation} 

These particle numbers allow us to define propensity functions corresponding to diffusive jumps between, or reactions within, the auxiliary regions. 
%\textcolor{red}{A propensity function $\alpha(t)$ corresponding to an event, is defined by \begin{equation*}
%\alpha(t)\delta t = \mathbb{P}(\text{Event occurs in }[t,t+\delta t))
%\end{equation*} for any small time-step $\delta t$.}
For diffusive jumps between the two auxiliary regions, the propensity functions for species $k$ within the PDE and Brownian auxiliary regions are (respectively): 
\begin{align}
\smallsub{\alpha}{P}^k(t) &= d_k\smallsub{N}{PA}^k(t)\quad\text{ for }\quad\smallsub{\Omega}{PA}, \label{eqn:propensity_functions_PDE}\\
\smallsub{\alpha}{B}^k(t) &= d_k\smallsub{N}{BA}^k(t)\quad\text{ for }\quad\smallsub{\Omega}{BA}. \label{eqn:propensity_functions_Brown}
\end{align} We note here that if $\smallsub{N}{PA}^k(t) < 1$, we set $\smallsub{\alpha}{P}^k(t) = 0$ to prevent the possibility of negative density. While it may be a problem if this scenario occurs persistently, practically speaking, we should choose the position of the interface such that density is always large enough that this does not happen. An adaptive interface will allow us to satisfy this criteria (see Section \ref{sect:Results_P4_Interface}), and hence this problem would not occur when using such an interface. \label{xr:Negative_Particles}

When a particle jumps from $\smallsub{\Omega}{BA}$ to $\smallsub{\Omega}{PA}$, a particle within the Brownian auxiliary region is chosen uniformly at random to be removed, and a particle's worth of mass is added to the PDE solution uniformly across $\smallsub{\Omega}{PA}$ for the species, $k$, which has changed: \begin{equation}
c_k(x,t) = c_k(x,t) + \frac{1}{h_a}\mathds{1}_{[x\in\smallsub{\Omega}{PA}]},
\label{eqn:add_particle_PDE}
\end{equation} where $\mathds{1}_{[x\in A]}$ is the indicator function for $x\in A$. \label{xr:Indicator} Similarly, if a jump is enacted in the opposite direction, from $\smallsub{\Omega}{PA}$ to $\smallsub{\Omega}{BA}$, we first remove a particle's worth of mass uniformly from $\smallsub{\Omega}{PA}$ for the appropriate species $k$: \begin{equation}
c_k(x,t) = c_k(x,t) - \frac{1}{h_a}\mathds{1}_{[x\in\smallsub{\Omega}{PA}]},
\label{eqn:remove_particle_PDE}
\end{equation} and a new particle is initialised within the Brownian auxiliary region, $\smallsub{\Omega}{BA}$, with position chosen uniformly at random. 

\subsection{Reaction implementation}

Throughout $\smallsub{\Omega}{P}$, all reactions are implemented using the reaction operator $\bs{f}(\bs{c})$.
The method we employ to implement reactions within $\smallsub{\Omega}{B}$ depends on the location of the reactant particles. Let $\mathcal{R}$ denote the set of reaction pathways (with $|\mathcal{R}|=R$). Define the subset of reactions
 $\mathcal{R}^*(t)$ at time $t$ as follows: 

$$\mathcal{R}^*(t)=\{\text{all reactions for which at least one set of reactant particles lies exclusively within }\smallsub{\Omega}{BA}\}.$$

Reactions between molecules for which at least one of the reactive molecules lies within $\smallsub{\Omega}{B}\backslash\smallsub{\Omega}{BA}$ are implemented using an appropriate microscopic approach, such as the $\lambda$-$\rho$ method \citep{erban2009smr,lipkova2011abd}. However, if at least one set of participating particles lie in $\smallsub{\Omega}{BA}$ (i.e. $r\in \mathcal{R}^*$), care needs to be taken over the interaction of such particles and the mass on the other side of the interface in $\smallsub{\Omega}{P}$. As explained below we will implement the reactions $r\in \mathcal{R}^*$ for these reactant particles using the compartment-based method.%(for example, a reversible dissociation reaction that produces multiple products that must be placed outside the common reactive radius). Brownian particles undergoing bimolecular reactions close to the interface should, in theory, be allowed to react with mass in the PDE regime, which would be difficult to implement practically. 

For illustrative purposes, consider a  reversible second-order reaction involving species $A$, $B$ and $C$: 
\begin{equation}
A +B \xrightleftharpoons[\kappa_2]{\kappa_1} C.
\label{eqn:second_order_reaction}
\end{equation} 
Under the $\lambda-\rho$ method \citep{erban2009smr} and its later modification \citep{lipkova2011abd}, for the forward reaction, a particle of species $A$ and a particle of species $B$ are required to be within a distance $\rho$ of one another in order to react. They then react with a rate $\lambda$, where $\lambda$ is a function of both the reaction radius $\rho$ and the reaction rate $\kappa_1$. Imagine that an $A$ particle (without loss of generality) in $\smallsub{\Omega}{B}$ is close enough to the interface that the reaction radius $\rho$ is larger than the distance between itself and the interface. For consistency with the Brownian representation, the $A$ particle should be allowed to react with a $B$ particle in the PDE region. The implementation of such reactions would be extremely difficult. Instead, by ensuring bimolecular reactions within the auxiliary region are implemented according to the mesoscopic compartment-based method, we avoid such issues (provided that the width of the auxiliary region is chosen to be larger than the interaction radius $\rho$).

According to the backwards reaction, two particles are created after the reaction has occurred. These particles are placed a certain distance away from each other (called the dissociation radius) in order to achieve a specified probability of geminate recombination (a recombination of any pair of $A$ and $B$ particle that were initialised from the same $C$ particle). \label{xr:Geminate} If this radius intersects with the PDE regime, then there is the potential for individual particles to be initialised within $\smallsub{\Omega}{P}$. By again employing the mesoscopic representations for reactions we resolve this issue. All product particles are assumed to be placed uniformly throughout the Brownian auxiliary region. Particles that are products of the backwards dissociation reaction in $\smallsub{\Omega}{B}\backslash\smallsub{\Omega}{BA}$ are extremely unlikely to be placed in $\smallsub{\Omega}{P}$ (again, providing that the auxiliary region is larger than the dissociation radius).

For these reasons, all of the reactions $r\in \mathcal{R}^*$ (for which at least one set of participating particles lie in $\smallsub{\Omega}{BA}$) are implemented using the compartment-based method, in which reactions are incorporated as events in the associated Markov chain, according to the RDME.  
We can write the following propensity functions for reactions within $\smallsub{\Omega}{BA}$:
\begin{equation}
\alpha_{r}(t)=g_r(\smallsub{\bs{N}}{BA}(t))\kappa_r h_a^{1-\nu},
\label{eqn:propensity_functions_reactions}
\end{equation} 
for any reaction channel $r\in\mathcal{R}^*(t)$ of order $\nu$ and corresponding reaction rate $\kappa_r$, where $\smallsub{\bs{N}}{BA}(t)=(\smallsub{N}{BA}^1(t),...,\smallsub{N}{BA}^K(t))^T$ and $g_r$ is the appropriate number of possible combinations of the reactants for reaction $r$ from the particles that lie within $\smallsub{\Omega}{BA}$. Recall, however, that in $\smallsub{\Omega}{B}\backslash\smallsub{\Omega}{BA}$, any such reactions are implemented according to the chosen microscopic reaction method \citep{erban2009smr,lipkova2011abd,doi1976std}.

\subsection{Simulation specifics}
The Gillespie SSA \citep{gillespie1977ess} is used to simulate the above-described reactions in $\smallsub{\Omega}{BA}$, as well as the diffusive fluxes over the interface. The SSA requires the computation of an exponential random variable which gives the time, $\tau$, until the next event, and can be found by transforming a uniform random variable $u_1\sim \text{Unif}(0,1)$ via the following equation \begin{equation}
\tau = \frac{1}{\alpha^0(t)}\ln\left(\frac{1}{u_1}\right).
\label{eqn:next_reaction_time}
\end{equation} Here, $\alpha^0(t)$ is the sum of all of the propensity functions: 
\begin{equation}
\alpha^0(t) = \smallsub{\alpha}{P}^0(t)+\smallsub{\alpha}{B}^0(t) + \sum_{r\in\mathcal{R}^*(t)}{\alpha_r(t)},
\label{eqn:a0}
\end{equation} 
where 
\begin{equation}
\smallsub{\alpha}{P}^0(t)= \sum_{k=1}^K{\smallsub{\alpha}{P}^k(t)},\label{eqn:sum_PDE_props}
\end{equation}
and
\begin{equation}
 \smallsub{\alpha}{B}^0(t)=\sum_{k=1}^K{\smallsub{\alpha}{B}^k(t)}.\label{eqn:sum_Brownian_props}
\end{equation}

The PDE solutions and Brownian dynamics are implemented using the same discrete time-step, $\Deltat$, and the diffusive jumps across the interface (and any required reactions, $r\in \mathcal{R}^*$) are implemented in an event-driven manner, according to the Gillespie SSA. Event-driven time-steps are implemented until the putative time for the next event passes the next Brownian/PDE update time, at which point the PDE and Brownian dynamics are updated. Pseudocode for the ARM is given in Algorithm \ref{alg:ARM}.

\begin{Algorithm}{Auxiliary region method (ARM)}\label{alg:ARM}
\item Initialise time $t=0$, set final time $T$, PDE/Brownian update time-step, $\Deltat$, the PDE discretisation grid size, $\Delta x$, and the auxiliary region spatial step, $h_a$. Initialise particles in both $\smallsub{\Omega}{P}$ and $\smallsub{\Omega}{B}$ as required. Calculate the time until the next PDE and Brownian update step $t_\Delta = \Deltat$.

\item Calculate the number of particles $\smallsub{N}{PA}^k$ and $\smallsub{N}{BA}^k$ in the auxiliary regions, for each species  $k\in\{1,2,...,K\}$,  using formulae \eqref{eqn:particle_numbers_PDE_AR} and \eqref{eqn:particle_numbers_BD_AR}. Consequently, calculate the corresponding propensity functions, $\smallsub{\alpha}{P}^k(t)$ and  $\smallsub{\alpha}{B}^k(t)$ as per equations \eqref{eqn:propensity_functions_PDE} and \eqref{eqn:propensity_functions_Brown}, and their sums according to equations \eqref{eqn:sum_PDE_props} and \eqref{eqn:sum_Brownian_props}. Calculate $\alpha_{r}(t)$, for $r\in\mathcal{R}^*$, using equation \eqref{eqn:propensity_functions_reactions} and finally compute $\alpha^0(t)$ according to equation \eqref{eqn:a0}. \label{item:while_loop_arm}

\item Calculate the time, $\tau$, until the next auxiliary region event according to equation (\ref{eqn:next_reaction_time}). Update the auxiliary region time $t_a = t+\tau$.

\item If $t_a<t_\Delta$
	\begin{enumerate}[(i)]
	\item Draw three random numbers $u_1,u_2,u_3\sim \text{Unif}(0,1)$.
	\item If $u_1\alpha^0(t) < \smallsub{\alpha}{P}^0(t)$ (corresponding to a jump from $\smallsub{\Omega}{PA}$ to $\smallsub{\Omega}{BA}$):
		\begin{itemize}
		\item Use $u_2$ to determine the species, $k$, which the jump affects, with each species selected with probability proportional to its propensity function.
		\item Remove a particle from the PDE auxiliary region for species $k$ via equation \eqref{eqn:remove_particle_PDE}.
		\item Initialise a new particle of species $k$ within $\smallsub{\Omega}{BA}$ at position $y^* =u_3h_a + I$.
		\end{itemize}
		Else if $u_1\alpha^0(t) < \smallsub{\alpha}{P}^0(t)+\smallsub{\alpha}{B}^0(t)$ (corresponding to a jump from $\smallsub{\Omega}{BA}$ to $\smallsub{\Omega}{PA}$):
		\begin{itemize}
		\item Use $u_2$ to determine the species, $k$, which the jump affects, with each species selected with probability proportional to its propensity function.
		\item Choose a particle of species $k$ uniformly at random from within the Brownian auxiliary region and remove it from the system. We do this by selecting an index $q$ such that $q=\lceil u_3 \smallsub{N}{BA}^k \rceil$, where $\lceil x \rceil$ denotes the smallest integer larger than $x$.
		\item Add a new particle into the PDE auxiliary region for species $k$ via equation \eqref{eqn:add_particle_PDE}.
		\end{itemize}
		Else (corresponding to a reaction in $\smallsub{\Omega}{BA}$)
		\begin{itemize}
		\item Use $u_2$ to choose the reaction $r\in \mathcal{R}^*(t)$ to be implemented with probability proportional to its propensity function. 
		\item Update particle numbers (and initialise positions, if appropriate) in the Brownian representation accordingly.
		\end{itemize}
	\item Set $t = t_a$.
% 	and update particle numbers $\smallsub{n}{P}^k$ and $\smallsub{N}{HB}^k$ for the species $k$.
	\end{enumerate}
	Else
	\begin{enumerate}[(i)]
	\item Update the PDE system \eqref{eqn:PDE} using an appropriate numerical method (see Section \ref{sect:ARM_description}).
	\item Update the positions of the Brownian particles according to equation \eqref{eqn:BD}.
	\item Implement any reactions using an appropriate method (see Section \ref{sect:ARM_description}). Note that production reactions should be implemented after any degradation reactions in order to prevent particles being created and destroyed in the same time-step.
	\item Set $t = t_\Delta$, update $t_{\Delta} = t + \Deltat$.
	\end{enumerate}
\item If $t<T$, return to \ref{item:while_loop_arm}, otherwise stop.
\end{Algorithm}

\section{Results} \label{sect:Results}
%Qualitative comparisons
%Figures - Intermediate time, Final time, Error plot
%	Uniform
%	LHS
%	RHS - this could go into supplementary material.
%	Morphogen gradient.

Within this section, we present four test problems which are used to demonstrate that the ARM correctly simulates reaction-diffusion systems. Two of these problems are models of pure diffusion with different initial conditions and will demonstrate that the fluxes over the interface are consistent with the expected behaviour of the fully Brownian simulations. The third problem is the formation of a morphogen gradient, which demonstrates the successful implementation of reactions in the ARM. Despite the fact that our method is valid for higher-order reactions, the first three test problems consider reactions up to first order. For such systems, no moment closure assumptions are required in deriving the mean-field reaction-diffusion PDE and hence its behaviour agrees with the mean behaviour of the individual-based models. This allows us to efficiently verify accuracy by comparing the mean behaviour of our hybrid method to the known mean-field behaviour. Finally, in test problem four, we implement a second-order reaction system in higher dimensions, indicating the applicability of the method to more complicated examples.

For each of the first three test problems, we use $\smallsub{\Omega}{P} = (-1,0)$ and $\smallsub{\Omega}{B}=(0,1)$, meaning that the interface is the single point at $0$. We take the value of the fixed PDE and Brownian update steps to be $\Delta t=0.02$, the auxiliary regions have width $h=0.05$ and the diffusion constant is $D = 0.025$. We will quantify the qualitative comparisons, presented here through density comparison snapshots, in Section \ref{sect:Error}. All simulations will comprise only a single species, so henceforth, all sub- or super-scripts, $k$, pertaining to species will be removed.

\subsection{Test problem 1: maintaining equilibrium} \label{sect:Results_P1}
For the first test problem, we simulate pure diffusion in the form of a simple Brownian motion with reflecting boundary conditions, which has Fokker-Planck equation given by the diffusion PDE and corresponding boundary conditions: 
\begin{align}
\text{PDE}&:\quad \partder{p}{t}=D\secpartder{p}{x};\quad x\in(-1,1);\quad t\in(0,T),\label{eqn:pure_diffusion_mean_field_model}\\
\text{BCs}&:\quad \partder{p}{x}(x,t) = 0; \quad x=-1,1; \quad t\in(0,T), \label{eqn:pure_diffusion_mean_field_model_BC} \\
\text{IC}&:\quad p(x,0) = p_0(x); \quad x\in [-1,1], \label{eqn:pure_diffusion_mean_field_model_IC}
\end{align} where $p_0(x)$ denotes the initial condition.
Note that $p(x,t)$ here represents the mean-field solution across the whole domain, whereas $c(x,t)$ represents the PDE solution in $\smallsub{\Omega}{P}$ in the hybrid method.
We initialise particles uniformly across the computational domain, so that $p_0(x) \equiv N/2$, where $N$ is the (constant) number of particles in the system.
\begin{figure}[ht!]
	\begin{center} 
	\subfigure[][]{
		\includegraphics[width=0.31\textwidth]{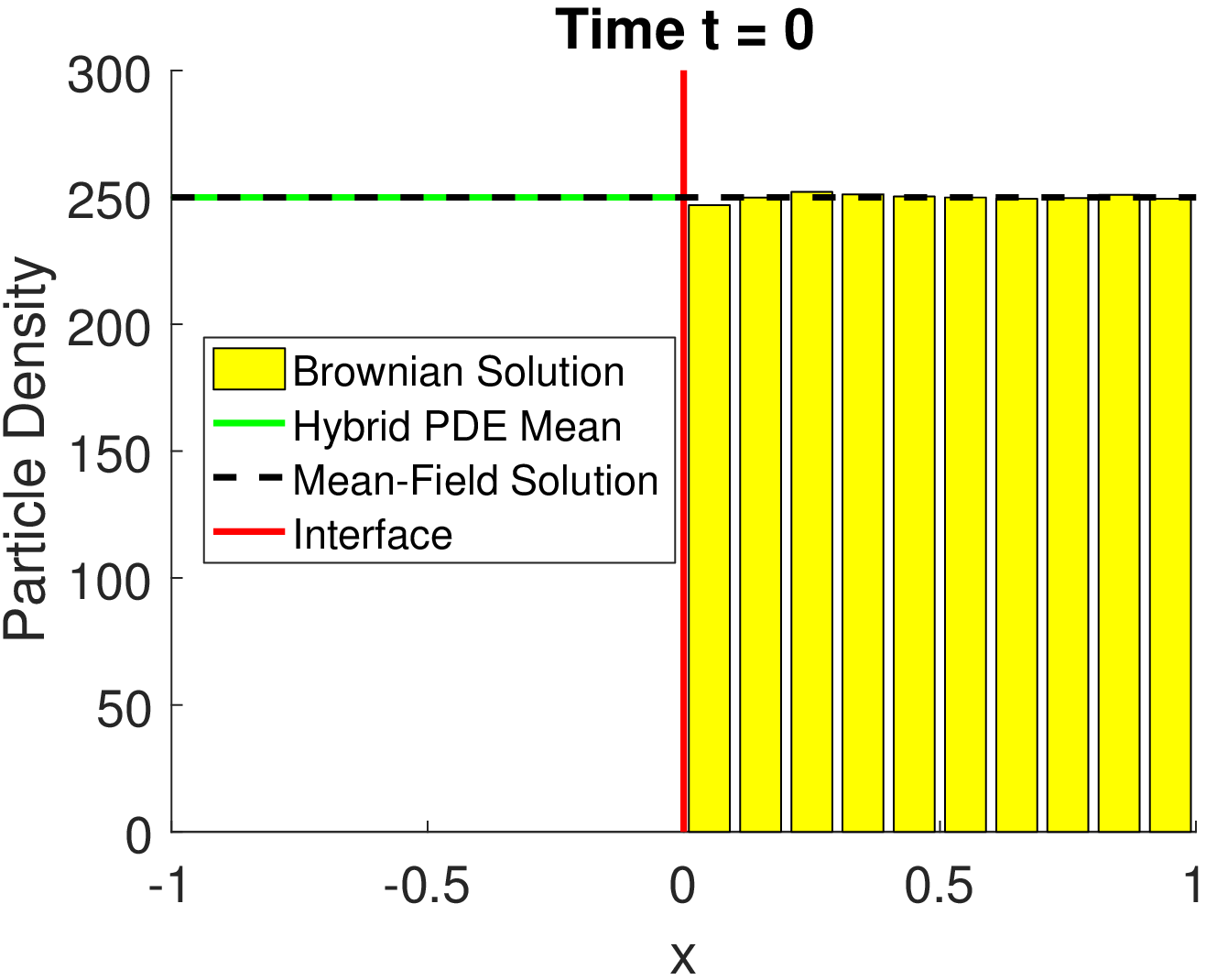}
		\label{fig:Plots_Unif_1}
	}
	\subfigure[][]{
		\includegraphics[width=0.31\textwidth]{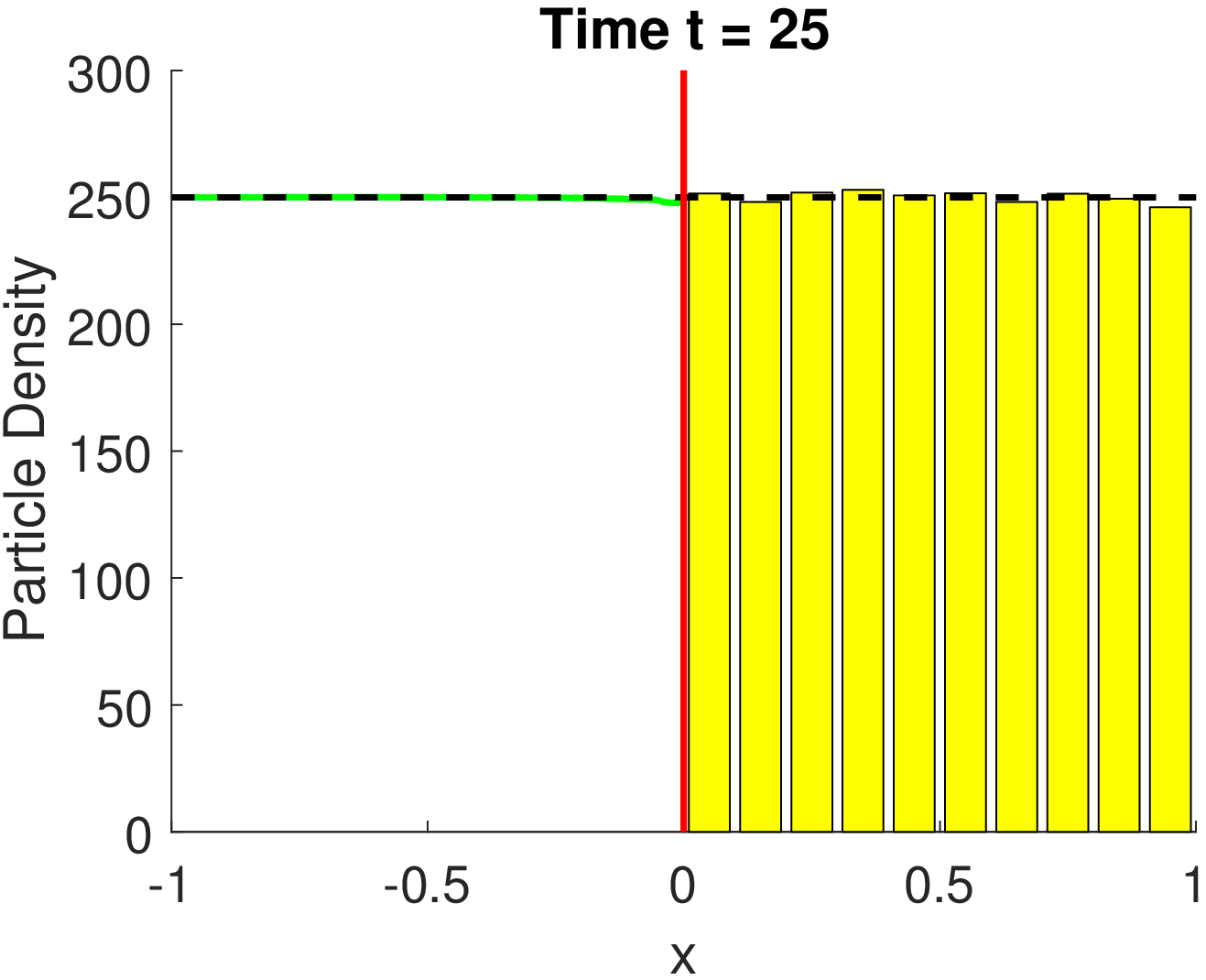}
		\label{fig:Plots_Unif_2}
	}
	\subfigure[][]{
		\includegraphics[width=0.31\textwidth]{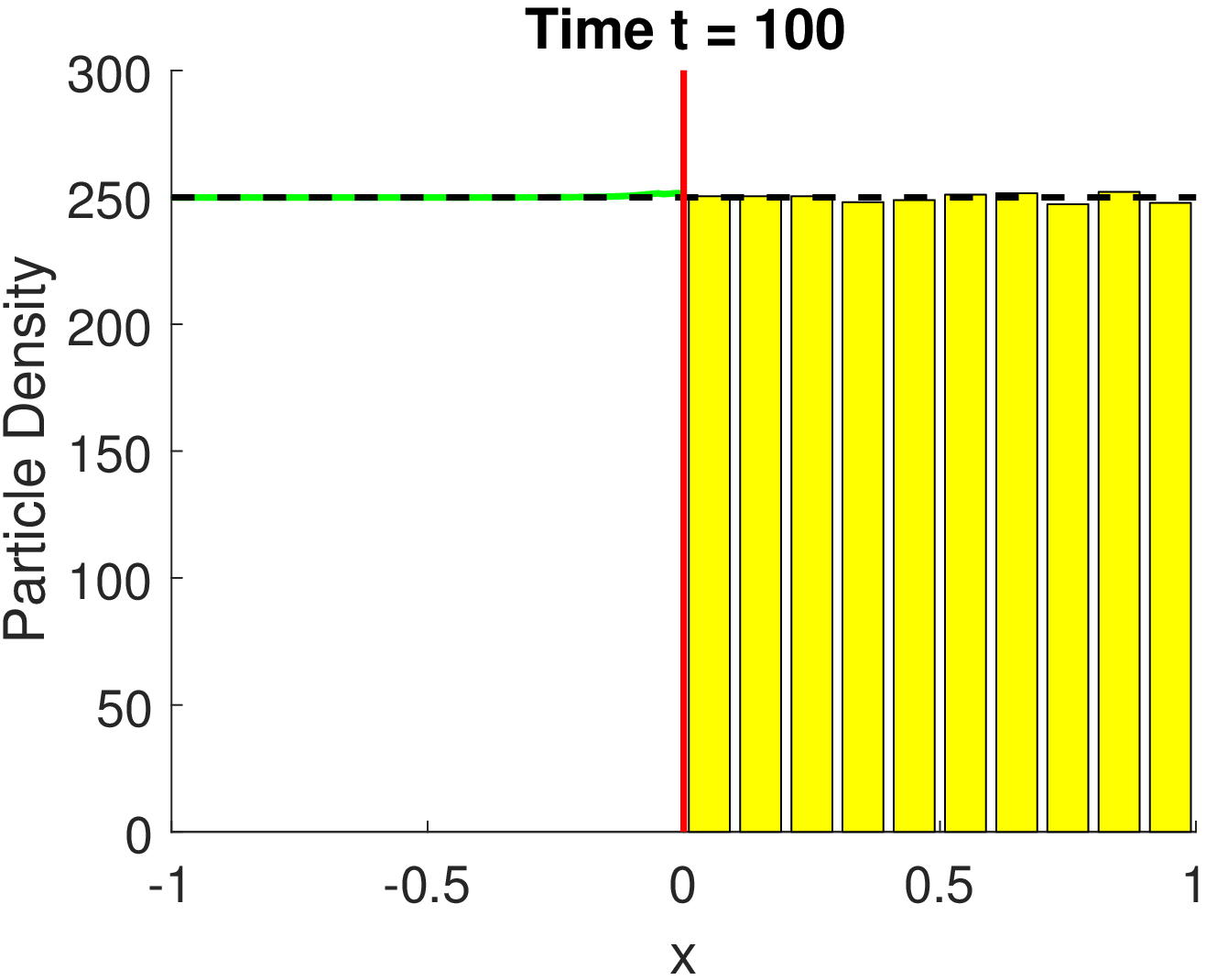}
		\label{fig:Plots_Unif_3}
	}
	\caption{The evolution of test problem 1 at times \subref{fig:Plots_Unif_1} 0, \subref{fig:Plots_Unif_2} 25 and \subref{fig:Plots_Unif_3} 100. The green line is the PDE part of the hybrid method, the yellow bars represent the Brownian dynamics for the hybrid method (appropriately binned for visualisation purposes), the red line is the interface and the black dashed line is the solution of the mean-field model \eqref{eqn:pure_diffusion_mean_field_model} under the given initial condition. Results shown contain 500 particles and are averaged over 1000 repeats. We solve the PDE with the $\theta$-method, with a value of $\theta=0.51$. All other parameters are given within the text.}
	\label{fig:Plots_Unif}
\end{center}
\end{figure}
Figure \ref{fig:Plots_Unif} shows that the ARM passes the most basic test by maintaining the steady state without causing an accumulation of mass on either side of the interface. For this test problem, we also include a plot which displays the variance in the density of particles (Figure \ref{fig:Vars}). In order to calculate this variance, we have binned the spatial domain onto a mesh of size $h_a$ (the same as the auxiliary region width) and calculated the variance of the density in each bin over a number of identically initialised (up to random allocation of particles in $\smallsub{\Omega}{B}$) repeats. This demonstrates a problem that occurs with all hybrid methods which contain an interface coupling a stochastic to a deterministic region. The variance is damped close to the interface in the stochastic part of the domain, due to the deterministic nature of the solver on the opposite side. Specifically, the PDE effectively has a stochastic boundary condition at the interface, caused by the diffusive jumps between the auxiliary regions. This causes a higher level of variance than would be expected if it was a purely deterministic regime. However, when a particle jumps from the PDE to the Brownian dynamics auxiliary region, since the PDE region is mostly deterministic, it contributes less variance than would be expected than if the stochastic method was employed across the entire domain. There are methods that can be used in order to fix this problem, such as the use of an overlap region (e.g. \citep{harrison2016hac}) and replacing the PDE with an appropriate SPDE (e.g. \citep{alexander2002ars}). This is explored in more detail in the discussion (Section \ref{sect:Discussion}). 

\begin{figure}[ht!]
	\centering
	\includegraphics[width=0.5\textwidth]{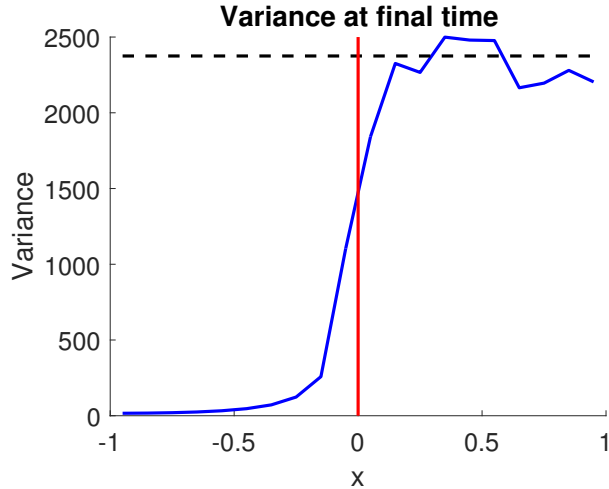}
	\caption{The plot of the variance in the density of particles at time $t = 100$ for the parameter values used to produce Figure \ref{fig:Plots_Unif}. The spatial domain is partitioned into a series of bins of width $h_a$, and the particle density variance is calculated in each bin over $S = 1000$ repeats. The blue line is the variance from the hybrid method, the black dashed line is the expected variance from the fully Brownian model, and the red line is the position of the interface. The variance can be seen to be damped in the stochastic domain close to the interface, as discussed in the text.}
	\label{fig:Vars}
\end{figure}

\subsection{Test problem 2: flux over the interface} \label{sect:Results_P2}
\begin{figure}[ht!]
	\begin{center} 
	\subfigure[][]{
		\includegraphics[width=0.31\textwidth]{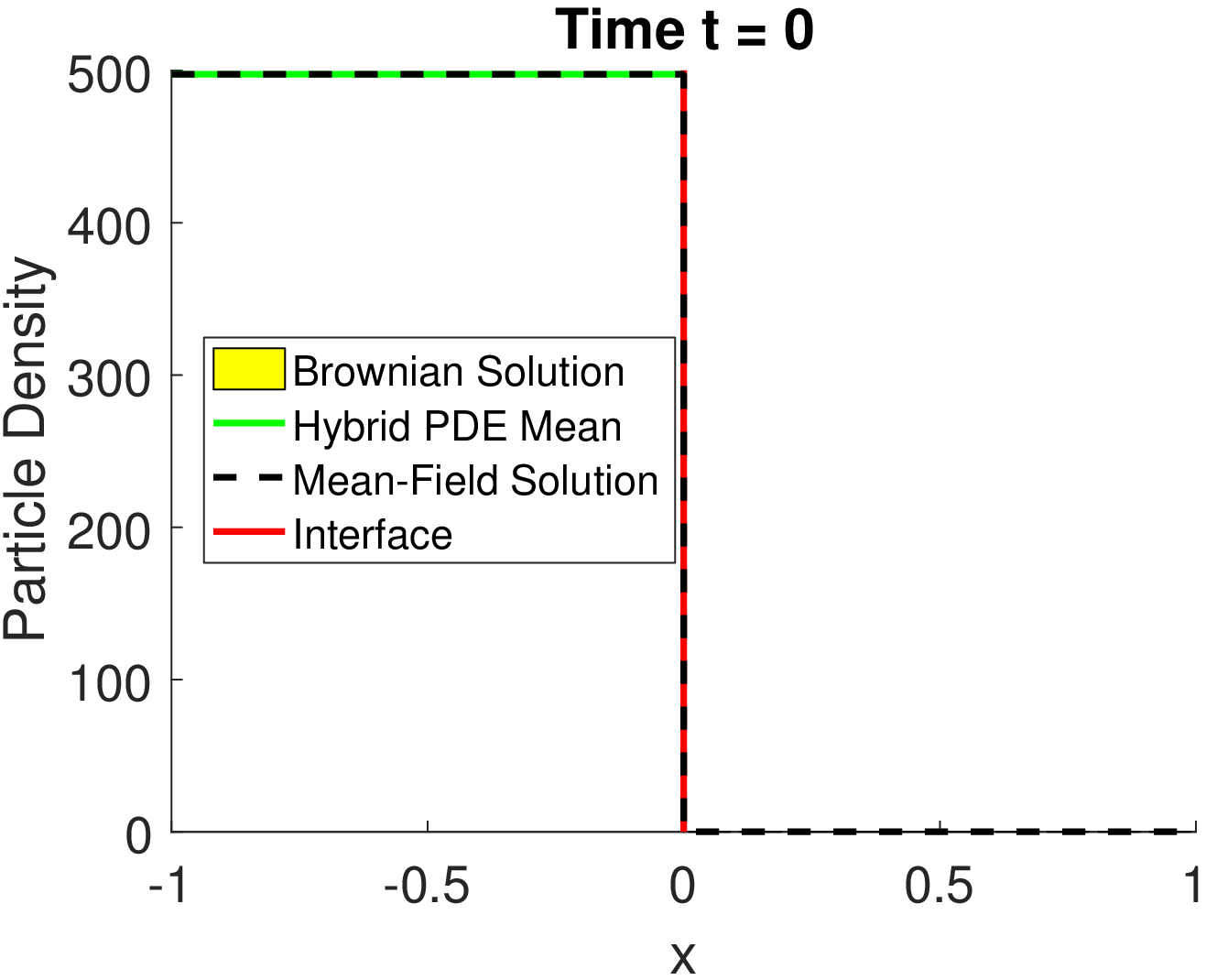}
		\label{fig:Plots_Left_1}
	}
	\subfigure[][]{
		\includegraphics[width=0.31\textwidth]{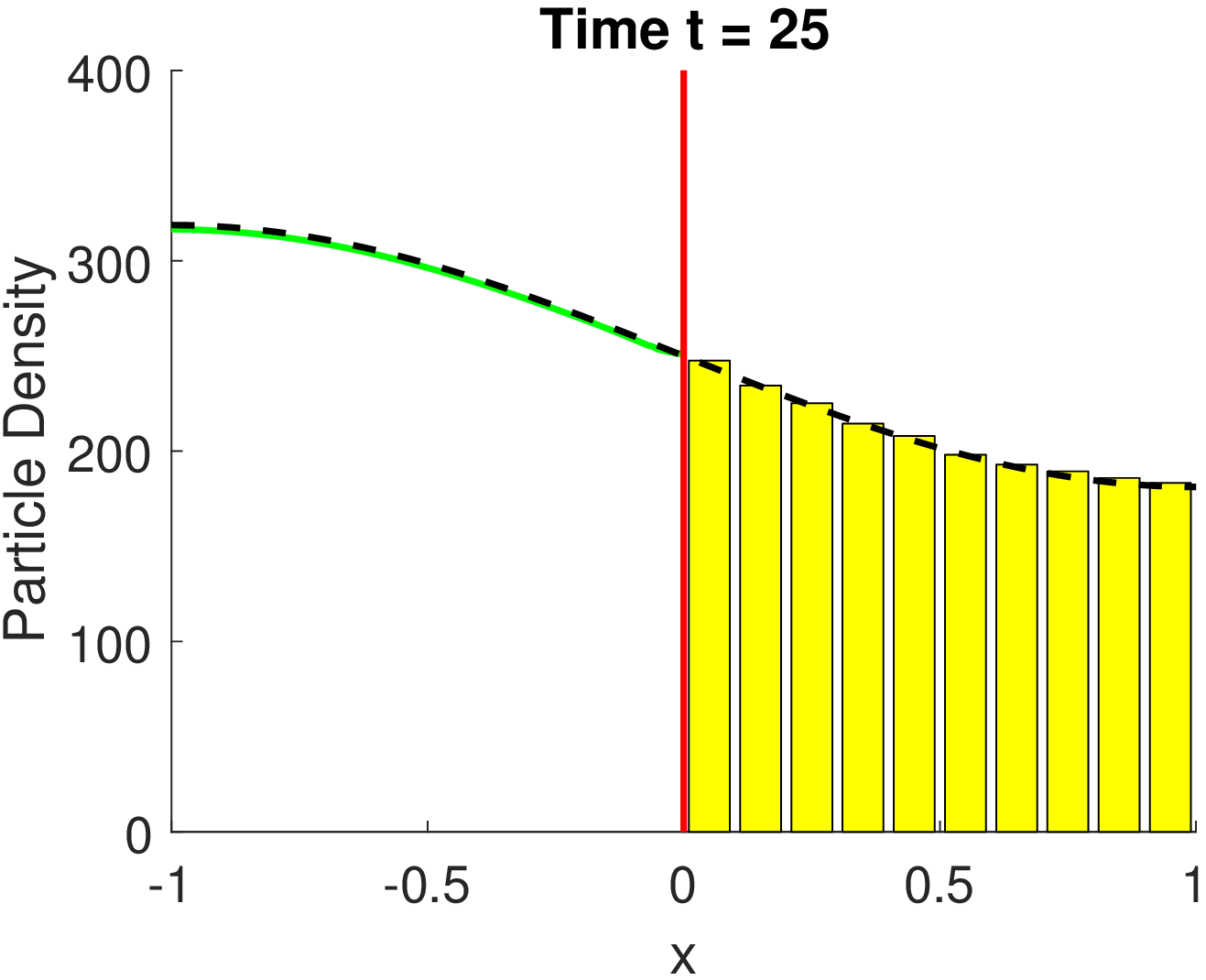}
		\label{fig:Plots_Left_2}
	}
	\subfigure[][]{
		\includegraphics[width=0.31\textwidth]{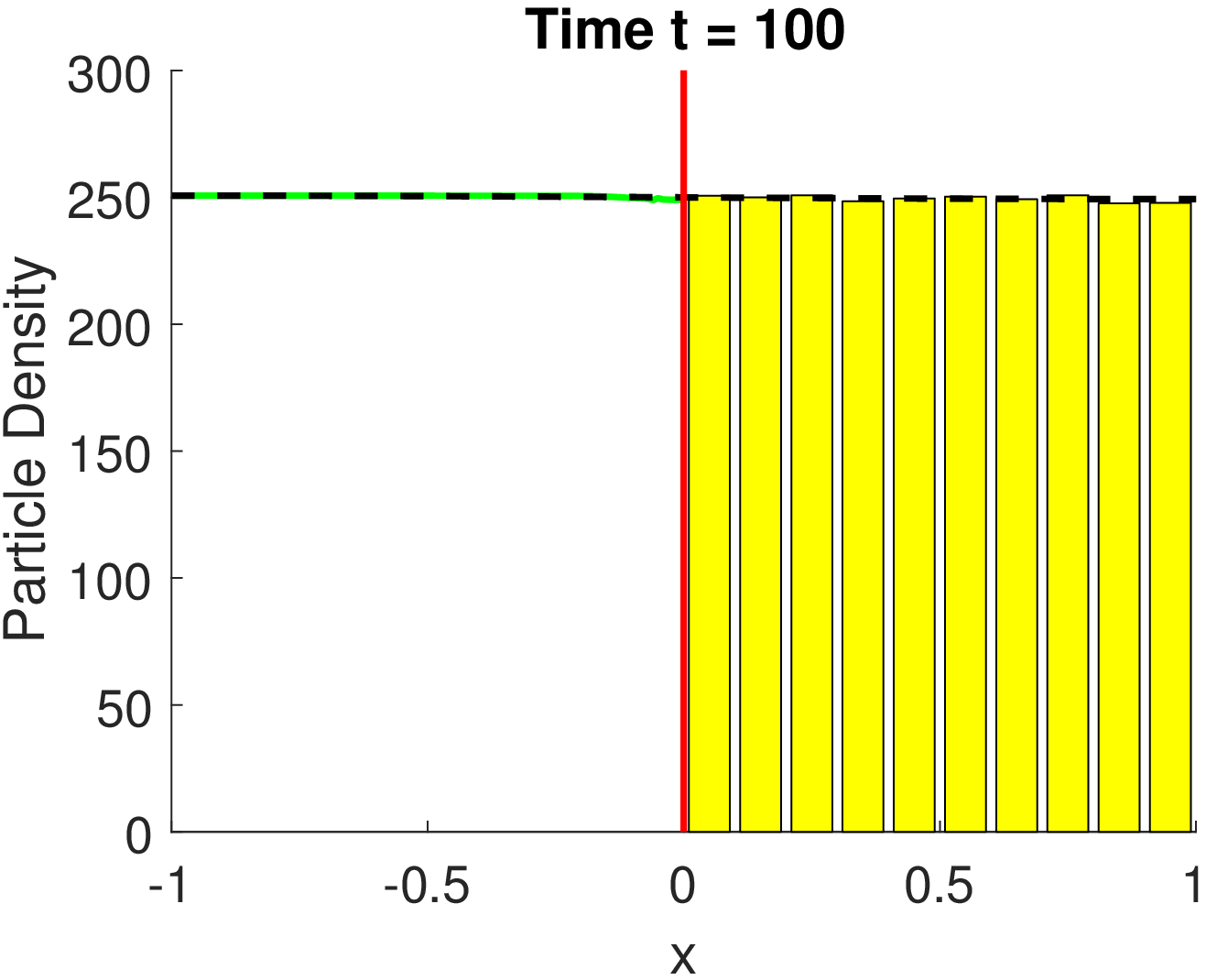}
		\label{fig:Plots_Left_3}
	}
	\caption{The evolution of test problem 2 at times \subref{fig:Plots_Left_1} 0, \subref{fig:Plots_Left_2} 25 and \subref{fig:Plots_Left_3} 100. All figure descriptions and parameter values are as in Figure \ref{fig:Plots_Unif}.}
	\label{fig:Plots_Left}
\end{center}
\end{figure}

The second test problem is a stress test for the interfacial flux. For the PDE part of the hybrid method we solve the same diffusion equation \eqref{eqn:pure_diffusion_mean_field_model}-\eqref{eqn:pure_diffusion_mean_field_model_IC} as in Section \ref{sect:Results_P1}. However this time we initialise by placing all particles uniformly within the PDE domain, $\smallsub{\Omega}{P}$, which results in \begin{equation*}
p_0(x) = \left\{\begin{array}{ll}
N & x\in\smallsub{\Omega}{P} \\ 
0 & x\in\smallsub{\Omega}{B}
\end{array} \right.
\end{equation*} The results from this simulation are displayed in Figure \ref{fig:Plots_Left}.
As with the uniform initial condition in test problem 1, we see from Figure \ref{fig:Plots_Left} that the hybrid method agrees with the solution of the mean-field model, indicating that the method simulates flux over the interface accurately. We have also tested our hybrid method with all the mass initialised uniformly across $\Omega_{\text{B}}$ and found a similarly good agreement between the hybrid method and the mean-field solution (figures not shown).

\subsection{Test problem 3: morphogen gradient}
For the third test problem, we investigate the formation of a morphogen gradient from a uniform initial condition. The gradient is formed by allowing particles to diffuse throughout the domain as well as to degrade at a rate $\mu$. We also have particles entering at the left-hand boundary, $x=-1$, at rate $D\lambda$, and a zero-flux condition at $x=1$. Thus, the PDE half of the hybrid domain is governed by the mean-field model representing the expected behaviour of the fully Brownian dynamics: 
\begin{align}
\text{PDE}&:\quad \partder{c}{t}=D\secpartder{c}{x} - \mu c;\quad x\in(-1,0);\quad t\in(0,T),\label{eqn:morphogen_gradient_mean_field_model}\\
\text{BCs}&:\quad \partder{c}{x}(-1,t) = -\lambda;\quad \partder{c}{x}(0,t) = 0; \quad t\in(0,T), \notag \\
\text{IC}&:\quad c(x,0) = c_0(x); \quad x\in [-1,0]. \notag
\end{align}
For the corresponding microscopic dynamics we implement Brownian motion for the diffusion of particles and a time-based method in order to enact the degradation reactions. We note that production of particles is not implemented within the microscopic domain since it occurs at $x=-1$. $N=500$ particles are initialised uniformly across the domain.
\begin{figure}[ht!]
	\begin{center} 
	\subfigure[][]{
		\includegraphics[width=0.31\textwidth]{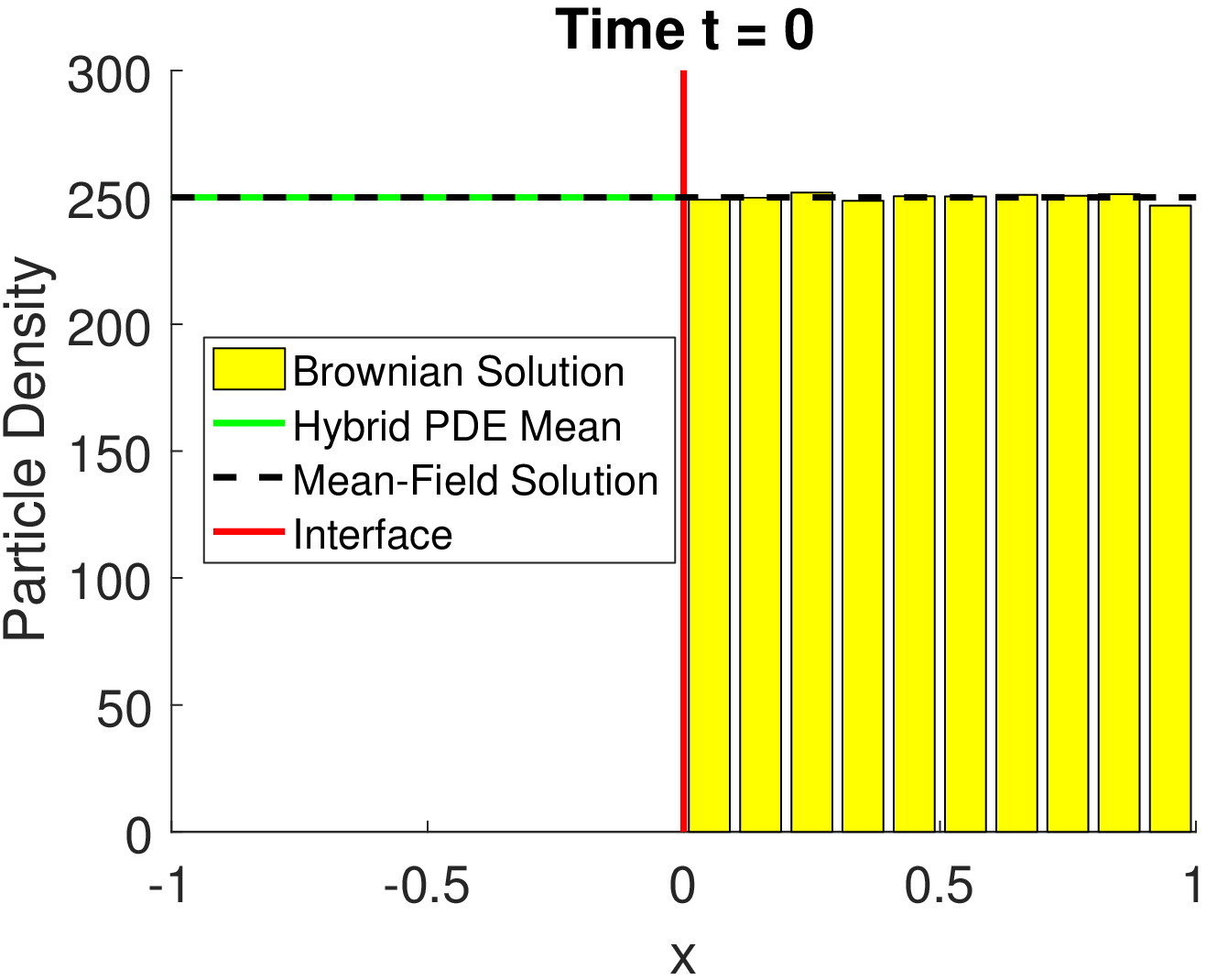}
		\label{fig:Plots_Morph_1}
	}
	\subfigure[][]{
		\includegraphics[width=0.31\textwidth]{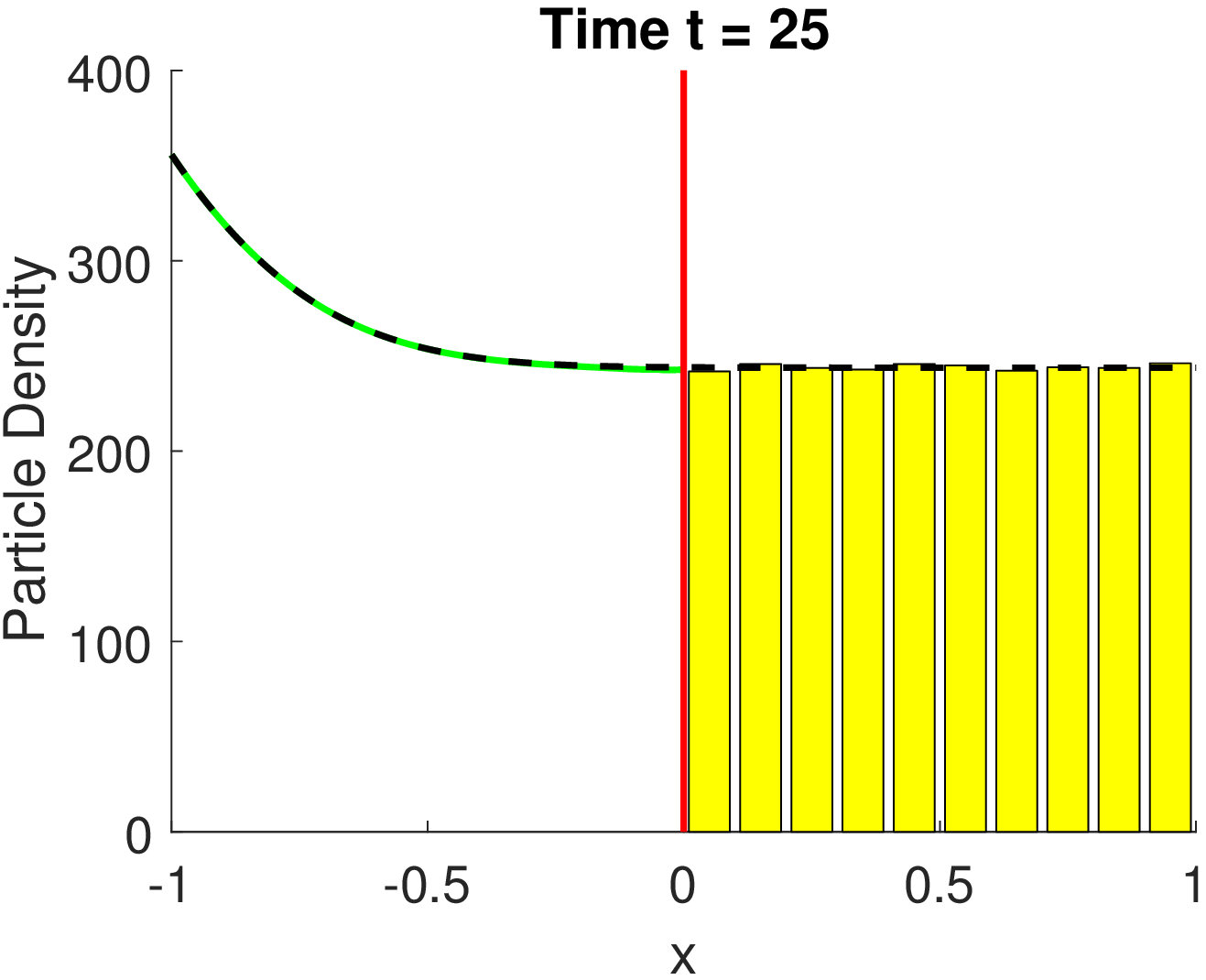}
		\label{fig:Plots_Morph_2}
	}
	\subfigure[][]{
		\includegraphics[width=0.31\textwidth]{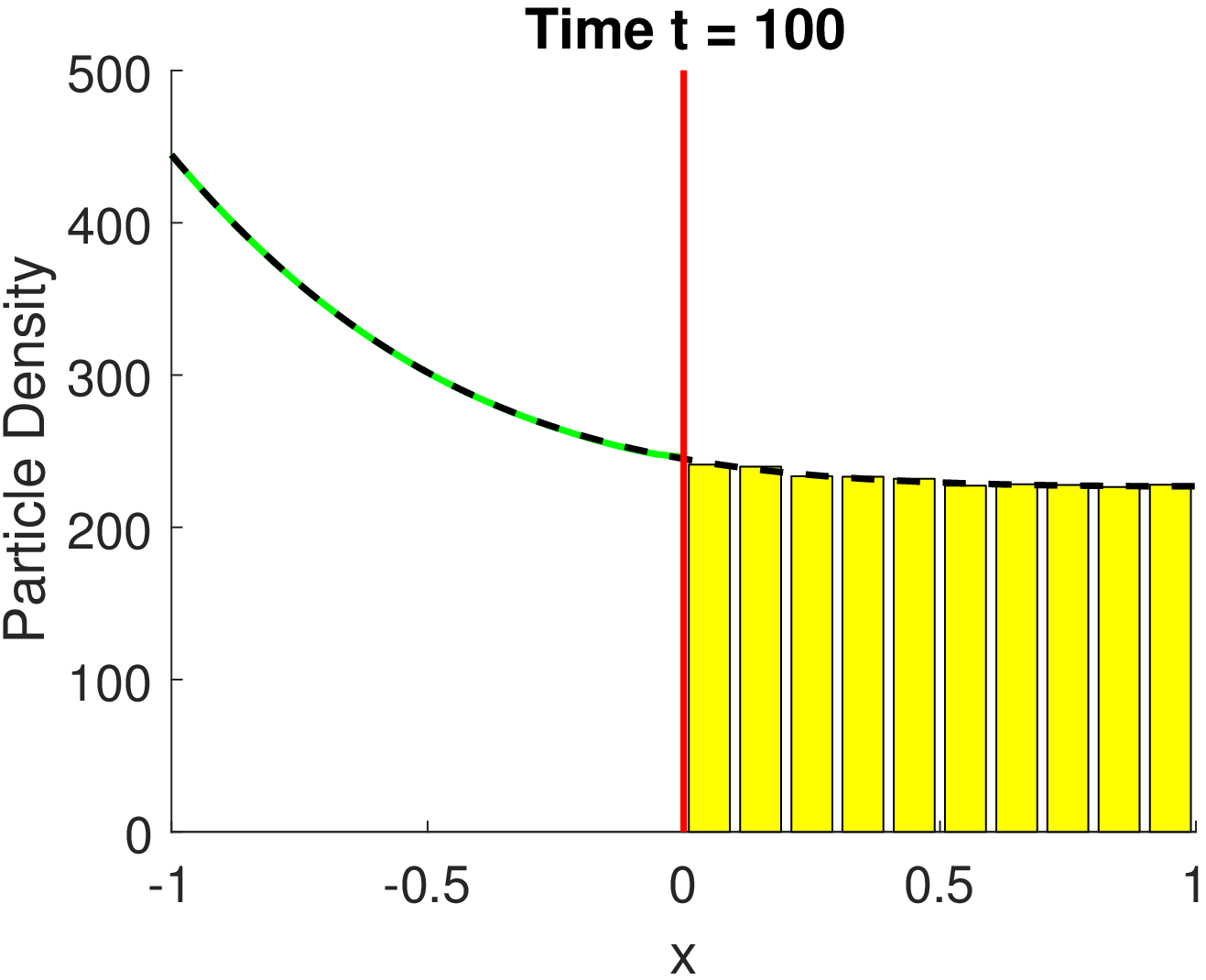}
		\label{fig:Plots_Morph_3}
	}
	\caption{The evolution of test problem 3 at times \subref{fig:Plots_Morph_1} 0, \subref{fig:Plots_Morph_2} 25 and \subref{fig:Plots_Morph_3} 100. The value for the diffusion coefficient is $D=0.025$, the production rate is $\lambda=400$, the degradation rate is $\mu=0.001$ and the time-step is $\Deltat = 0.005$. All other figure descriptions and parameter values are the same as Figure \ref{fig:Plots_Unif}.}
	\label{fig:Plots_Morph}
\end{center}
\end{figure}

As demonstrated in Figure \ref{fig:Plots_Morph} the solution of the hybrid method matches that of the corresponding mean-field model, as with the previous two test problems.

\subsection{Test problem 4: Higher-order systems} \label{sect:Results_P4}
For our final test problem, we look at the reaction system: \begin{equation}
2A \xrightarrow{\kappa_1} \emptyset,\quad \emptyset \xrightarrow{\kappa_2} A, \label{eqn:TP4_System}
\end{equation} which takes place in a three-dimensional cuboid $\Omega\subseteq\mathbb{R}^3$ of volume $V$, where $\Omega=(x_0,x_1)\times(y_0,y_1)\times(z_0,z_1)$. We further split this domain by firstly defining the position of the adaptive planar interface $I(t) \in (x_0,x_1)$ which is to be implemented for this test problem (see Section \ref{sect:Results_P4_Interface}). In an analogous way to in the one-dimensional case, we then define the time-dependent PDE and individual-based subdomains, $\smallsub{\Omega}{P}(t)$ and $\smallsub{\Omega}{B}(t)$, with volumes $\smallsub{V}{P}(t)$ and $\smallsub{V}{B}(t)$ respectively. These subdomains and volumes depend on $t$ due to the adaptive interface position. The interface will move according to the local density profile within the PDE and Brownian dynamics auxiliary regions $\smallsub{\Omega}{PA}(t)$ and $\smallsub{\Omega}{BA}(t)$, which are explicitly defined to be:\begin{align*}
\smallsub{\Omega}{PA}(t) &= (I(t)-h_a,I(t))\times(y_0,y_1)\times(z_0,z_1),\\
\smallsub{\Omega}{BA}(t) &= (I(t),I(t)+h_a)\times(y_0,y_1)\times(z_0,z_1).
\end{align*} 

We will firstly find a PDE in one dimension that will form the deterministic part of our domain. We do this by considering the reaction system \eqref{eqn:TP4_System} and forming an ODE to simulate this system in three dimensions. We then include isotropic diffusion to obtain a three-dimensional PDE, and finally impose a constraint on the initial condition to simplify this to a one-dimensional PDE.  We then briefly describe the process we use to evolve the individual-level behaviour, before introducing an adaptive interface. We will finish this subsection with the results of some simulations of this system. Note that from now on, we will drop the dependence on $t$ for any of the subdomains, their volumes and the interface position for brevity, unless they are explicitly needed.

\subsubsection{PDE model} \label{sect:Results_P4_PDE}

We will use the chemical master equation (CME) for the reaction system in order to derive a PDE that approximates the system \eqref{eqn:TP4_System} in $\smallsub{\Omega}{P}$. Let $p_n(t)=\mathbb{P}(A(t)=n)$, where $A(t)$ is the number of particles at time $t$. Then the CME for the evolution of this probability is given by: \begin{equation*}
\ordder{p_n}{t} = \frac{\kappa_1}{\smallsub{V}{P}}\left[(n+2)(n+1)p_{n+2} - n(n-1)p_n\right] + \kappa_2\smallsub{V}{P}\left[p_{n-1}-p_n\right].
\end{equation*} If we now define the $k^\text{th}$ central moment $\langle A^k\rangle := \sum_{n=0}^\infty n^kp_n$, we can multiply the CME by $n$ and sum over all $n\in\mathbb{N}_0$ to yield the mean equation: \begin{equation}
\ordder{\langle A \rangle}{t} = \frac{2\kappa_1}{\smallsub{V}{P}}\left[\langle A\rangle - \langle	A^2 \rangle\right] + \kappa_2\smallsub{V}{P}. \label{eqn:TP4_ODE_Exact}
\end{equation} The ODE \eqref{eqn:TP4_ODE_Exact} is currently exact, but depends on the second moment of $A$. Furthermore, the ODE for every moment of $A$ depends on higher moments still --- the system is not closed. In order to close the system, we follow \citet{erban2009smr} and apply Poisson moment closure, which implies: \begin{equation}
\Var(A) = \mathbb{E}[A] \implies \langle A^2 \rangle = \langle A \rangle + \langle A \rangle ^2. \label{eqn:Moment_Closure}
\end{equation} Applying the moment closure \eqref{eqn:Moment_Closure} to the ODE \eqref{eqn:TP4_ODE_Exact}, and setting $c = \langle A \rangle/\smallsub{V}{P}$ gives us the closed ODE \begin{equation*}
\ordder{c}{t} = \kappa_2 - \kappa_1c^2. 
\end{equation*} Finally, including isotropic diffusion through the usual Laplace operator yields the three-dimensional PDE: \begin{equation}
\partder{c}{t} = D\nabla^2c - \kappa_1c^2 + \kappa_2;\quad (x,y,z)\in \Omega;\quad t\in(0,T). \label{eqn:TP4_PDE_3D}
\end{equation} We will enforce an initial condition which is translationally invariant in both the $y$ and $z$ co-ordinates, which means that the dynamics will remain translationally invariant for all time. As such, $c$ is simply a function of $x$ and $t$, and the dynamics can be represented by a one-dimensional equivalent of this PDE by implementing zero-flux boundaries on all boundaries and using the transformation: \begin{equation*}
\bar{C}(x,t) = \int_{z_0}^{z_1}{\int_{y_0}^{y_1}{c(x,t)\ dy}\ dz} = L_yL_zc(x,t),
\end{equation*}where $L_y=y_1-y_0$ and similarly for $L_z=z_1-z_0$. This gives: \begin{align}
\text{PDE}&:\quad \partder{\bar{C}}{t}=D\secpartder{\bar{C}}{x}-\frac{\kappa_1}{L_yL_z}\bar{C}^2 + \kappa_2L_yL_z;\quad x\in(x_0,I);\quad t\in(0,T),\label{eqn:TP4_PDE}\\
\text{BCs}&:\quad \partder{\bar{C}}{x}(x,t) = 0; \quad x=x_0,I; \quad t\in(0,T), \notag \\
\text{IC}&:\quad \bar{C}(x,0) = \bar{C}_0(x); \quad x\in [x_0,I]. \notag
\end{align} 

\subsubsection{Individual-based formulation} \label{sect:Results_P4_Individual}

We now turn our attention to the individual-based system. In order to simulate the three-dimensional individual-based model, we will follow the $\lambda$-$\rho$ method \citep{erban2009smr}. In the context of this system, whenever two particles are within the reaction radius $\rho$, they react with a probability $P_\lambda$, which is a function of the kinetic rate $\kappa_1$, the time-step $\Delta t$, and the diffusion coefficient $D$. For more information on how $P_\lambda$ is chosen, we refer the reader to \citet{erban2009smr}. The zeroth-order reaction is completed by initialising a particle uniformly throughout the individual-based domain $\smallsub{\Omega}{B}$ with probability $\kappa_2\Delta t\smallsub{V}{B}$, which we ensure is below 1 by choosing $\Delta t$ to be sufficiently small.

\subsubsection{Adaptive interface} \label{sect:Results_P4_Interface}

Test problems 1--3 have been simulated using a static interface. However, this requires \textit{a priori} knowledge of where the interface should be for all time. When the finer scale modelling regime is required in order to resolve a specific area of space in more detail (for example, the region around ion channels \citep{dobramysl2015pbd}), the interface position will be known. However, if the purpose of the interface is to split regions of space in which there are high and low particle numbers, a different approach is required. In this case, the interface (or interfaces) need to move with the density of particles to maintain the computational savings they are designed to provide. We now describe a method, adapted from \citet{robinson2014atr} which allows the interface to move adaptively.

The interface at time $t$, which we shall denote by $I(t)$, moves according to local particle numbers in the auxiliary regions around it. We set two thresholds $\beta_u > \beta_l$, and move the interface towards the PDE subdomain if $\smallsub{N}{PA}(t) < \beta_l$ and towards the individual-based subdomain if $\smallsub{N}{BA}(t) > \beta_u$ (borrowing the notation from Section \ref{sect:ARM}). The two threshold values are designed to prevent the interface from rapidly oscillating between two values, which is a possibility when $\beta_u=\beta_l$ due to the stochastic nature of the system. We enforce that when the interface moves, it moves a distance $h_a$, the width of the auxiliary region, in the chosen direction.

If the interface moves towards the PDE subdomain (i.e. $\smallsub{N}{PA}(t) < \beta_l$), we convert the PDE auxiliary region into particles, initialising each one uniformly. As $\smallsub{N}{PA}(t)$ is not necessarily an integer, we treat the fractional part ($\smallsub{N}{PA}(t)\ \text{mod}\ 1$) to be the probability of initialising one extra particle within the newly formed individual-based region. We then scale the rest of the PDE subdomain to ensure that we conserve mass. During an interface movement towards the individual-based subdomain (i.e. $\smallsub{N}{BA}(t) > \beta_u$), the Brownian auxiliary region is converted to PDE mass by initialising a density of $\smallsub{N}{BA}(t)/h_a$ uniformly across the new PDE mesh points created by moving the interface. For a more detailed description of a similar method, we direct the interested reader to \citet{robinson2014atr}. 

\subsubsection{Results} \label{sect:Results_P4_Results}

\begin{figure}[ht!]
	\begin{center} 
	\subfigure[][]{
		\includegraphics[width=0.48\textwidth]{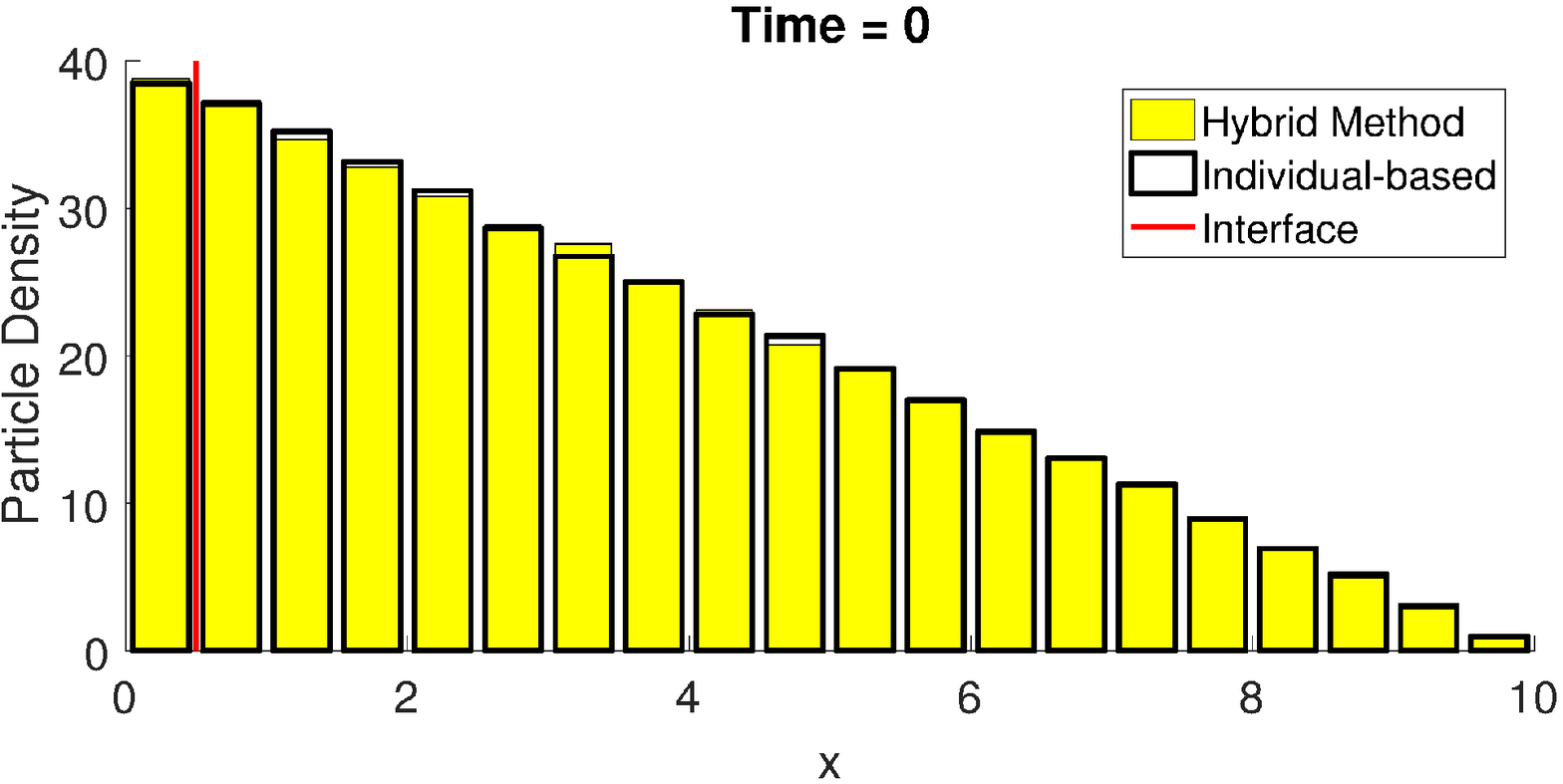}
		\label{fig:Plots_Second_1}
	}
	\subfigure[][]{
		\includegraphics[width=0.48\textwidth]{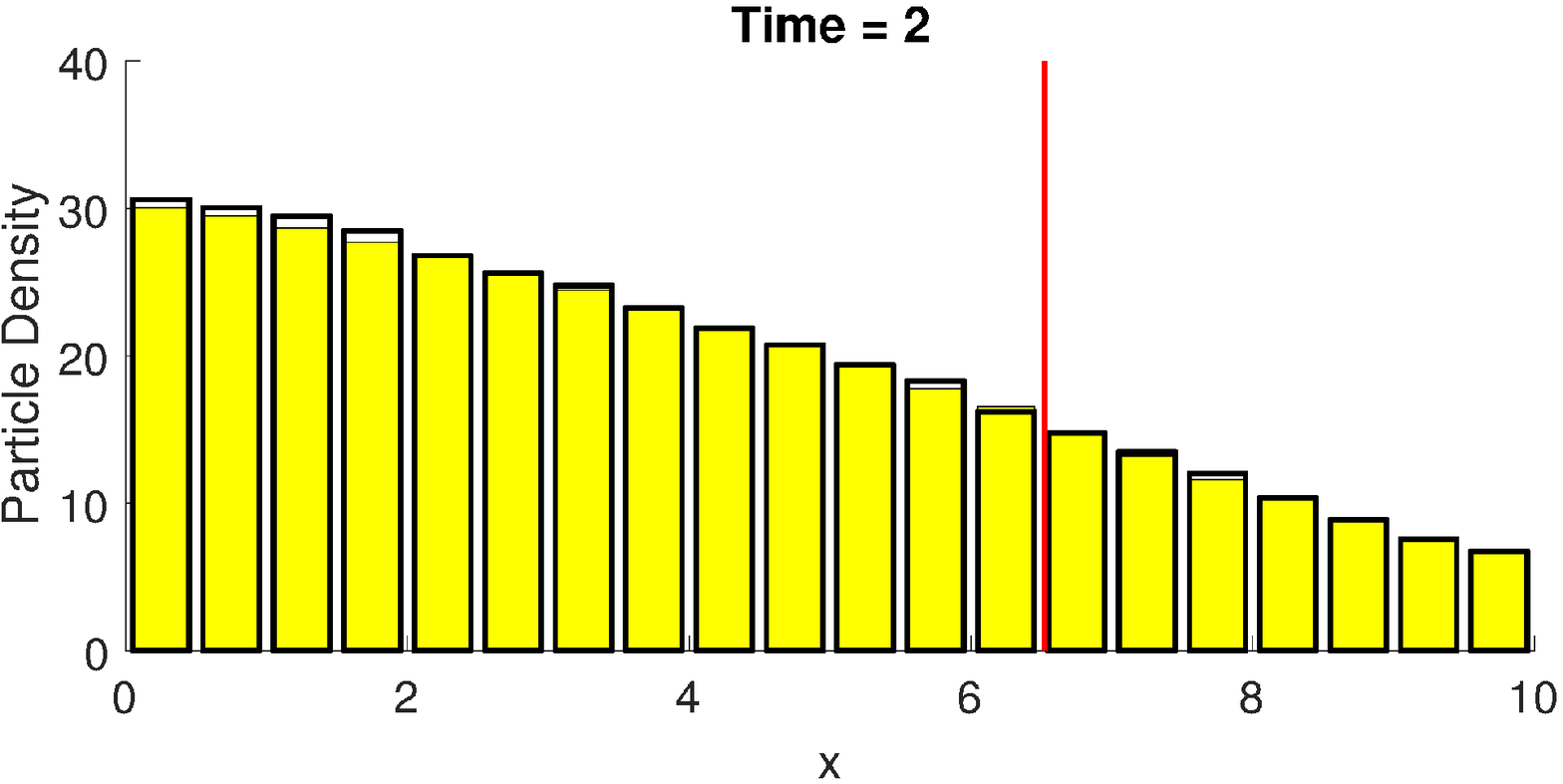}
		\label{fig:Plots_Second_2}
	}
	\subfigure[][]{
		\includegraphics[width=0.48\textwidth]{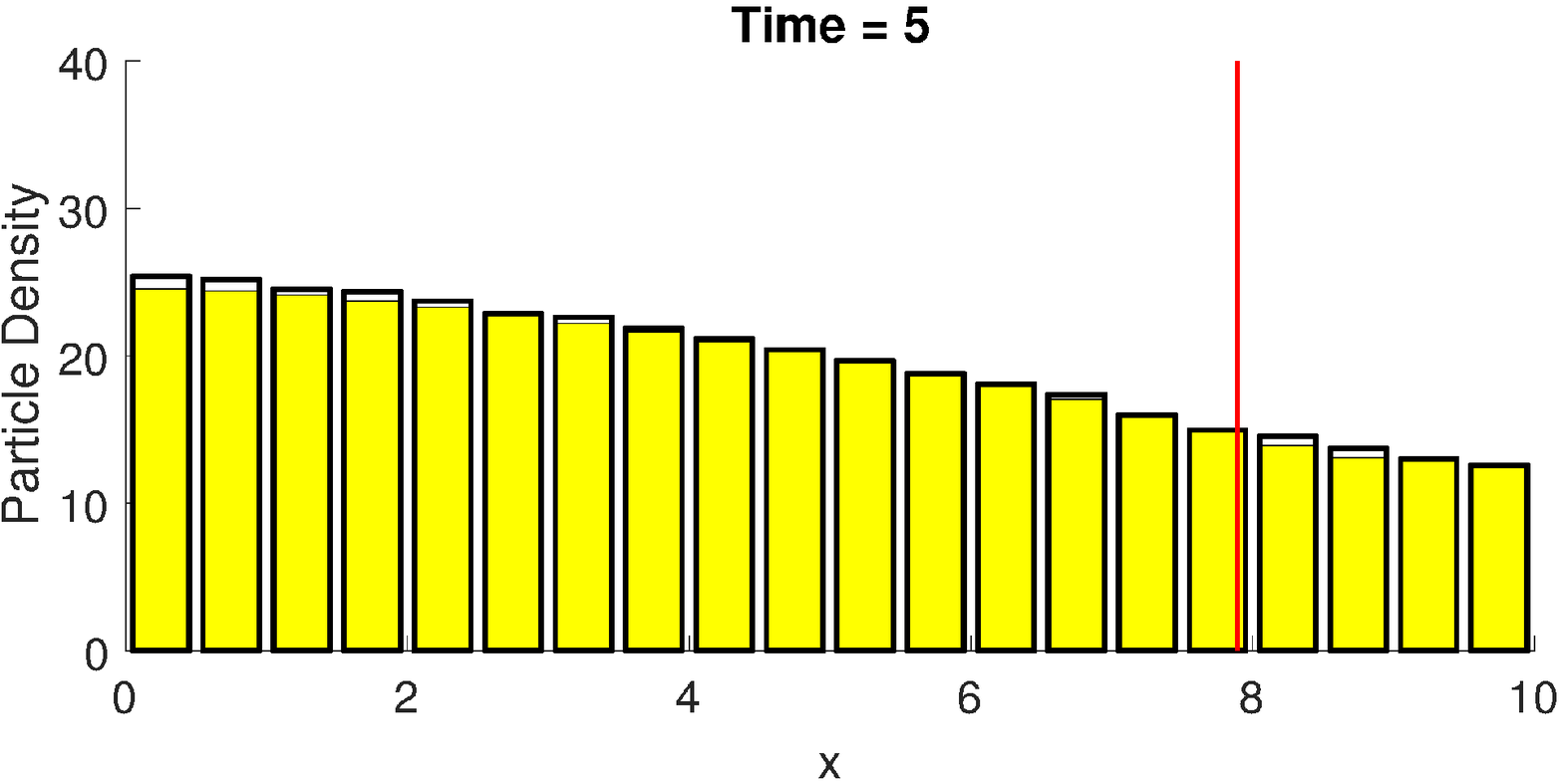}
		\label{fig:Plots_Second_3}
	}
	\caption{The evolution of test problem 4 at times \subref{fig:Plots_Second_1} 0, \subref{fig:Plots_Second_2} 2 and \subref{fig:Plots_Second_3} 5. The yellow bars represent the hybrid solution, binned onto a mesh with width $h_a$, and the black outline bars are the fully individual-based solution, which has been binned onto the same mesh as the hybrid solution. The vertical red line is the average position of the interface over the $S=1000$ repeats. All parameters are as in Table \ref{tab:Params}.}
	\label{fig:Plots_Second}
\end{center}
\end{figure}

We consider $N$ particles initialised throughout $\Omega$ with a constant negative gradient so that the density of particles at position $x_1$ is equal to zero. This ensures that the interface will move as the dynamics progress. The results can be seen in Figure \ref{fig:Plots_Second}, in which the hybrid method has been averaged over $S=1000$ repeats. The hybrid density in the case of the moving interface is represented as yellow bars throughout the domain. This is because the interface position changes with each repeat, and so very few regions of space are solely represented by one or the other modelling paradigm over all repeats.

\begin{table}[ht!]
	\centering
	\begin{tabular}{|c|c|c|c|c|c|}
		\hline
		\multicolumn{2}{|>{\centering}m{2.5cm}|}{\textbf{Space}} & \multicolumn{2}{|>{\centering}m{2.5cm}|}{\textbf{Experimental}} & \multicolumn{2}{|>{\centering}m{2.5cm}|}{\textbf{Model}} \\ \hline
		$x_0$ & 0 & $D$ & 0.2 & $\Delta t$ & 0.01 \\
		$x_1$ & 10 & $\kappa_1$ & 0.01 & $h_p$ & 0.1 \\
		$y_0$ & 0 & $\kappa_2$ & 0.5 & $h_a$ & 0.5 \\
		$y_1$ & 2 & $\rho$ & 0.1 & $I_0$ & 0.5 \\
		$z_0$ & 0 &  &  & $N$ & 200 \\	
		$z_1$ & 2 &  &  & $T$ & 5 \\
		$V$ & 40 &  &  & $\beta_u$ & 9.5 \\ 
		 &  &  &  & $\beta_l$ & 4 \\ \hline
	\end{tabular}
	\caption{Table of parameter values for test problem 4.}
	\label{tab:Params}
\end{table}
We can see good agreement between the hybrid method and the fully individual-based method throughout the domain, with the only discrepancy close to the left hand boundary at 0 caused by the difference between the PDE and individual-based methods due to moment closure. We compare our hybrid method to the fully individual-based method here, in contrast to the PDE solution used in test problems 1-3, due to the inaccuracy introduced in the PDE by the moment closure required for the second order reaction. 

\section{Error analysis} \label{sect:Error}
%Quantitative comparisons.
% Justification of the widths.
We have seen in Section \ref{sect:Results} that the solutions provided by the hybrid method visually match the mean-field solution. Within this section we quantify the difference between the solutions of these test problems. We compare the mass in the PDE and Brownian regions of the domain between the two methods. Separately we compare the density profile across the whole domain using the histogram distance error (HDE). We then proceed to investigate the dependence of the accuracy of the hybrid method on the two free simulation parameters ($\Deltat$ and $h_a$).

\subsection{Quantitative comparisons}

\begin{figure}[ht!]
	\begin{center}
	\subfigure[][]{
		\includegraphics[width=0.31\textwidth,trim={0pt 0pt 0pt 0pt},clip]{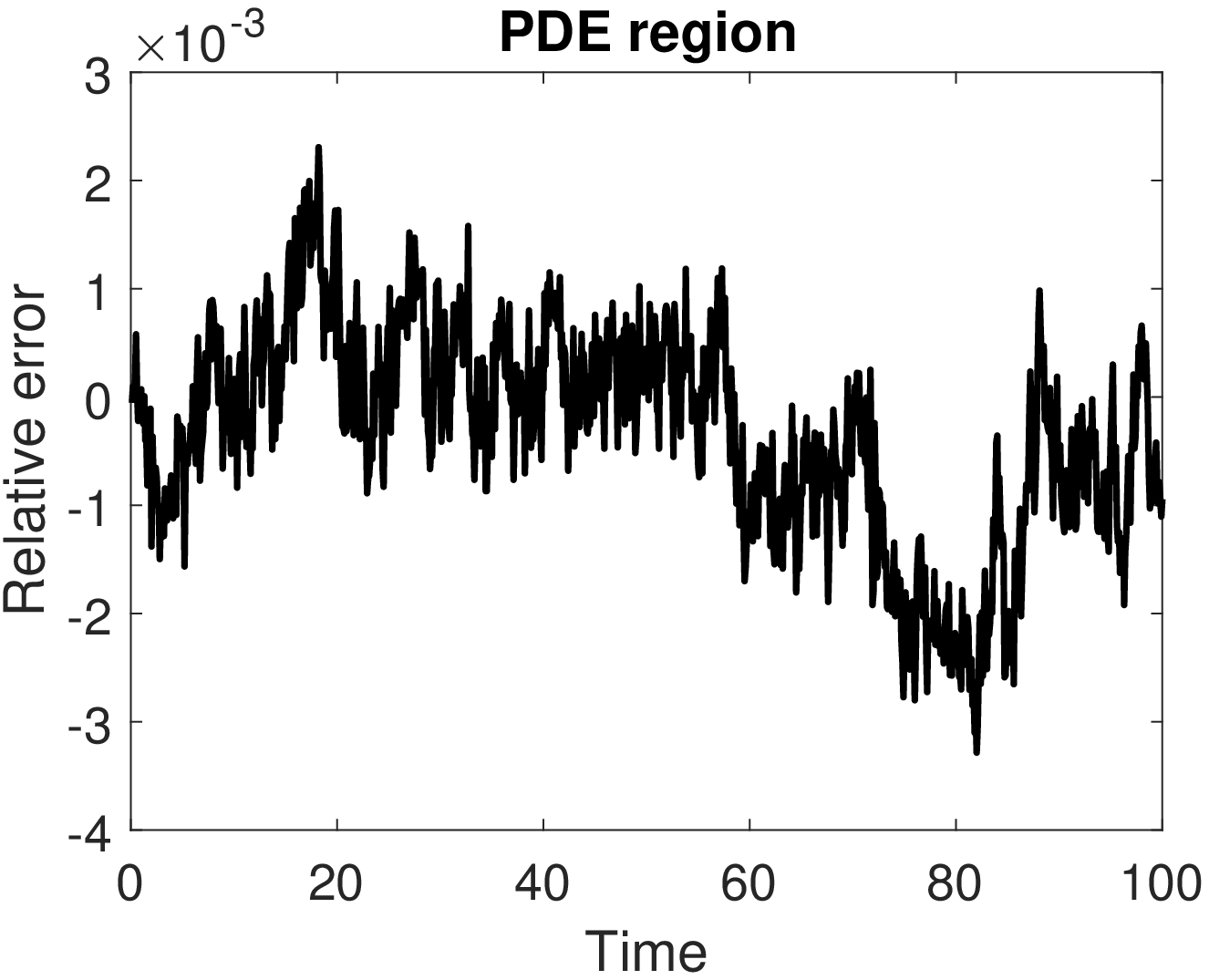}
		\label{fig:Errors_Unif_PDE}
	}
	\subfigure[][]{
		\includegraphics[width=0.31\textwidth,trim={0pt 0pt 0pt 0pt},clip]{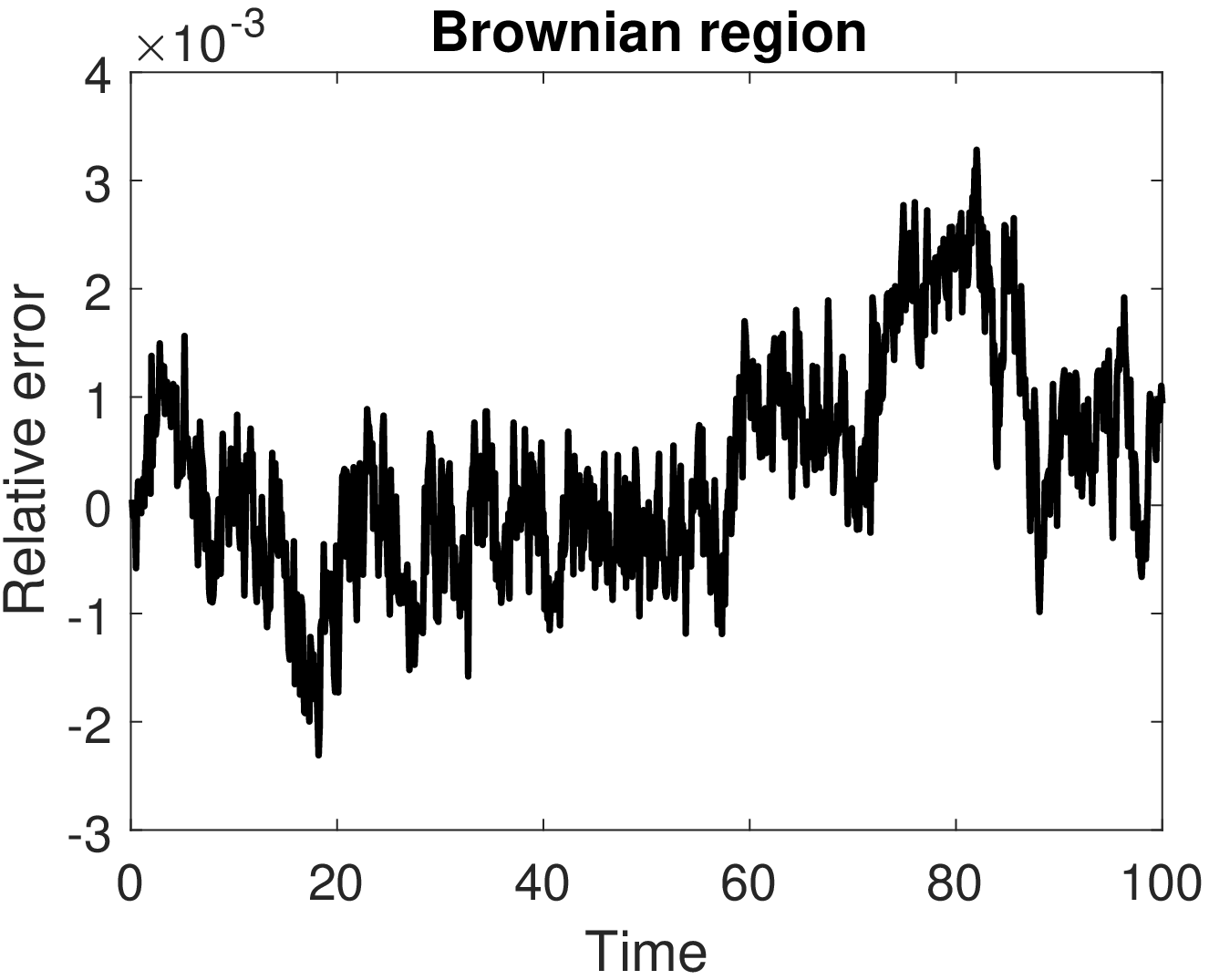}
		\label{fig:Errors_Unif_Brown}
	}
		\subfigure[][]{
		\includegraphics[width=0.31\textwidth,trim={0pt 0pt 0pt 0pt},clip]{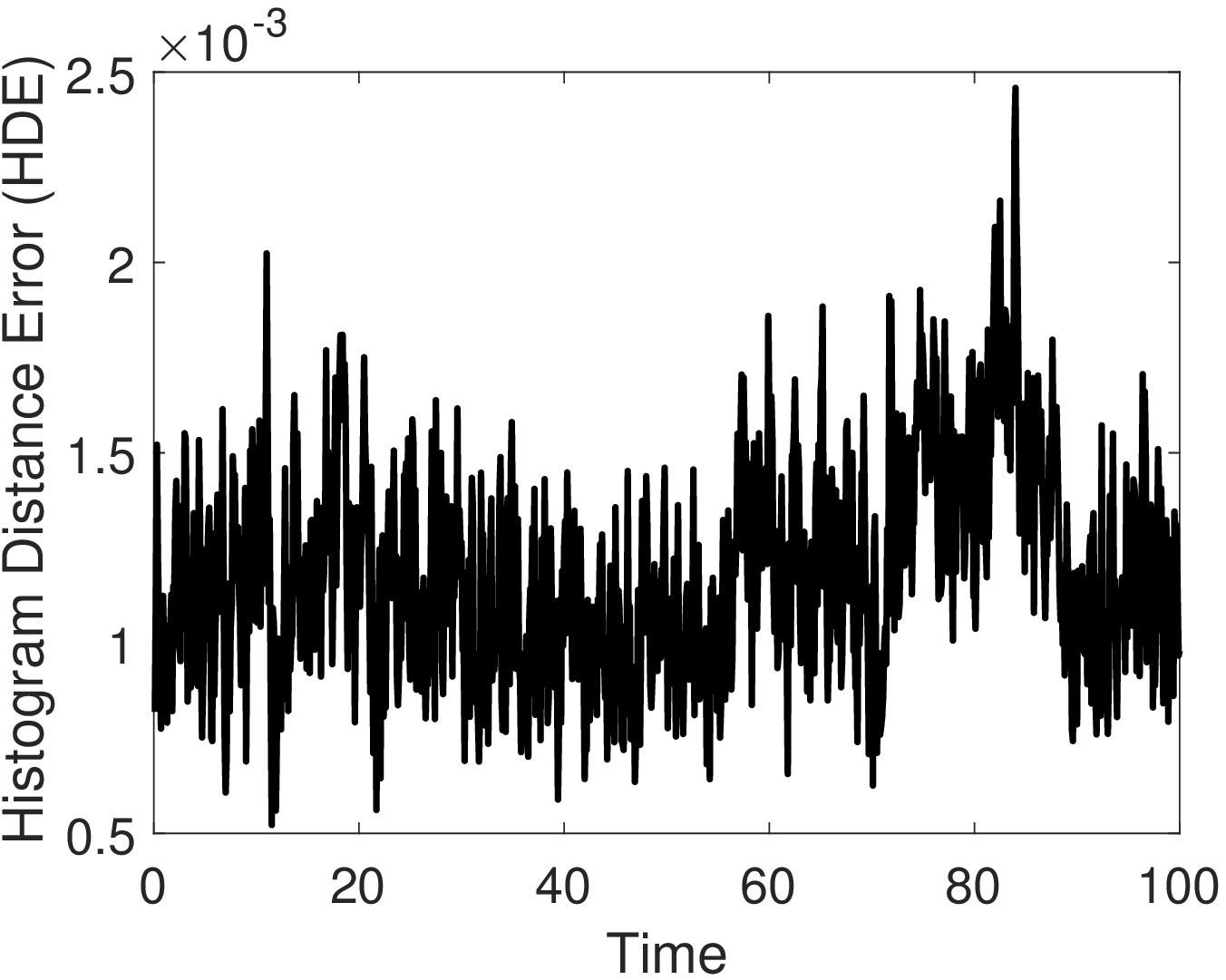}
		\label{fig:Errors_Unif_HDE}
	}
	\subfigure[][]{
		\includegraphics[width=0.31\textwidth,trim={0pt 0pt 0pt 0pt},clip]{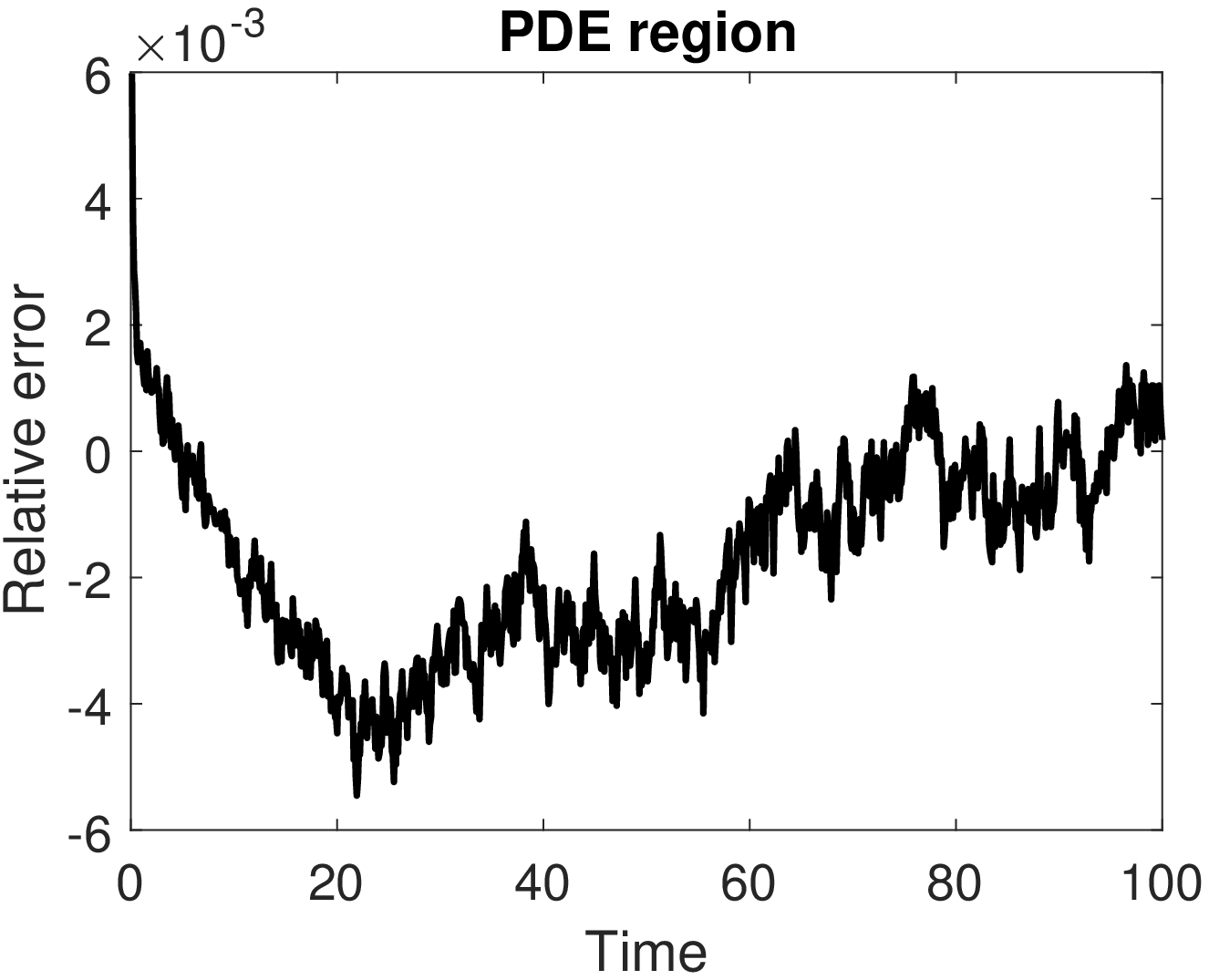}
		\label{fig:Errors_Left_PDE}
	}
	\subfigure[][]{
		\includegraphics[width=0.31\textwidth,trim={0pt 0pt 0pt 0pt},clip]{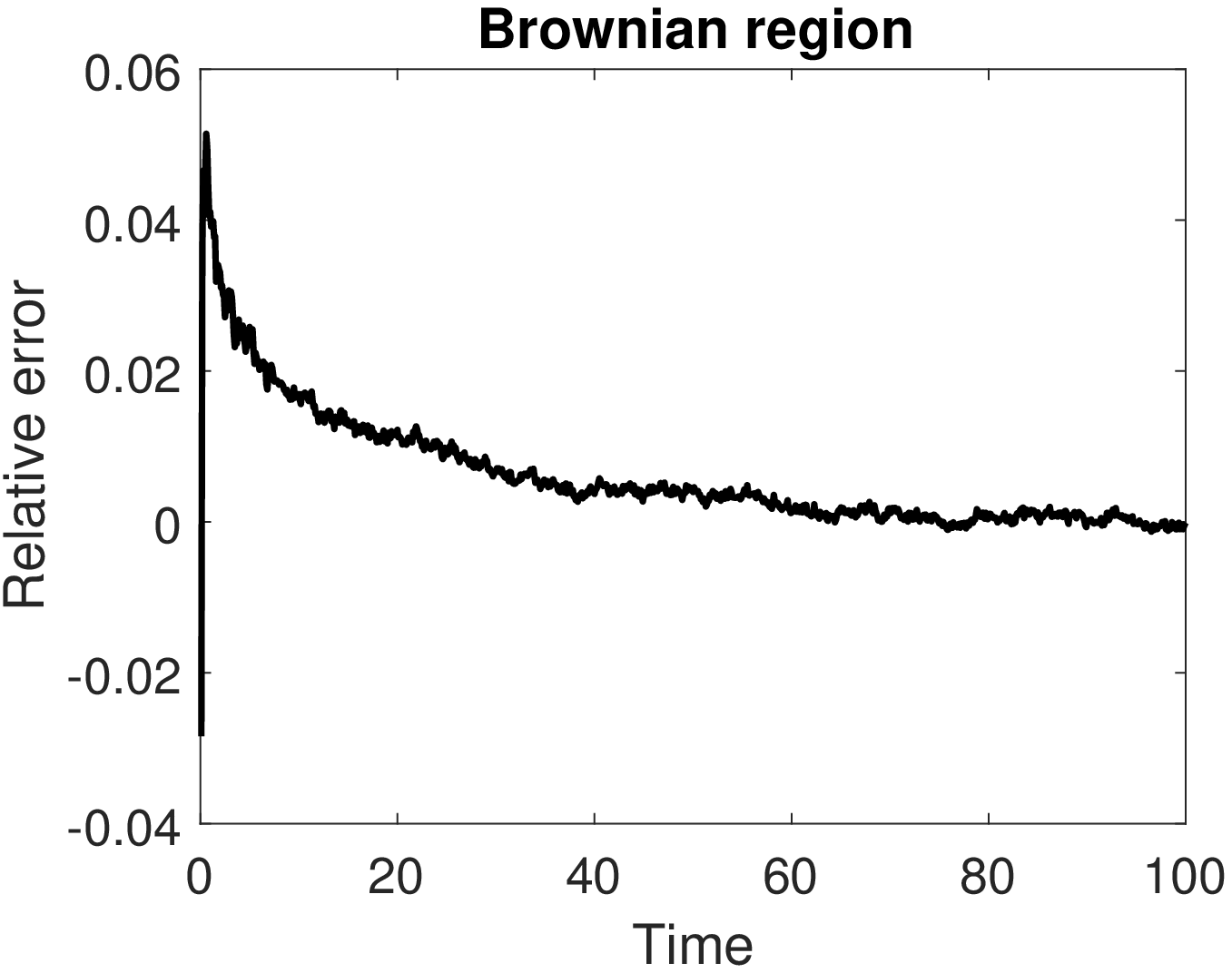}
		\label{fig:Errors_Left_Brown}
	}
		\subfigure[][]{
		\includegraphics[width=0.31\textwidth,trim={0pt 0pt 0pt 0pt},clip]{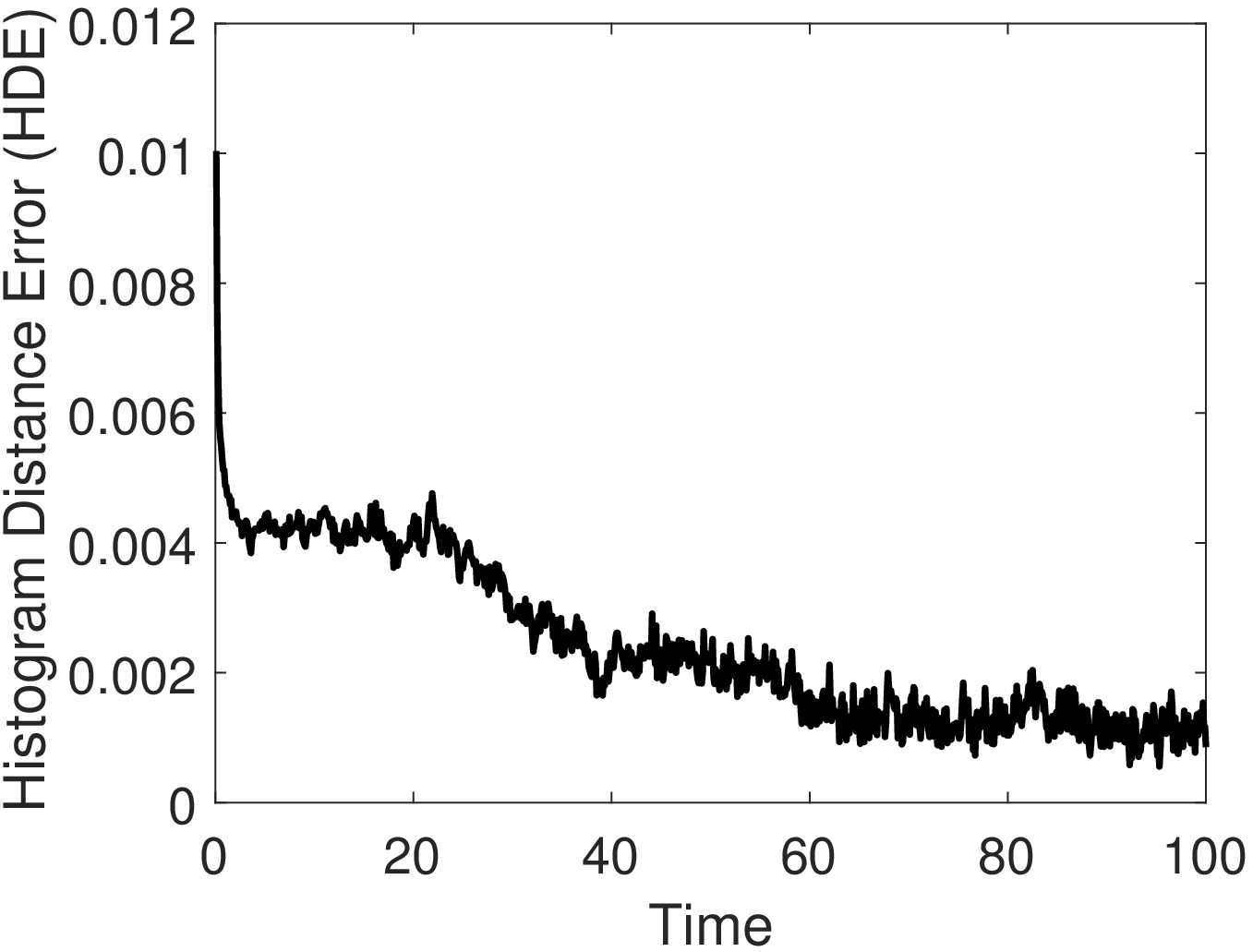}
		\label{fig:Errors_Left_HDE}
	}
	\subfigure[][]{
		\includegraphics[width=0.31\textwidth,trim={0pt 0pt 0pt 0pt},clip]{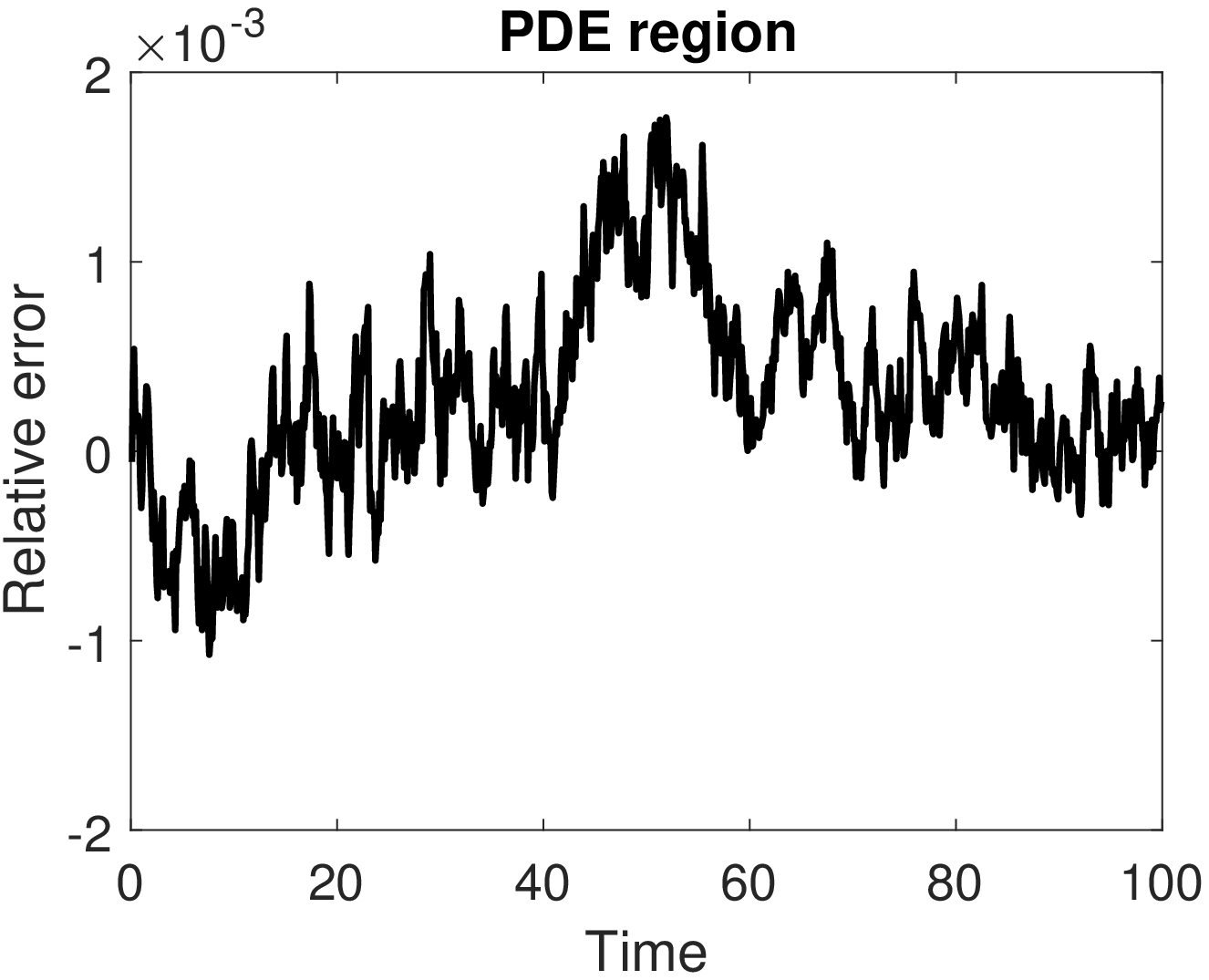}
		\label{fig:Errors_Morph_PDE}
	}
	\subfigure[][]{
		\includegraphics[width=0.31\textwidth,trim={0pt 0pt 0pt 0pt},clip]{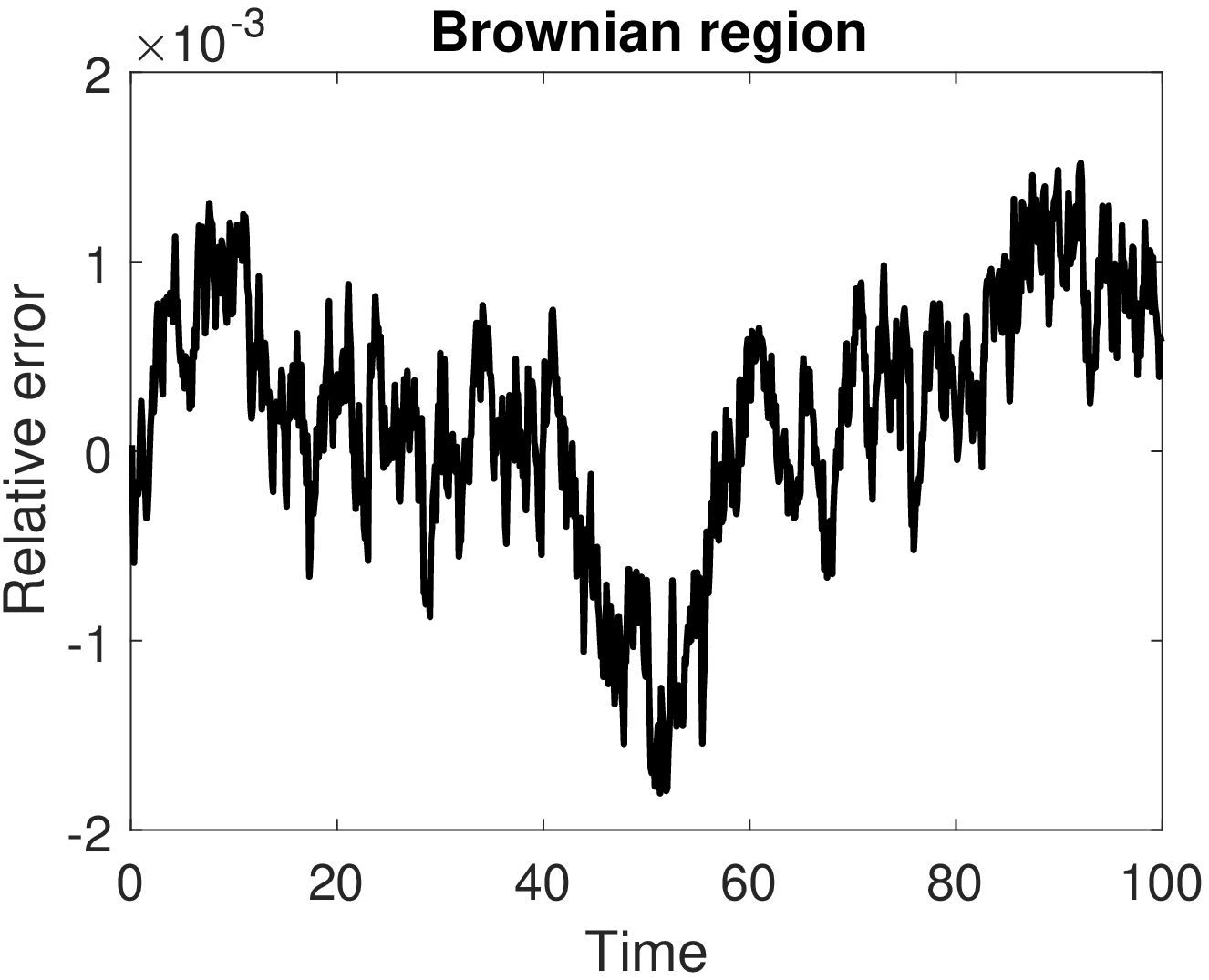}
		\label{fig:Errors_Morph_Brown}
	}
		\subfigure[][]{
		\includegraphics[width=0.31\textwidth,trim={0pt 0pt 0pt 0pt},clip]{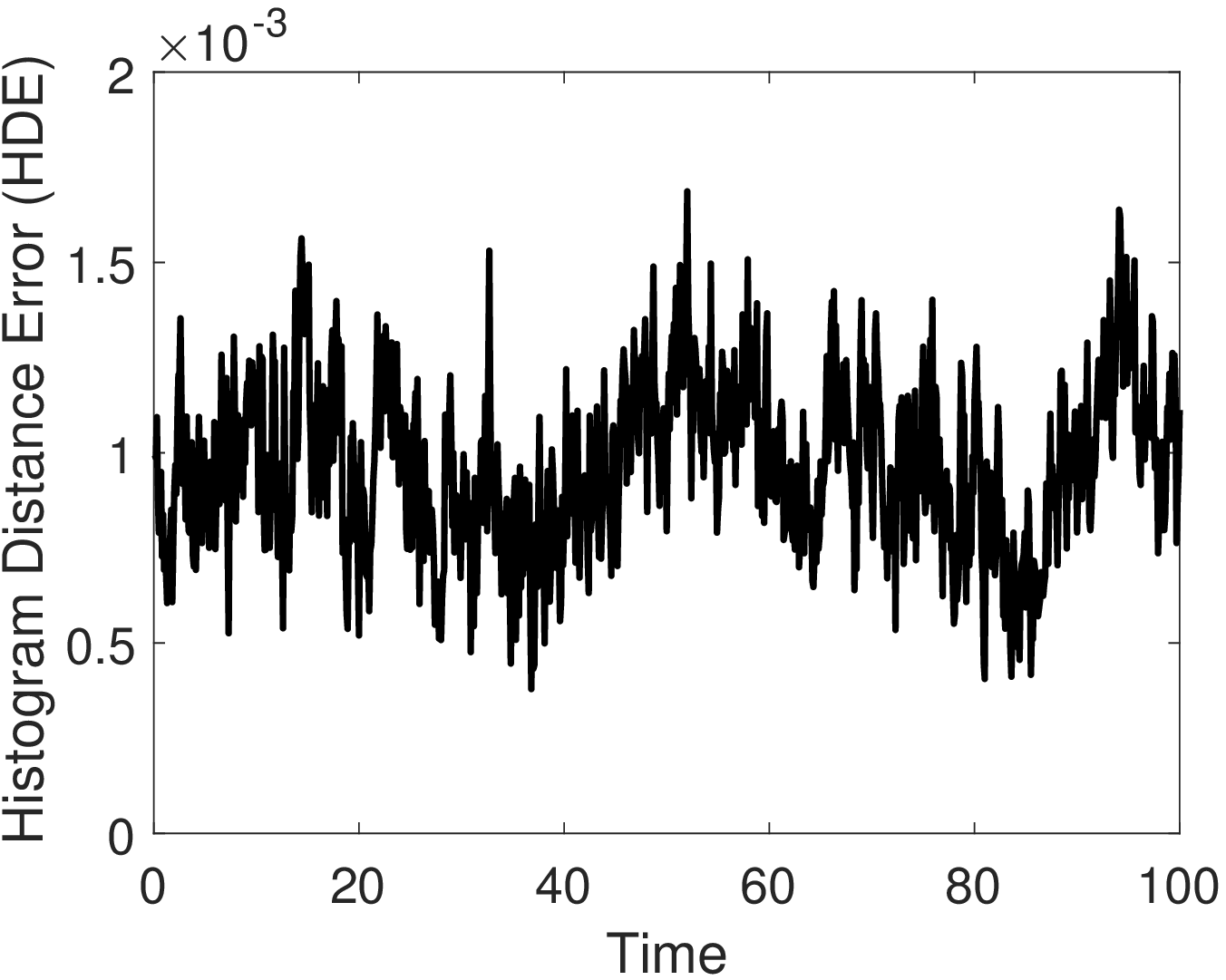}
		\label{fig:Errors_Morph_HDE}
	}
	\caption{Error plots for test problems 1 \subref{fig:Errors_Unif_PDE}-\subref{fig:Errors_Unif_HDE}, 2 \subref{fig:Errors_Left_PDE}-\subref{fig:Errors_Left_HDE} and 3 \subref{fig:Errors_Morph_PDE}-\subref{fig:Errors_Morph_HDE}. The first column contains the relative errors $(\smallsub{N}{HP}-\smallsub{N}{MP})/\smallsub{N}{MP}$, while the second column contains the relative errors $(\smallsub{N}{HB}-\smallsub{N}{MB})/\smallsub{N}{MB}$, and the third column contains the HDE comparison using equation \eqref{eqn:HDE}.}
	\label{fig:Errors}
\end{center}
\end{figure}

In order to evaluate the accuracy of the ARM for test problems 1,2 and 3, we compare its mean behaviour (averaged over $S=1000$ repeat simulations) to the mean-field model for which we compute the analytical solution across the entire domain $\Omega$, for each of our test problems. Figure \ref{fig:Errors} contains nine plots which demonstrate the error for the first three test problems above; \subref{fig:Errors_Unif_PDE}-\subref{fig:Errors_Unif_HDE} are for test problem 1, \subref{fig:Errors_Left_PDE}-\subref{fig:Errors_Left_HDE} are for test problem 2 and \subref{fig:Errors_Morph_PDE}-\subref{fig:Errors_Morph_HDE} are for test problem 3. The first and second columns show particle number comparisons between the hybrid and analytical solutions.  
Specifically, in the first column we compare
\begin{equation}                                                                                                                                                                                                                                                                                                                                                                                                                                                                                                                                                                                                                                                                                                                                                                                                                                                                                                                                                                                                                                                                                                                                            \smallsub{N}{MP}(t)=\int_{-1}^0p(x,t)\ud x,
 \end{equation}
the expected number of particles in $\smallsub{\Omega}{P}$ in the mean-field model to  
\begin{equation}
\smallsub{N}{HP}(t)=\frac{1}{S}\displaystyle\sum_{s=1}^S\int_{-1}^0 c^s(x,t)\ud x,
\end{equation}
the expected number of particles in $\smallsub{\Omega}{P}$ in the hybrid method. Here, as before, $p(x,t)$ represents the mean-field PDE solution at position $x$ at time $t$ and $c^s(x,t)$ represents the PDE part of the solution in the hybrid method for repeat $s$ of $S$.
Explicitly, we plot $(\smallsub{N}{HP}-\smallsub{N}{MP})/\smallsub{N}{MP}$, which shows no bias around zero for any of the three test problems. For completeness, in the second column we also compare 
\begin{equation}
\smallsub{N}{MB}(t)=\int_{0}^1p(x,t)\ \ud x,
 \end{equation}
the expected number of particles in $\smallsub{\Omega}{B}$ in the mean-field model to
\begin{equation}
\smallsub{N}{HB}(t)= \frac{1}{S}\displaystyle\sum_{s=1}^S \smallsub{N}{HB}^s(t),
\end{equation}
the expected number of particles in $\smallsub{\Omega}{B}$ in the hybrid methods. Here, $\smallsub{N}{HB}^s(t)$ is the number of particles in the Brownian region of the hybrid method at time $t$ for repeat $s$ of $S$.
Explicitly, we plot $(\smallsub{N}{HB}-\smallsub{N}{MB})/\smallsub{N}{MB}$, which again shows no bias around zero for any of the three test problems.

The last column of Figure \ref{fig:Errors} contains the histogram distance error (HDE), which is defined by 
\begin{equation}
\text{HDE}(t) = \frac{1}{2}\sum_{\ell=1}^{L}{\left|c_\ell^{H}(t)-c_\ell^{P}(t)\right|},
\label{eqn:HDE}
\end{equation}where $\ell$ indexes a common mesh on which the solutions are compared. $c_\ell^{H}(t)$ is the normalised solution of the hybrid method at mesh point $\ell$ and time $t$, and $c_\ell^{P}(t)$ is the normalised solution of the mean-field model at the same common mesh point and time, where 
$$\sum_{\ell=1}^L{c_\ell^{H}(t)}=\sum_{\ell=1}^L{c_\ell^{P}(t)}=1\quad\forall t\geq 0.$$
This ensures a value of the HDE between 0 and 1. Here, 0 means that the two solutions are exactly the same, and 1 corresponds to the two solutions having non-overlapping supports. All figures were produced using the same number of repeats ($S=1000$).

% \textcolor{red}{Figures \ref{fig:Errors_Unif_PDE} and \ref{fig:Errors_Unif_Brown} are compared against the theoretical numbers of particles for the uniform initial condition rather than the fully Brownian simulation. It should also be noted that the HDE for all particles initialised uniformly in the PDE domain (Figure \ref{fig:Errors_Left_HDE}) starts off high due to a discrepancy between discrete and continuous space. [[[[[MAY NOT NEED THIS PARAGRAPH ANY MORE]]]]}

In all cases, the relative errors between the mean-field and hybrid methods, in Figure \ref{fig:Errors}, are low with no discernible bias about zero. 
Similarly, all HDE plots in Figure \ref{fig:Errors} are low for the majority of the simulations. This demonstrates numerically that the hybrid scheme presented in this paper is correctly reproducing the behaviour of the Brownian model in the mean-field. These error plots confirm the visual concurrence shown in Figures \ref{fig:Plots_Unif}--\ref{fig:Plots_Morph}.

For the fourth test problem, we use a different error measurement due to the disparity between the mean-field PDE and individual-based systems. Consequently, we choose to compare the number of particles in the final compartment for both the hybrid method and the individual-based method. We motivate this in two ways. Firstly, using this measure of error, we are able to minimise the influence of the extra error caused by the difference between the mean-field PDE and the individual-based method. Secondly, several biological systems require detailed knowledge of the particle concentrations at the end of the domain. Apical growth of filamentous cells such as fungi \citep{goriely2008mmh} is such an example. If we define $\smallsub{N}{H}(t)$ to be the average number of particles in the region $(x_1-h_a,x_1)\times(y_0,y_1)\times(z_0,z_1)$ (where we recall that $\Omega = (x_0,x_1)\times(y_0,y_1)\times(z_0,z_1)$), when simulating the hybrid method at time $t$, and the quantity $\smallsub{N}{M}(t)$ to be the same for the fully microscopic simulation, we can obtain a measurement of error given by \begin{equation}
\smallsub{E}{Rel}(t) = \frac{\smallsub{N}{M}(t)-\smallsub{N}{H}(t)}{\smallsub{N}{M}(t)}. \label{eqn:Error_Second}
\end{equation}

\begin{figure}[ht!]
	\centering
	\includegraphics[width=0.8\textwidth]{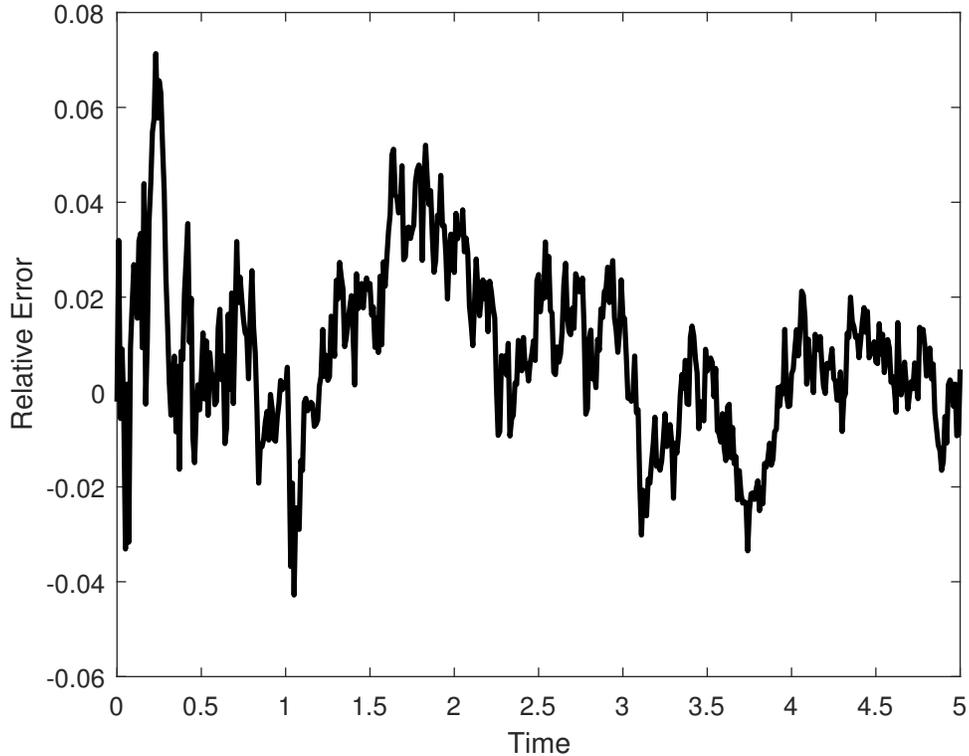}
	\caption{The error measurement given in equation \eqref{eqn:Error_Second} for the system simulated in Figure \ref{fig:Plots_Second}.}
	\label{fig:Error_Second}
\end{figure}
The relative error shows no long-term bias in either direction, and oscillates around zero, indicating a close agreement between our hybrid method and the ground truth individual-based method. The hybrid method completed its 1000 repeats in 485.5 seconds, while the fully individual-based method took 1047.4 seconds. 

\subsection{Parameter choice} \label{sect:Error_Parameter} 

Within the ARM, there are two free parameters that need to be chosen -- the width of the auxiliary regions $h_a$ and the time-step for the PDE and Brownian updates $\Deltat$. These need to be chosen so that the quantity $D\Deltat/h_a^2$ remains small enough that the particle numbers in the auxiliary regions do not become overly equilibrated between PDE/Brownian update steps. That is to say, if there is a gradient across the interface, $\Delta t$ should be small enough that the closed system of the two auxiliary regions should not reach steady state between PDE/Brownian update steps. 

In order to demonstrate why $D\Deltat/h_a^2$ must be small, we consider the evolution of particle numbers in the two auxiliary regions between PDE/Brownian update steps. We form an ODE for particle numbers in one of these boxes (using the fact that particle numbers are conserved between PDE/Brownian updates). 

Let $\nu_0$ be the (constant) number of particles in the two auxiliary regions combined, $\smallsub{M}{P}(t),\ \smallsub{M}{B}(t)$ be the mean number of particles in the PDE and Brownian auxiliary regions respectively at time $t$, and $\smallsub{\mu}{P},\ \smallsub{\mu}{B}$ be the number in the PDE and Brownian auxiliary regions respectively at time 0, which will represent the beginning of a time-step. Then, the equation for the mean number of particles in the PDE auxiliary region can be calculated from a simple probability master equation as
\begin{equation*}
\ordder{\smallsub{M}{P}}{t} = d\smallsub{M}{B} - d\smallsub{M}{P} = d(\nu_0-\smallsub{M}{P})-d\smallsub{M}{P} = d\nu_0 -2d\smallsub{M}{P},
\end{equation*} 
where we recall that $d$ is the jumping rate between the two auxiliary regions and is linked to the diffusion constant, $D$, via equation \eqref{eqn:link_d_D}. Solving this ODE gives 
\begin{equation}
\smallsub{M}{P}(t) = \frac{1}{2}\left[\nu_0-(\nu_0-2\smallsub{\mu}{P})e^{-2dt}\right].
\label{eqn:ODE_MP_Sol}
\end{equation}  
Assuming a small time-step, $\Deltat$, we can approximate $\smallsub{M}{P}(\Deltat)$, the number of particles after a time-step has occurred, by Taylor expanding equation \eqref{eqn:ODE_MP_Sol} to first order:
\begin{align*}
\smallsub{M}{P}(\Deltat) &= \smallsub{M}{P}(0) + \Deltat \smallsub{M}{P}'(0) + o(\Deltat)\\
&\approx\frac{1}{2}\left[\nu_0-(\nu_0-2\smallsub{\mu}{P})\right] + \frac{\Deltat}{2}\left[2d(\nu_0-2\smallsub{\mu}{P})\right]\\
&=(1-2d\Deltat)\smallsub{\mu}{P} + d\Deltat\nu_0.
\end{align*} 
Fixing the value of $D$ and using equation \eqref{eqn:link_d_D}, we find that $$\smallsub{M}{P}(\Deltat)\approx\left(1-\frac{2D\Deltat}{h_a^2}\right)\smallsub{\mu}{P}+\frac{D\Deltat}{h_a^2}\nu_0.$$
We require the change in the number of particles over the small time-step to be small, and so would like $\smallsub{M}{P}(\Deltat) \approx \mu_P$. Thus we need to choose our parameters such that the quantity $D\Deltat/h_a^2$ small. This elucidates an important relationships between the fixed and free parameters of the model. If the diffusion coefficient is large then we must choose a small update time-step or a larger auxiliary region length to compensate.

\begin{figure}[ht!]
\centering
\includegraphics[width=0.6\textwidth]{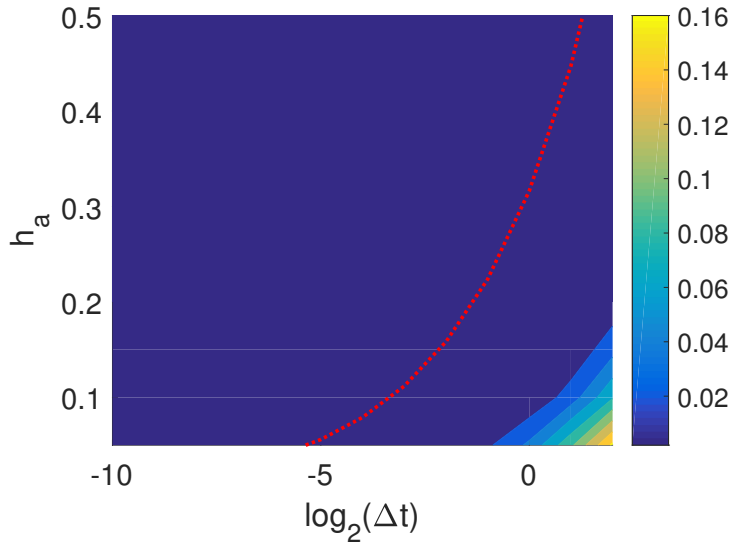}
\caption{A contour plot for the HDE at time $T=10$ for a series of simulations initialised with all particles uniformly distributed within the PDE region of the domain. Dark colours indicate low error. The red line is the representative contour $D\Deltat/h_a^2=1/2$. Here, $D=0.05$ and all simulations are averaged over $100$ repeats.}
\label{fig:HDE_Error}
\end{figure} 
Figure \ref{fig:HDE_Error} shows that a large region of the $\Deltat-h_a$ space has a very low histogram distance error, meaning that our method is robust to parameter change, and only breaks down once the value of $D\Deltat/h_a^2$ becomes very large. The plot also shows that, given any choice of the width of the auxiliary regions, $h_a$, there is a value for the time-step, $\Deltat$, which will give a low level of error. Also, depending on our choice of $\Delta t$, we can adjust $h_a$ to make the simulation more accurate.

\section{Discussion} \label{sect:Discussion}
%Developing this hybrid is important because of application areas (see Franz).
%Method of Franz et al works, but is of limited scope.
%We have come up with a new method which we have shown to be robust across a range of different simulations and parameter values. 
%We have carried out error analysis on the method and specified broad regions of parameter space in which our method will work.
%
%Future work
%- Adaptive interfaces,
%- Variance/consider an overlap region or SPDE model.
%- Applications - talk about calcium puffs, but also about some other examples borrowed from introductions of other papers.
We have presented a new spatially coupled hybrid method for coupling a Brownian dynamics representation of a reaction-diffusion system to its corresponding mean-field PDE description. By bridging the gap in spatial scales with intermediate auxiliary regions, we have produced an algorithm that is not only accurate, but is also robust to the choice of the free parameters within the problem, namely the width of the auxiliary regions, $h_a$, and the fixed time-step, $\Delta t$ used to update both the PDE and Brownian dynamics. This is in direct contrast to a previously presented PDE-to-Brownian hybrid, which we demonstrated to be extremely parameter-sensitive. In order to make the ARM even more robust, applicable and efficient, we now discuss several areas for possible extension, which will be addressed in future works.

In the interests of completeness we should point out that, as with the pseudo-compartment method of \citet{yates2015pcm}, the auxiliary region method requires that the mass in the PDE auxiliary region $\Omega_\text{PA}$ be sufficient for a step function, corresponding to the mass of a particle, to be removed uniformly from across the auxiliary region. \label{xr:Negative_Particles_Prior} This will lead to difficulties in situations in which particle numbers are low around the interface. Arguably though, we should not employ such hybrid methods in situations for which particle density is low around the interface as the PDE will be a poor model of the true stochastic, microscopic dynamics in these regions. A possible solution to this inconvenience, is the incorporation of an adaptive interface, which we have employed in test problem 4. Such interfaces evolve with the simulation dynamics, ensuring the appropriate model is used for the corresponding particle density \citep{robinson2014atr}. 

A related issue is that of multiple interfaces. Multiple interfaces will allow the efficient simulation of stochastic reaction diffusion systems in which multiple regions of high and low concentration are expected. Such patterns will require interfaces to be dynamic in number and transient in nature. Although we have not implemented such interfaces in this work we expect it to be a relatively straightforward extension. While we have presented an example in which the system is simulated in a cuboid with a planar interface (test problem 4), non-planar interfaces, such as those which have corners or are curved, and complex domain geometries, present deeper challenges that we hope to address in a future publication. 

% Firstly, for a dynamical biological system, the regions of space in which there are low copy numbers may change over time, meaning that prescribing an interface location at the beginning of simulation and leaving it there throughout is unlikely to be efficient. In order to address this problem, a dynamic interface which moves based on local particle numbers could be implemented \citep{robinson2014atr}. Further to this, many biological objects change in size dynamically. The study of cell migration in the developing embryo is a classic example, in which the domain grows during the process of cellular organisation. In order to make our hybrid method as applicable as possible we will also extend it to domains which change in size over time.

Failing to maintain stochastic variation is a problem which is common amongst many spatially coupled hybrid methods. As a result of the deterministic nature of the PDE, the noise in the Brownian dynamics region of the domain is damped in comparison to the fully microscopic model (see Figure \ref{fig:Vars}). In the literature, two approaches have been used in order to rectify this. The first is an overlap region, which has been employed in several papers \citep{harrison2016hac,franz2012mrd,flekkoy2001cpf}. These methods introduce a region of space which lies in the intersection of the two domains. In these regions, mass is simultaneously represented using both scales of description. The second is to replace the deterministic PDE with an appropriately chosen stochastic partial differential equation (SPDE). \citet{alexander2002ars} consider such a coupling and demonstrate they can indeed fix the discrepancy by using an SPDE as their continuum macro-scale model. We will address both the use of SPDEs and overlap regions (in which the region between the PDE and the Brownian dynamics regions is simulated using a purely compartment-based method) in forthcoming work. 

The auxiliary region method provides a simple yet accurate method to couple an individual Brownian dynamics representation of a reaction-diffusion system to a corresponding PDE representation. Our hybrid algorithm will be of particular interest to researchers modelling reaction-diffusion systems whose concentrations vary significantly across the spatial domain. By reducing the computational expense of simulations, the ARM will facilitate the investigation of stochastic effects in such systems, in some cases, making the difference between being able to interrogate the system and not. In particular, we suggest that our method will be useful for the investigation of stochastic Turing patterns \citep{flegg2016srk}, Fisher waves \citep{breuer1995mls,breuer1994few}, oscillatory dynamics \citep{hoffmann2014omp} and excitatory dynamics \citep{gerisch2013mar} with applications in embryogenesis \citep{mort2016rdm}, intracellular dynamics \citep{khan2011scd} and pattern formation \citep{flegg2016srk} amongst others. It may also be worthwhile to interface the methods presented here with commonly used Brownian dynamics simulation software packages such as Smoldyn \citep{andrews2004ssc}.

\section*{Acknowledgements}
Cameron Smith is supported by a scholarship from the EPSRC Centre for Doctoral Training in Statistical Applied Mathematics at Bath (SAMBa), under the project EP/L015684/1. Dr Christian Yates would like to thank the CMB/CNCB preprint club for constructive and helpful comments on a preprint of this paper. 

\begin{appendices}

\section{Comparing to PDE-assisted Brownian dynamics} \label{sect:Appendix}

\begin{figure}[ht!]
	\begin{center} 
	\subfigure[][]{
		\includegraphics[width=0.31\textwidth]{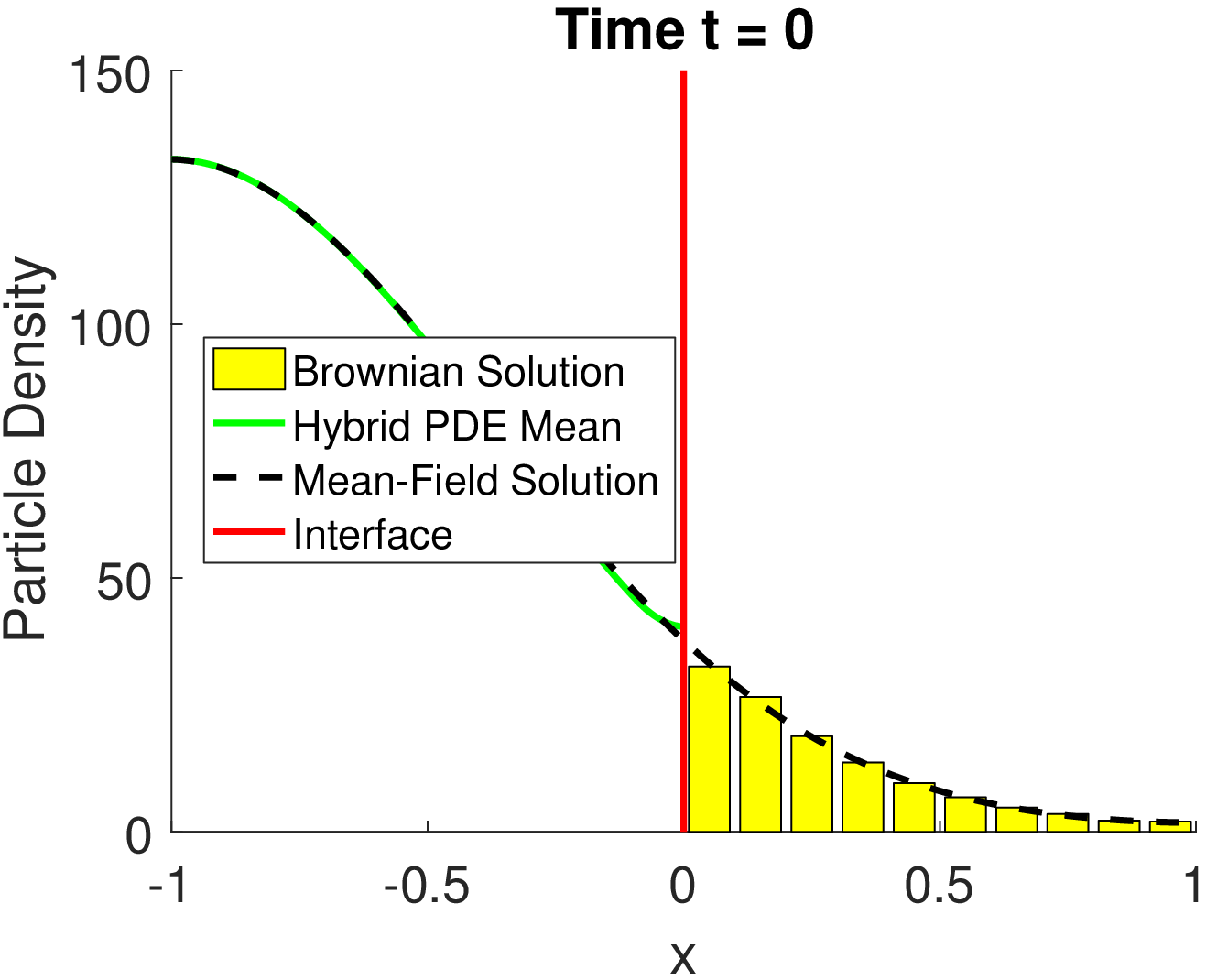}
		\label{fig:Plots_Comp_1}
	}
	\subfigure[][]{
		\includegraphics[width=0.31\textwidth]{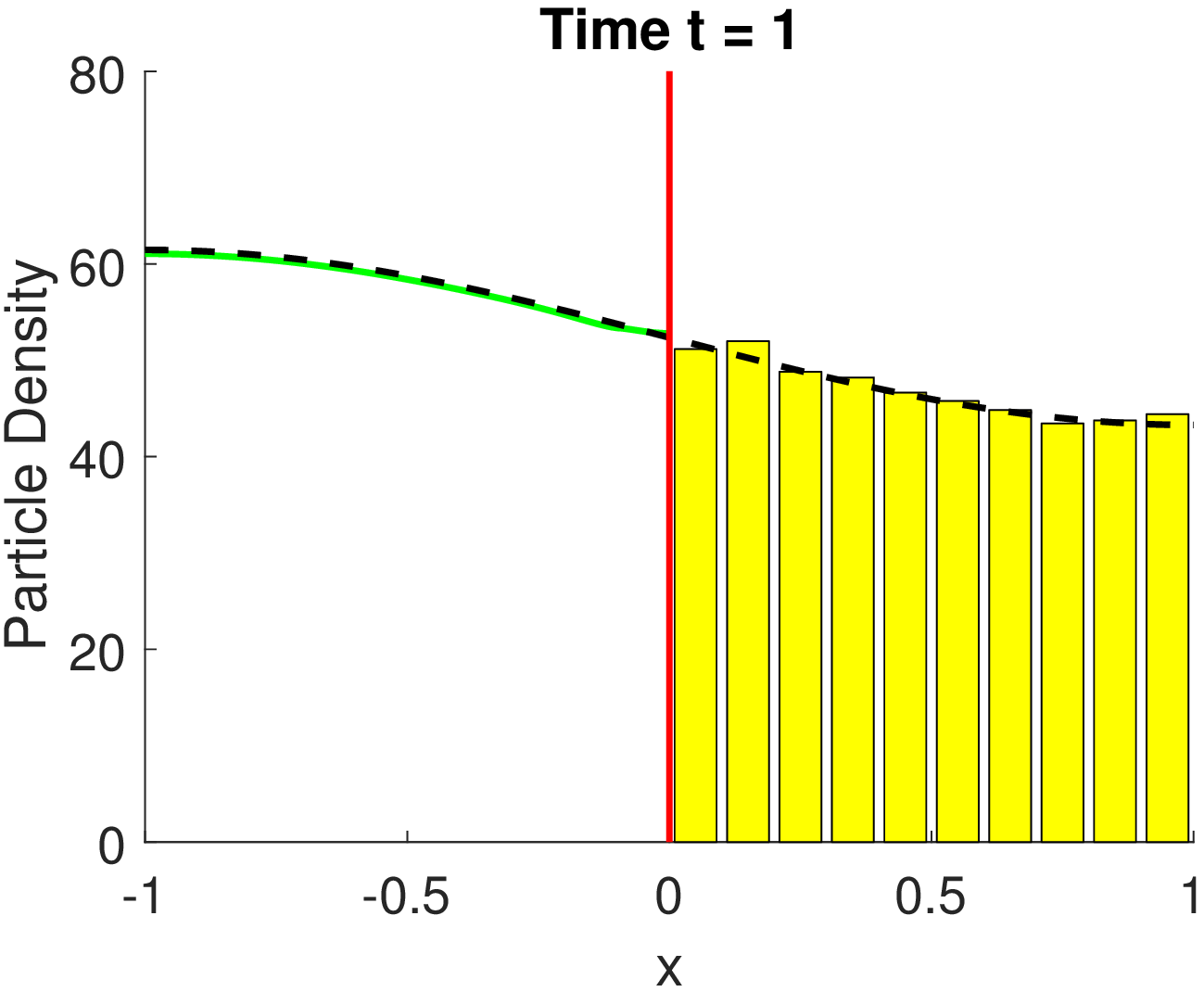}
		\label{fig:Plots_Comp_2}
	}
	\subfigure[][]{
		\includegraphics[width=0.31\textwidth]{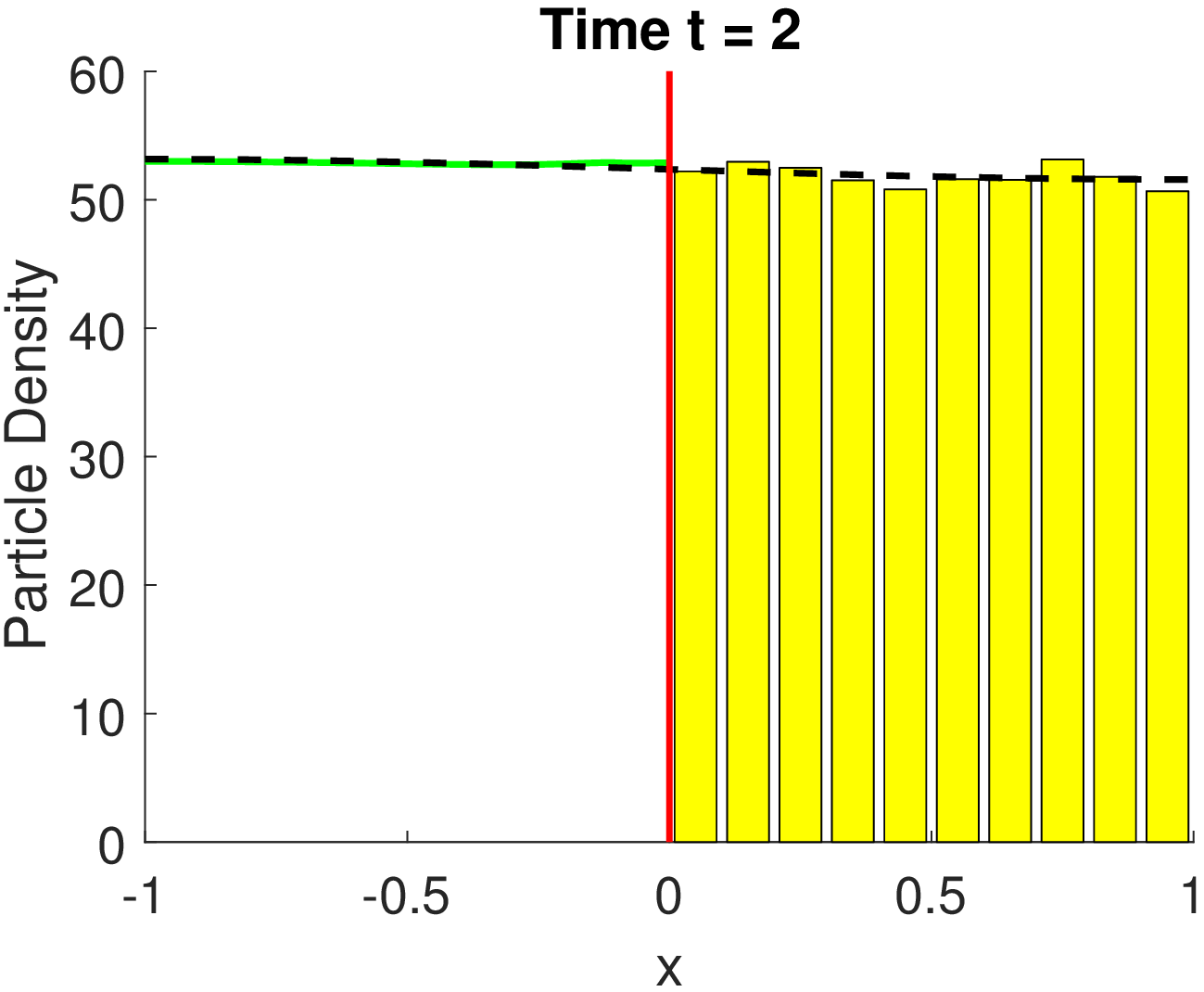}
		\label{fig:Plots_Comp_3}
	}
	\caption{The evolution of the system corresponding to the reproduced figure from \citet{franz2012mrd} (Figure \ref{fig:Plots_franz}), simulated using the ARM, at times \subref{fig:Plots_Comp_1} 0.2, \subref{fig:Plots_Comp_2} 1 and \subref{fig:Plots_Comp_3} 2. The colours and parameters are the same as in Figure \ref{fig:Plots_franz}, with the auxiliary region size being $h_a=0.1$.}
	\label{fig:Plots_Comp}
\end{center}
\end{figure}

Within this section, we apply the same parameter values as used in Section \ref{sect:Previous} in order to demonstrate that the ARM can accurately simulate the problem that PDE assisted Brownian dynamics could not (see Figure \ref{fig:Plots_franz}). Recall, that we use $\smallsub{\Omega}{P}=(-1,0)$, $\smallsub{\Omega}{B} = (0,1)$ with the interface placed at $I=0$. The only additional parameter that is to be defined is the auxiliary region width, which we set here to be $h_a=0.1$. The results can be seen in Figure \ref{fig:Plots_Comp}.

As can be seen from this figure, the agreement between the mean-field and hybrid solutions is much closer than that of the PDE assisted Brownian dynamics \citep{franz2012mrd}. This indicates an improvement over the previous method. We also present the error plots which are described in Section \ref{sect:Error} --- namely the relative errors in particle numbers and the histogram distance errors.

\begin{figure}[ht!]
	\begin{center} 
	\subfigure[][]{
		\includegraphics[width=0.31\textwidth]{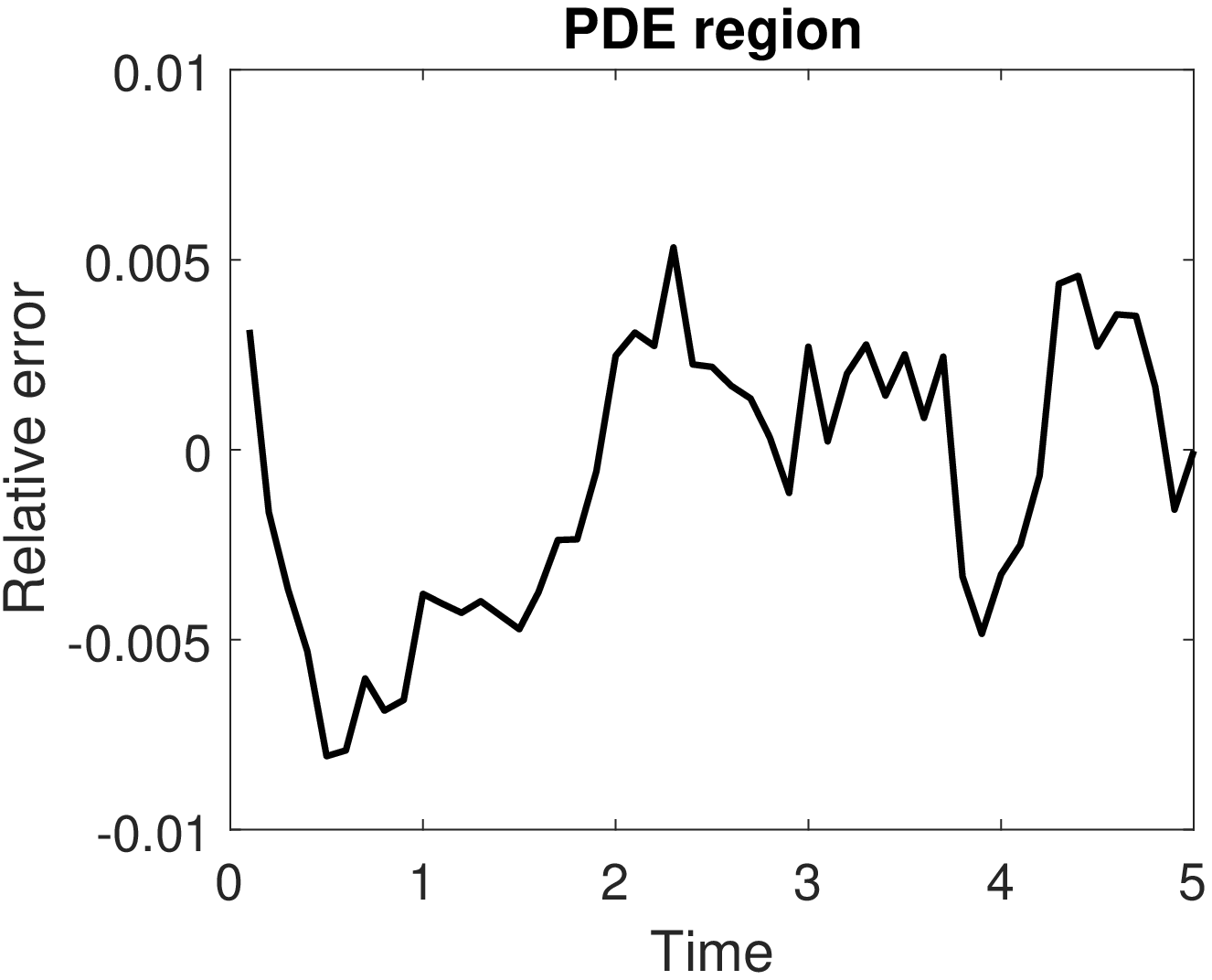}
		\label{fig:Errors_Comp_PDE}
	}
	\subfigure[][]{
		\includegraphics[width=0.31\textwidth]{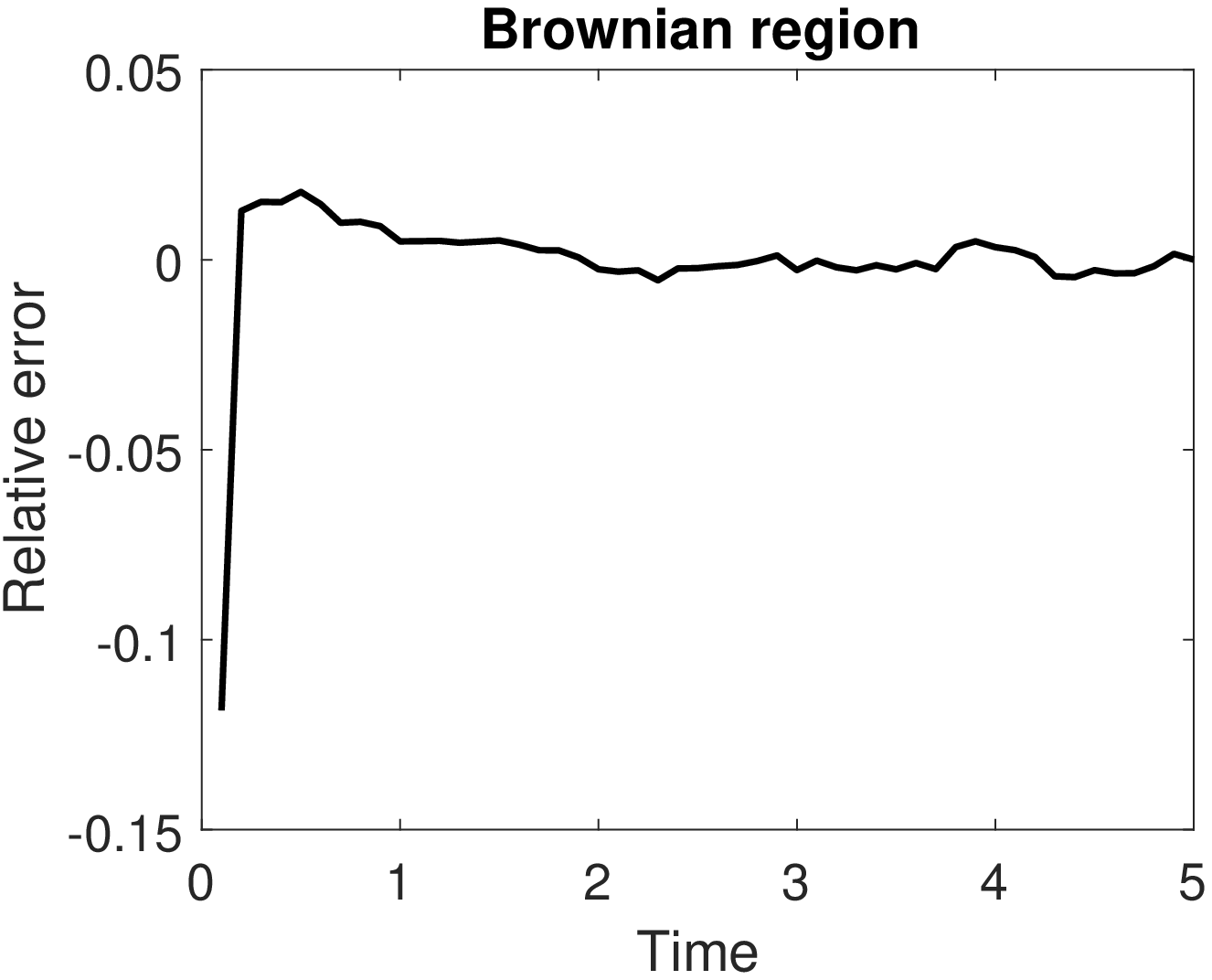}
		\label{fig:Errors_Comp_Bro}
	}
	\subfigure[][]{
		\includegraphics[width=0.31\textwidth]{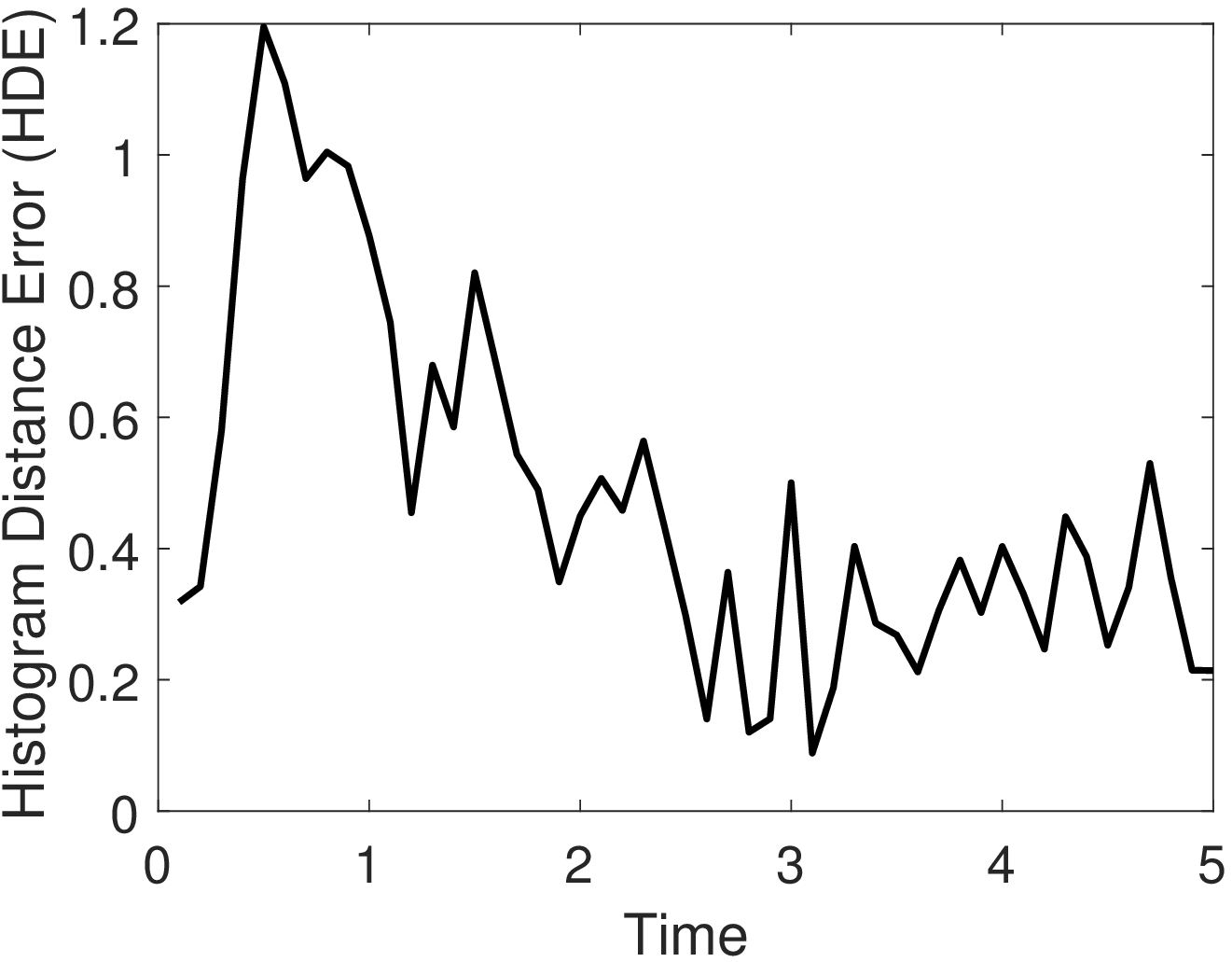}
		\label{fig:Errors_Comp_HDE}
	}
	\caption{The error plots for the example in Figure \ref{fig:Plots_Comp}. The first two panels show relative errors in particle numbers in the  \subref{fig:Errors_Comp_PDE} PDE and \subref{fig:Errors_Comp_Bro} Brownian dynamics subdomains. The histogram distance error is displayed in \subref{fig:Errors_Comp_HDE}.}
	\label{fig:Errors_Comp}
\end{center}
\end{figure}

Once again, the relative error plots (Figures \ref{fig:Errors_Comp}\subref{fig:Errors_Comp_PDE}-\subref{fig:Errors_Comp_Bro}) appear to show no long-term bias in either direction and the histogram distance error (Figure \ref{fig:Errors_Comp_HDE}) is small. 

\end{appendices}

\bibliographystyle{plainnat}
\bibliography{../arm}

\end{document}